%% file: main.tex
\documentclass[12pt,twoside,english]{article}
\usepackage[T1]{fontenc} 			
\usepackage{amssymb}
\usepackage{a4wide}
\usepackage{graphicx}
\usepackage{fancyhdr}
\usepackage{amsmath, amssymb}
\usepackage{subfigure}
\usepackage{setspace}
\usepackage{verbatim} 
\usepackage[english]{babel}
\usepackage{caption}
\usepackage{float}
\usepackage{booktabs}

\usepackage{siunitx}
\usepackage{xfrac}

\usepackage{multirow}

\makeatletter
\usepackage{hyperref}

\hypersetup{
	colorlinks = false,
	linktocpage = true,
	allbordercolors = {0.8 0.8 0.8}
}

\addto\captionsenglish{}
\addto\captionsenglish{}
\addto\extrasenglish{}
\addto\extrasenglish{}
\addto\extrasenglish{}
\addto\extrasenglish{}
\addto\extrasenglish{}
\addto\extrasenglish{}
\addto\extrasenglish{}

\newcommand{\lyxaddress}[1]{
\par {\raggedright #1
\vspace{1.4em}
\noindent\par}
}

\numberwithin{equation}{section}

\newcommand{\vek}[1]{\mathchoice{\displaystyle\boldsymbol#1}
{\textstyle\boldsymbol#1}{\scriptstyle\boldsymbol#1}
{\scriptscriptstyle\boldsymbol#1}}
\newcommand{\mat}[1]{\mathchoice{\displaystyle\mathbf#1}
{\textstyle\mathbf#1}{\scriptstyle\mathbf#1}
{\scriptscriptstyle\mathbf#1}}

\newcommand{\ScalProd}{\boldsymbol{\cdot}}
\newcommand{\FrobProd}{\boldsymbol{:}}
\newcommand{\jumpl}{[\![}
\newcommand{\jumpr}{]\!]}

\makeatother

\graphicspath{{FiguresNew/}}

\begin{document}

\title{Higher-order, mixed-hybrid finite elements for Kirchhoff--Love shells}

\author{Jonas Neumeyer, Michael Wolfgang Kaiser, Thomas-Peter Fries}
\maketitle

\lyxaddress{\begin{center}
Institute of Structural Analysis\\
Graz University of Technology\\
Lessingstr. 25/II, 8010 Graz, Austria\\
\texttt{www.ifb.tugraz.at}\\
\texttt{jonas.neumeyer@tugraz.at}
\end{center}}

\vspace{-0.2cm}
\begin{abstract}
\input{Abstract}\\
Keywords: \input{Keywords}
\end{abstract}
\newpage{}\tableofcontents{}\newpage{}

\input{Body}

\section*{Acknowledgments}
\input{Acknowledgments}
\appendix
\input{Appendix}

\bibliographystyle{schanz}
\addcontentsline{toc}{section}{\refname}\bibliography{NeumeyerRefs}
 
\end{document}

%% file: Abstract.tex
A novel mixed-hybrid method for Kirchhoff--Love shells is proposed that enables the use of classical, possibly higher-order Lagrange elements in numerical analyses. In contrast to purely displacement-based formulations that require higher continuity of shape functions as in IGA, the \emph{mixed} formulation features displacements and moments as primary unknowns. Thereby the continuity requirements are reduced, allowing equal-order interpolations of the displacements and moments. \emph{Hybridization} enables an element-wise static condensation of the degrees of freedom related to the moments, at the price of introducing (significantly less) rotational degrees of freedom acting as Lagrange multipliers to weakly enforce the continuity of tangential moments along element edges. The mixed model is formulated coordinate-free based on the Tangential Differential Calculus, making it applicable for explicitly and implicitly defined shell geometries. All mechanically relevant boundary conditions are considered. Numerical results confirm optimal higher-order convergence rates whenever the mechanical setup allows for sufficiently smooth solutions; new benchmark test cases of this type are proposed.

%% file: Keywords.tex
Kirchhoff--Love shell, Surface FEM, PDEs on manifolds, higher-order convergence studies, mixed finite elements, hybrid methods

%% file: Body.tex
\section{Introduction}\label{sec: Introduction}

For simulations in structural mechanics, dimensional reduction is often an essential step towards efficient solutions through the reduction of complexity and number of degrees of freedoms involved in the analysis. For example, a thin-walled structure with two dominating dimensions may be modeled with focus on the mid-surface. Thin, \emph{curved} structures in the three-dimensional space are usually called \emph{shells}. The applications of shells are versatile and relevant in almost every field of engineering, such as civil, biomedical, mechanical, and aerospace engineering. An overview of classical shell theory may be found in \cite{Bischoff_2017a, Calladine_1983a, Chapelle_2011a, Zienkiewicz_2014a, Basar_1985a, Zingoni_2017a}.

The classical theory of shells is based on parametrizations using (local) curvilinear coordinates, the shell surface is then defined explicitly. Alternatively, geometric quantities and surface differential operators may also be defined with respect to a (global) Cartesian coordinate system, thus applying to explicitly (via parametrizations) \emph{and} implicitly (via level sets) defined shell geometries. This is, for example, the case in Tangential Differential Calculus (TDC), see \cite{Delfour_2011a, Dziuk_2013a} for further details. Using coordinate-free formulations of shell models in terms of the TDC may be seen as a modern and more general approach and is also done herein. Displacement-based shell models in the frame of the TDC may be found in \cite{Schoellhammer_2019a} for the Kirchhoff--Love shell and in \cite{Schoellhammer_2019b} for the Reissner--Mindlin shell; for the case of pure membranes, see \cite{Hansbo_2014a, Fries_2020a}.

A significant distinction for shell models is the consideration of transverse shear deformations. The Reissner--Mindlin shell theory \cite{Reissner_1945a, Mindlin_1951a} considers these shear deformations and is applicable for thin to moderately thick shells. The governing equations are usually a system of two (vector-valued) partial differential equations (PDEs), with the displacement of the middle surface and the rotation of the normal vector as unknowns. Due to the fact that the PDEs only involve up to second-order derivatives, the FEM analysis is straightforward as $C^0$-continuous shape functions, e.g., furnished by standard Lagrange elements, are permitted. A significant disadvantage of Reissner--Mindlin shells is the presence of transverse shear locking effects for thinner shells. Neutralizing those effects requires special treatment, see \cite{Echter_2013a, Bischoff_2017a, Zou_2020a} for further details. In contrast, the Kirchhoff--Love shell theory  \cite{Kirchhoff_1850a, Love_1888a} neglects shear deformation, applies to thin shells, and avoids this type of locking. Its numerical implementation, however, is more challenging due to the fact that, in the displacement-based formulation, up to fourth-order derivatives in the governing equations are present, leading to increased continuity requirements in the FEM analysis as at least $C^1$-continuous shape functions are required.

Isogeometric analysis (IGA) is a viable option to meet those increased continuity requirements. Introduced in \cite{Hughes_2005a}, IGA uses non-uniform rational B-splines (NURBS), mainly known from Computer-Aided Design (CAD), as shape functions, naturally featuring higher-order continuity. Strictly speaking, this is only true for shell geometries that are defined by a \emph{single} NURBS-patch. Those typically feature two opposite sides and corner nodes, resulting from a map of a rectangular parameter space (similar to the map of one quadrilateral reference element to some surface element in the classical FEM). For more general shell geometries, however, multiple patches may be needed and, across the patches, the shape functions only feature $C^0$-continuity, again requiring special consideration in the displacement-based formulation of Kirchhoff--Love shells, see, e.g., \cite{Kiendl_2009a, Kiendl_2010a}.

Alternatively, a \emph{mixed} (rather than displacement-based) formulation can be used where the components of the moment tensor are introduced as primary variables, in addition to the displacements. Consequently, the continuity requirements are lowered to $C^0$-continuity, enabling the use of standard Lagrange elements in the FEM analysis. This approach is common for Kirchhoff--Love \emph{plates} as shown in various publications, see, e.g., \cite{Johnson_1973a, Brezzi_1974a, Arnold_1985a, Comodi_1989a, Rafetseder_2018a, Walker_2022a}, and was recently used for Kirchhoff--Love \emph{shells} in \cite{Rafetseder_2019a, Neunteufel_2019a, Neunteufel_2024a}. According to \cite{Arnold_1985a}, higher convergence rates for the stress field, in our case, the moment tensor, may be achieved in mixed methods as the moment tensor is calculated directly as a primary variable and is not derived from the displacements. However, significant disadvantages regarding the solvability are that it leads to an indefinite system of equations and additional degrees of freedom (DOFs). Hybridization of the moment tensor field can avert both of these problems \cite{Arnold_1985a, Brezzi_1985a, Cockburn_2009a}.

For the hybridization scheme described in \cite{Boffi_2013a}, the continuity of the moment tensor is broken between elements and reinforced weakly by a Lagrange multiplier living only on the edges of neighboring elements. This first adds even more DOFs which may seem counterintuitive. However, as discontinuous, local function spaces are now used for the components of the moment tensor, an element-wise static condensation is enabled for the DOFs related to the moments. The origins of mixed-hybrid FEM, sometimes called hybridized mixed FEM, can be traced back to the 1960s, see, e.g., \cite{FraeijsDeVeubeke_1965a}. The method was analyzed in \cite{Arnold_1985a, Brezzi_1985a, Cockburn_2004a} among others and was used in structural mechanics in \cite{Andelfinger_1993a, Pechstein_2011a, Echter_2013a, Pechstein_2017a} to alleviate locking. Hybridization techniques also became relevant in discontinuous Galerkin (DG) methods. Introduced in \cite{Cockburn_2009a}, so-called hybridizable DG methods were developed for problems in structural mechanics \cite{Soon_2009a, Kabaria_2015a} as well as in fluid dynamics \cite{Nguyen_2010a, Nguyen_2011a, Giorgiani_2013a} and are closely related to mixed-hybrid methods as mentioned in \cite{Cockburn_2023a}.

Recently, mixed-hybrid elements were used in \cite{Neunteufel_2019a, Neunteufel_2024a} for a geometrically non-linear, $C^0$-continuous Kirchhoff--Love shell model, relying on $\mathcal{H}$(div~div) elements. In contrast, our formulation herein does not need any special function spaces: equal-order interpolations of the displacements and moments in the frame of the isoparametric FEM are enabled. The proposed formulation considers all mechanically relevant boundary conditions for the Kirchhoff-Love shells, including the delicate topic of corner and \emph{effective} boundary forces.

The main contributions of this paper are summarized as follows:
\begin{itemize}
    \item A novel mixed-hybrid formulation of the Kirchhoff--Love shell is presented that enables standard finite element spaces for the discretization, typically based on Lagrange elements.
    \item The mechanical model applies to explicit and implicit geometry descriptions because the governing equations are formulated coordinate-free in terms of the TDC.
    \item All mechanical Dirichlet and Neumann boundary conditions are adequately discussed, including inhomogeneous cases.
    \item A higher-order convergence of $\mathcal{O}(p+1)$ is shown not only for the displacements $\vek{u}$ but also for the moment tensor $\mat{m}_\Gamma$.
    \item A series of new benchmark test cases are presented featuring smooth solutions and thus enabling higher-order convergence rates.
\end{itemize}

The structure of this paper is as follows: In Section~\ref{sec: Geometrical setup and differential operators on surfaces}, the geometrical description and definition of surface differential operators in the frame of the TDC are briefly introduced. Section~\ref{sec: Mechanical model for the Kirchhoff-Love shell} covers the mechanical model of the mixed Kirchhoff--Love shell, leading to the strong form of the governing equations. The corresponding discretized weak form and the subsequent hybridization are presented in Section~\ref{sec: A mixed-hybrid FEM for Kirchhoff--Love shells}. Numerical results are discussed in Section~\ref{sec: Numerical results}. Here, we combine traditional benchmarks for shells to verify our model and new innovative test cases to demonstrate higher-order convergence.

\section{Geometrical setup and differential operators on surfaces}\label{sec: Geometrical setup and differential operators on surfaces}

The geometries of shells are curved, two-dimensional surfaces $\Gamma$ embedded in the three-dimensional space. Let $\Gamma$ be sufficiently smooth, orientable, connected, and bounded by $\partial \Gamma$. Such geometries may be defined \textit{explicitly} based on parametrizations or \textit{implicitly}, e.g., by level sets. For both cases, differential operators are needed to set up mechanical models for the shells. We define them conveniently based on the TDC following \cite{Delfour_2011a, Dziuk_2013a, Fries_2018a, Jankuhn_2017a, Schoellhammer_2019a}.

\subsection{Geometric quantities}\label{sec: Geometric quantities}

For the case of an explicit description of the shell surface through a parametrization, there is a mapping from a reference space $\hat{\Omega}$  to the real domain $\Gamma$, $\vek{x}(\vek{r}): \hat{\Omega} \rightarrow \Gamma$, with $\hat{\Omega}\subset\mathbb{R}^2$, $\Gamma\subset\mathbb{R}^3$, and $\vek{r} = [r, s]^\text{T}$ as the coordinates in the reference space. The unit normal vector $\vek{n}(\vek{x})\in\mathbb{R}^3$, $\vek{x}\in\Gamma$, of a surface is then defined as
\begin{equation}\label{eq: explicit normal vector}
	\vek{n}(\vek{x}) = \frac{\vek{n}^\star}{\| \vek{n}^\star \|} \quad \text{with} \quad \vek{n}^\star = 
	\begin{bmatrix}
		\partial_r x\\
		\partial_r y\\
		\partial_r z
	\end{bmatrix} \times
	\begin{bmatrix}
		\partial_s x\\
		\partial_s y\\
		\partial_s z
	\end{bmatrix},
\end{equation}
which is the cross product of the columns of the Jacobi matrix $\mat{J}(\vek{r}) = \partial_{\vek{r}} \vek{x}$. It is noted that a parametrization is also available in the presence of a surface mesh discretizing the shell geometry; often, the usual isoparametric concept defines the corresponding map $\vek{x}(\vek{r})$ for every element.

\begin{figure}
	\centering
	
	\includegraphics[width=0.75\textwidth]{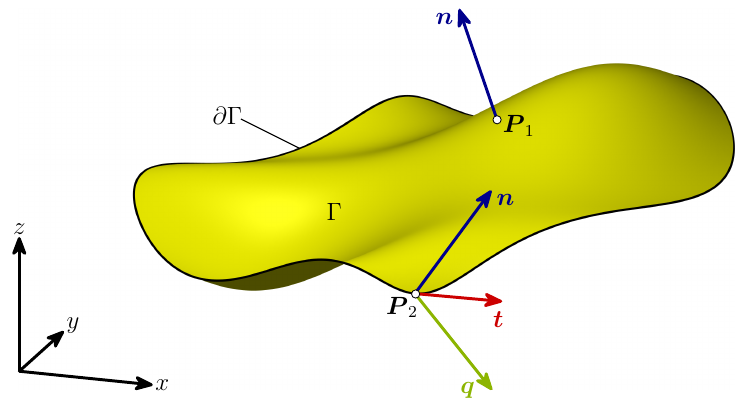}
	
	\caption{\label{fig: vector on surface}Some vectors on an arbitrary surface $\Gamma$ with boundary $\partial\Gamma$. The normal vector $\vek{n}$ on the inner point $\vek{P}_1$ is shown in blue. The normal vector $\vek{n}$, the tangential vector $\vek{t}$ and the conormal vector $\vek{q}$ on the boundary point $\vek{P}_2$ are shown in blue, red, and green, respectively.}
\end{figure}

In order to properly setup the boundary value problem (BVP) later, let us define tangential and conormal vector fields on the boundary $\partial\Gamma$. For the surface, there is a unit tangent vector $\vek{t}(\vek{x})\in\mathbb{R}^3$, $\vek{x}\in\partial\Gamma$, along the boundary $\partial\Gamma$, orientated in the direction of $\partial\Gamma$. This vector is obtained from the parameterization as follows: When $\hat{\Omega}$ is the domain that is mapped using $\vek{x}(\vek{r})$ and $\partial\hat{\Omega}$ its boundary, then the tangent vector $\hat{\vek{t}}\in \mathbb{R}^2$ on $\partial\hat{\Omega}$ is mapped as
\begin{equation}\label{eq: tangent vector}
	\vek{t}(\vek{x}) = \frac{\vek{t}^\star}{\| \vek{t}^\star \|} \quad \text{with} \quad \vek{t}^\star = \mat{J}(\vek{r})\cdot \hat{\vek{t}}.
\end{equation}
The unit conormal vector $\vek{q}(\vek{x})\in\mathbb{R}^3$ on $\partial\Gamma$ points outside of $\Gamma$ and is orthogonal to both, the normal vector $\vek{n}(\vek{x})$ and tangent vector $\vek{t}(\vek{x})$, following from the cross product
\begin{equation}\label{eq: conormal vector}
	\vek{q}(\vek{x}) = \vek{t} \times \vek{n}.
\end{equation}
Subsequently, the vectors $\vek{n}(\vek{x})$, $\vek{t}(\vek{x})$, and $\vek{q}(\vek{x})$ are essential for the formulation of integral theorems and the implementation of boundary conditions. An example of these vectors for some arbitrary surface is seen in Fig.~\ref{fig: vector on surface}.

\subsection{Differential operators on surfaces}\label{sec: Differential operators on surfaces}

Tangential or surface differential operators are introduced to define BVPs on curved surfaces later. Surface operators are marked with the subscript $\Gamma$, e.g., $\nabla_\Gamma$ for the tangential gradient. The TDC-based formulation avoids (local) curvilinear coordinates that are classically used in shell mechanics.

An important quantity for the formulation of tangential operators is the projector $\mat{P}(\vek{x}) \in \mathbb{R}^{3 \times 3}$, $\vek{x}\in\Gamma$,
\begin{equation}\label{eq: tangential projector}
	\mat{P}(\vek{x}) = \mat{I} - \vek{n} \otimes \vek{n},
\end{equation}
where $\mat{I}$ is the ($3 \times 3$)-identity matrix. $\mat{P}(\vek{x})$ is often used to project onto the tangent space $T_P \Gamma$ of the manifold.

The tangential gradient of a scalar function $f(\vek{x}): \Gamma \rightarrow \mathbb{R}$ is defined as
\begin{equation}\label{eq: tangential gradient of scalar}
	\nabla_\Gamma f = \mat{P} \cdot \nabla \tilde{f},
\end{equation}
with the conventional gradient operator $\nabla$ and $\tilde{f}$ being a smooth extension of $f$ in the neighborhood of $\Gamma$. According to \cite{Dziuk_2013a}, for parametrized surfaces based on the aforementioned map $\vek{x}(\vek{r}): \hat{\Omega} \rightarrow \Gamma$, the tangential gradient can also be computed  without explicitly constructing $\tilde{f}$ by
\begin{equation}\label{eq: tangential gradient of scalar vie parametrization}
    \nabla_\Gamma f = \mat{J} \cdot \mat{G}^{-1} \cdot \nabla_{\vek{r}}f(\vek{r}) = 
    \begin{bmatrix}
        \partial^\Gamma_x f\\
        \partial^\Gamma_y f\\
        \partial^\Gamma_z f
    \end{bmatrix},
\end{equation}
where $\mat{G} = \mat{J}^{\text{T}} \cdot \mat{J}$ is the first fundamental form and $\nabla_{\vek{r}}$ is the gradient with respect to the reference space $\hat{\Omega}$. This procedure is especially convenient in the context of finite elements since $f(\vek{r})$ may be some shape function in the reference element, while the tangential gradients are determined on the physical, curved element. Note that $\nabla_\Gamma f$ is in the tangent plane of the surface, i.e., $\nabla_\Gamma f \in T_P \Gamma$.

For a vector-valued function $\vek{v}(\vek{x}): \Gamma \times \Gamma \rightarrow \mathbb{R}^3$ with extension $\tilde{\vek{v}}$, the \textit{directional} and \textit{covariant} tangential gradients are distinguished. The directional gradient of $\vek{v}(\vek{x})$ is defined as
\begin{equation}\label{eq: directional gradient of vector}
    \nabla_\Gamma^{\text{dir}} \vek{v} = \nabla_\Gamma^{\text{dir}}
    \begin{bmatrix}
        v_x \\ v_y \\ v_z
    \end{bmatrix} = 
    \begin{bmatrix}
        (\nabla_\Gamma v_x)^{\text{T}} \\ (\nabla_\Gamma v_y)^{\text{T}} \\ (\nabla_\Gamma v_z)^{\text{T}}
    \end{bmatrix} = \nabla \tilde{\vek{v}} \cdot \mat{P},
\end{equation}
whereas the covariant gradient of $\vek{v}$ is defined as
\begin{equation}\label{eq: covariant gradient of vector}
	\nabla_\Gamma^{\text{cov}} \vek{v} = \mat{P} \cdot \nabla_\Gamma^{\text{dir}} \vek{v} = \mat{P} \cdot \nabla \tilde{\vek{v}} \cdot \mat{P}.
\end{equation}
Note that the directional gradient is typically not in-plane, whereas the covariant gradient is. The tangential \emph{divergence} operator of a vector function $\vek{v}$ is
\begin{equation}\label{eq: tangential divergence of vector}
	\text{div}_\Gamma \vek{v} = \text{tr}(\nabla_\Gamma^{\text{dir}} \vek{v}) = \text{tr}(\nabla_\Gamma^{\text{cov}} \vek{v}),
\end{equation}
and for the divergence of a second-order tensor-valued function $\mat{T}\in \mathbb{R}^{3 \times 3}$,
\begin{equation}\label{eq: tangential divergence of tensor}	
	\text{div}_\Gamma \mat{T} = 
	\begin{bmatrix}
		\text{div}_\Gamma [T_{11},T_{12},T_{13}]\\
		\text{div}_\Gamma [T_{21},T_{22},T_{23}]\\
		\text{div}_\Gamma [T_{31},T_{32},T_{33}]
	\end{bmatrix}.
\end{equation} 

\subsection{Weingarten map and mean curvature}\label{sec: Weingarten map and curvature}
The Weingarten map $\mat{H}(\vek{x}) \in \mathbb{R}^{3 \times 3}$, $\vek{x}\in\Gamma$, as in, e.g., \cite{Jankuhn_2017a, Delfour_2011a}, is a symmetric, in-plane tensor computed by
\begin{equation}\label{eq: Weingarten map}	
	\mat{H} = \nabla_\Gamma^{\text{dir}} \vek{n} = \nabla_\Gamma^{\text{cov}} \vek{n}.
\end{equation}
The two non-zero eigenvalues of $\mat{H}$, $\varkappa_1$ and $\varkappa_2$, are the principal curvatures. The mean (rather summed) curvature $\varkappa(\vek{x}) \in \mathbb{R}$, $\vek{x}\in\Gamma$, can be computed as $\varkappa = \text{tr}(\mat{H})=\varkappa_1+\varkappa_2$.

\subsection{Integral theorems}\label{sec: Integral theorems}
Integral theorems are later required to obtain a useful weak formulation of the shell BVP. As such, the divergence theorem for manifolds $\Gamma$, in the context of some scalar function $f$ and vector function $\vek{v}$, is:
\begin{equation}\label{eq: integral theorem 1}	
     \int_\Gamma f \cdot \text{div}_\Gamma \vek{v} \, \text{d}\Gamma = - \int_\Gamma \nabla_\Gamma f \ScalProd \vek{v} \, \text{d}\Gamma + \int_\Gamma \varkappa \cdot f \cdot \big(\vek{v} \ScalProd \vek{n}\big) \, \text{d}\Gamma + \int_{\partial\Gamma} f \cdot \vek{v} \ScalProd \vek{q} \, \text{d}\partial\Gamma.
\end{equation}
Extending Eq.~(\ref{eq: integral theorem 1}), the divergence theorem for a vector-valued function $\vek{v}$ and tensor-valued function $\mat{T}$ is
\begin{equation}\label{eq: integral theorem 2}	
    \int_\Gamma \vek{v} \ScalProd \text{div}_\Gamma \mat{T} \, \text{d}\Gamma = - \int_\Gamma \nabla_\Gamma^{\text{dir}} \vek{v} \FrobProd \mat{T} \, \text{d}\Gamma + \int_\Gamma \varkappa \cdot \vek{v} \ScalProd \big(\mat{T} \cdot \vek{n}\big) \, \text{d}\Gamma + \int_{\partial\Gamma} \vek{v} \ScalProd \big(\mat{T} \cdot \vek{q}\big) \, \text{d}\partial\Gamma,
\end{equation}
where $\vek{a} \ScalProd \vek{b} = \vek{a}^{\text{T}} \cdot \vek{b}$ refer to the inner product of two vectors and $\mat{A} \FrobProd \mat{B} = \text{tr}\big(\mat{A} \cdot \mat{B}^{\text{T}}\big)$ to the Frobenius inner product of two tensors. The second terms on the right-hand sides of Eqs.~(\ref{eq: integral theorem 1})~and~(\ref{eq: integral theorem 2}) vanish for the case of in-plane vectors $\vek{v} = \mat{P} \cdot \vek{v}$ and tensors $\mat{T} = \mat{P} \cdot \mat{T} \cdot \mat{P}$, respectively, because then $\vek{v} \ScalProd \vek{n} = 0$ and $\mat{T} \cdot \vek{n} = \vek{0}$.

\section{Mechanical model of Kirchhoff--Love shells}\label{sec: Mechanical model for the Kirchhoff-Love shell}
A reformulation of the Kirchhoff--Love shell model in a coordinate-free form based on the TDC is found in \cite{Schoellhammer_2019a}. There, the discretization of the weak form is based on the displacement-based formulation which requires $C^1$-continuous shape functions in the analysis as provided, e.g., through isogeometric analysis (IGA) \cite{Hughes_2005a, Kiendl_2009a}. Herein, the aim is to avoid the displacement-based formulation, hence the need for $C^1$-continuous shape functions, by using a mixed formulation in which the moment tensor $\mat{m}_\Gamma$ serves as an additional primary variable, next to the displacement vector $\vek{u}$. In the following, the strong form of this mixed formulation is derived concerning kinematics, constitutive relations, equilibrium, and boundary conditions of Kirchhoff--Love shells featuring infinitesimal deformations, leading to a geometrically linear model.

\subsection{Kinematics}\label{sec: Kinematics}
\begin{figure}
	\centering
	
	\includegraphics[width=1.0\textwidth]{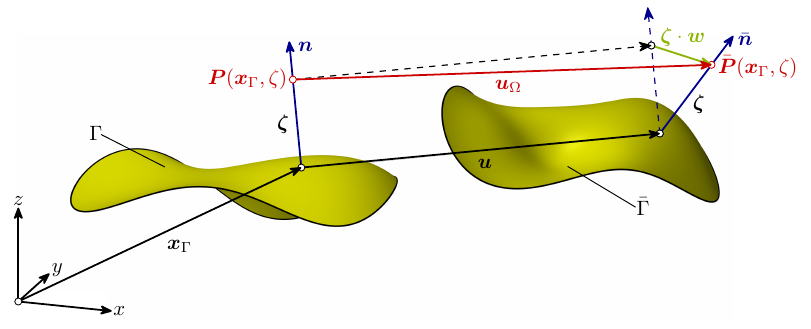}
	
	\caption{\label{fig: kinematic deformation}Sketch of the kinematic situation between the undeformed middle surface $\Gamma$ and the deformed middle surface $\bar{\Gamma}$ of a shell. The displacement $\vek{u}_{\Omega}$ of a point $\vek{P}$ in the continuum $\Omega$ and other related quantities are depicted.}
\end{figure}

Let $\Gamma$ be the middle surface of a shell. The three-dimensional continuum associated with $\Gamma$ is the domain $\Omega \subset \mathbb{R}^3$. Any point within the continuum can then be described by
\begin{equation}\label{eq: point in continuum}	
	\vek{x} = \vek{x}_\Gamma + \zeta \cdot \vek{n} \in \Omega,
\end{equation}
where $\vek{x}_\Gamma \in \Gamma$ represents a point directly on the surface and $\zeta \in \mathbb{R}$ is a control variable in the direction of the thickness of the shell restricted by $|\zeta|\leq \frac{t}{2}$. The displacement of some point in $\Omega$ before and after the deformation, $\vek{P}$ and $\bar{\vek{P}}$, respectively, is evaluated as
\begin{equation}\label{eq: displacement in continuum}	
	\vek{u}_\Omega(\vek{x}_\Gamma,\zeta) = \bar{\vek{P}}(\vek{x}_\Gamma,\zeta) - \vek{P}(\vek{x}_\Gamma,\zeta) = \vek{u}(\vek{x}_\Gamma) + \zeta \cdot \vek{w}(\vek{x}_\Gamma),
\end{equation}
with $\vek{u}(\vek{x}_\Gamma)$ representing the displacement vector field of the (curved) middle surface and $\vek{w}(\vek{x}_\Gamma) \in T_P\Gamma$ being the difference vector in the tangent space, describing the rotation of the normal vector. Fig.~\ref{fig: kinematic deformation} shows the kinematic setup of the displacement at the point $\vek{P}$. According to Bernoulli's hypothesis, the Kirchhoff--Love model disregards displacements resulting from transverse shear stresses. Consequently, the cross-section of the shell remains straight and normal to the middle surface after the deformation. As a result, the difference vector $\vek{w}$ only depends on the bending deformation of the surface and is defined as \cite{Delfour_1996a, Schoellhammer_2019a}
\begin{equation}\label{eq: difference vector}	
	\vek{w}(\vek{x}_\Gamma) = \mat{H} \cdot \vek{u} - \nabla_\Gamma\big(\vek{u} \ScalProd \vek{n}\big) = -(\nabla^\text{dir}_\Gamma \vek{u})^\text{T} \cdot \vek{n}.
\end{equation}

The linear in-plane strain tensor is defined as
\begin{equation}\label{eq: strain tensor}	
	\vek{\varepsilon}_\Gamma(\vek{x}) = \frac{1}{2}
	\big[
		\nabla^\text{cov}_\Gamma\vek{u}_\Omega + (\nabla^\text{cov}_\Gamma\vek{u}_\Omega)^\text{T}
	\big],
\end{equation}
and may be split into a membrane and a bending part
\begin{equation}\label{eq: strain tensor split}	
	\vek{\varepsilon}_\Gamma = \vek{\varepsilon}_{\Gamma,\text{Memb}} + \zeta \cdot \vek{\varepsilon}_{\Gamma,\text{Bend}},
\end{equation}
with
\begin{equation}\label{eq: strain tensor membrane}	
	\vek{\varepsilon}_{\Gamma,\text{Memb}}(\vek{u}) = \frac{1}{2}
	\big[
		\nabla^\text{cov}_\Gamma\vek{u} + (\nabla^\text{cov}_\Gamma\vek{u})^\text{T}
	\big],
\end{equation}
\begin{equation}\label{eq: strain tensor bending}
    \begin{split}
        \vek{\varepsilon}_{\Gamma,\text{Bend}}(\vek{u}) &= \frac{1}{2}
    	\big[
    		\mat{H} \cdot \nabla^\text{dir}_\Gamma\vek{u} + (\nabla^\text{dir}_\Gamma\vek{u})^\text{T} \cdot \mat{H} + \nabla^\text{cov}_\Gamma\vek{w} + (\nabla^\text{cov}_\Gamma\vek{w})^\text{T}
    	\big]
        \\ &= - \sum_{i=1}^3 \nabla^\text{cov}_\Gamma (\nabla_\Gamma u_i) \cdot n_i 
    \end{split}
\end{equation}
respectively. Both tensors are symmetric and in-plane.

\subsection{Constitutive relations}\label{sec: Constitutive relations}
The following constitutive relations are based on the assumptions of linear elasticity according to Hooke's law and isotropic material behavior. Under this premise, the in-plane stress tensor for the shell is defined as
\begin{equation}\label{eq: stress tensor}	
	\vek{\sigma}_\Gamma(\vek{\varepsilon}_\Gamma) = \mat{C}_\Gamma : \vek{\varepsilon}_\Gamma = 2 \mu \cdot \vek{\varepsilon}_\Gamma + \lambda \cdot \text{tr}(\vek{\varepsilon}_\Gamma) \cdot \mat{P},
\end{equation}
where $\mat{C}_\Gamma$ is the fourth-order surface material tensor and $\mu = \frac{E}{2(1+\nu)}$ and $\lambda = \frac{E\cdot\nu}{1-\nu^2}$ are the Lam\'e constants for plane stress with $E \in \mathbb{R}^+$ being Young's modulus and $\nu \in [0, 0.5)$ Poisson's ratio, see \cite{Mang_2018a}.

Similarly to the strain tensor, the stress tensor can be decomposed into a membrane and a bending part. Assuming constant thickness over $\vek{x}_\Gamma$, an analytic pre-integration in $\zeta$ is feasible, resulting in the definitions of the effective normal force tensor
\begin{equation}\label{eq: normal force tensor}	
	\tilde{\mat{n}}_\Gamma = \int_{-\frac{t}{2}}^{\frac{t}{2}}\vek{\sigma}_\Gamma \, \text{d}\zeta  = t \cdot \vek{\sigma}_\Gamma(\vek{\varepsilon}_{\Gamma,\text{Memb}})
\end{equation}
and the bending moment tensor
\begin{equation}\label{eq: moment tensor}	
	\mat{m}_\Gamma = \int_{-\frac{t}{2}}^{\frac{t}{2}} \zeta \cdot \vek{\sigma}_\Gamma \, \text{d}\zeta  = \frac{t^3}{12} \cdot \vek{\sigma}_\Gamma(\vek{\varepsilon}_{\Gamma,\text{Bend}}).
\end{equation}
Like the strain tensors, the effective normal force tensor and the moment tensor are in-plane and symmetric. Their two non-zero eigenvalues refer to the principle moments and effective normal forces.

The effective normal force tensor  $\tilde{\mat{n}}_\Gamma$ shall not be mistaken by the physical normal force tensor $\mat{n}_\Gamma^\text{real}$, which is determined by
\begin{equation}\label{eq: physical normal force tensor}	
	\mat{n}_\Gamma^\text{real} = \tilde{\mat{n}}_\Gamma + \mat{H} \cdot \mat{m}_\Gamma.
\end{equation}
The physical normal force tensor is in-plane but not symmetric in general.

For the desired mixed formulation, according to the Hellinger--Reissner principle, the inverse material law of the bending part has to be defined. This results in the bending strain tensor in terms of the moment tensor
\begin{equation}\label{eq: inverse material law bending}	
	\vek{\varepsilon}_{\Gamma,\text{Bend}}(\mat{m}_\Gamma) = \frac{12}{t^3} \cdot \mat{C}_\Gamma^{-1} : \mat{m}_\Gamma = \frac{12}{E \cdot t^3} \cdot
	\big[ 
		\big(1+\nu\big) \cdot \mat{m}_\Gamma - \nu \cdot \text{tr}(\mat{m}_\Gamma) \cdot \mat{P}
	\big].
\end{equation}
As the definitions of the bending strain tensor from Eqs.~(\ref{eq: strain tensor bending}) and (\ref{eq: inverse material law bending}) are equal, one can state that
\begin{equation}\label{eq: strong form equation 1}	
	-\vek{\varepsilon}_{\Gamma,\text{Bend}}(\mat{m}_\Gamma) + \vek{\varepsilon}_{\Gamma,\text{Bend}}(\vek{u}) = \mat{0},
\end{equation}
which is the first equation of the strong form for the mixed formulation, establishing a connection between the primal variables $\mat{m}_\Gamma$ and $\vek{u}$.

For further details on the constitutive relations of shells, see, e.g., \cite{Basar_1985a, Bischoff_2017a, Simo_1989a, Simo_1989b} for the classical theory and \cite{Schoellhammer_2019a, Gfrerer_2021a} for descriptions expressed in terms of the TDC.

\subsection{Equilibrium}\label{sec: Equilibrium}
In the context of the TDC, the equilibrium of forces derived by \cite{Schoellhammer_2019a} is defined as
\begin{equation}\label{eq: equilibrium}	
	\vek{n} \cdot \text{div}_\Gamma \big(\mat{P} \cdot \text{div}_\Gamma \mat{m}_\Gamma\big) + \mat{H} \cdot \text{div}_\Gamma \mat{m}_\Gamma + \text{div}_\Gamma \mat{n}_\Gamma^\text{real}(\mat{m}_\Gamma, \vek{u}) = -\vek{f},
\end{equation}
with $\vek{f}(\vek{x})$ being the body load vector per area on the surface $\Gamma$. Considering the definition of the physical normal force tensor from Eq.~(\ref{eq: physical normal force tensor}), Eq.~(\ref{eq: equilibrium}) can be rewritten as
\begin{equation}\label{eq: strong form equation 2}	
	\vek{n} \cdot \text{div}_\Gamma \big(\mat{P} \cdot \text{div}_\Gamma \mat{m}_\Gamma\big)
    + \mat{H} \cdot \text{div}_\Gamma \mat{m}_\Gamma
    + \text{div}_\Gamma \big(\mat{H} \cdot \mat{m}_\Gamma\big)
    + \text{div}_\Gamma \tilde{\mat{n}}_\Gamma(\vek{u})
    = -\vek{f},
\end{equation}
representing the second equation of the strong form.

\subsection{Boundary conditions}\label{sec: Boundary conditions}
\begin{figure}
	\centering
	
	\subfigure[Displacement quantities]
	{\includegraphics[width=0.475\textwidth]{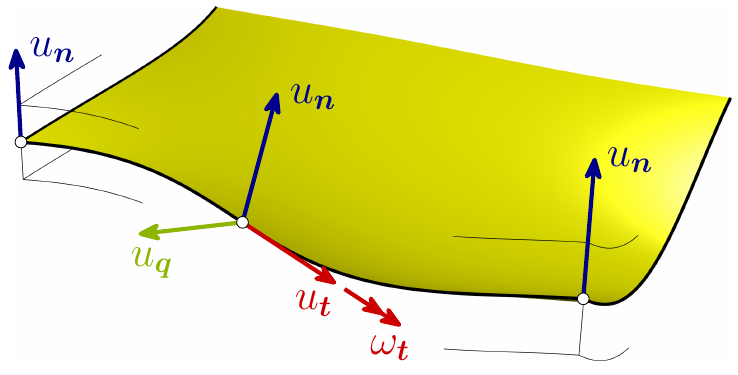}\label{fig: Dir BCs}}\hfill
	\subfigure[Force quantities]
	{\includegraphics[width=0.475\textwidth]{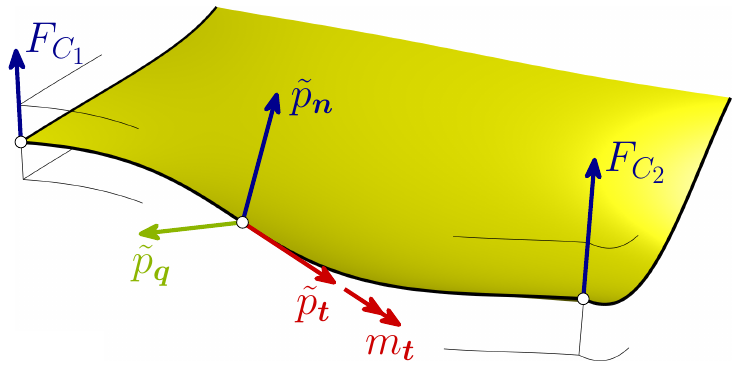}\label{fig: Neum BCs}}
	
	\caption{\label{fig: all BCs}Boundary quantities at an arbitrary point on the boundary $\partial\Gamma$ with displacement and force components in terms of the local triad $\vek{t}$, $\vek{n}$, and $\vek{q}$, and rotation and moment around the tangential direction. Normal displacements and Kirchhoff forces are depicted on the corners $C_1$ and $C_2$. Quantities in tangential, normal, and conormal directions are shown in red, blue, and green, respectively.} 
\end{figure}

As in the classical theory of Kirchhoff--Love shells, treating the boundary conditions requires particular attention. Fig.~\ref{fig: all BCs} provides an overview of the relevant boundary conditions. For the fields $\vek{u}$, representing general translational displacements, and $\omega_{\vek{t}}$, representing rotations around the tangential direction, there exist two non-overlapping segments of the boundary $\partial\Gamma$, respectively, i.e., $\partial\Gamma_{\text{D}, i} \cup \partial\Gamma_{\text{N}, i} = \partial\Gamma$ and $\partial\Gamma_{\text{D}, i} \cap \partial\Gamma_{\text{N}, i} = \emptyset$. In particular, one distinguishes between the Dirichlet boundaries $\partial\Gamma_{\text{D},\vek{u}}$ and $\partial\Gamma_{\text{D},\omega}$ and the Neumann boundaries $\partial\Gamma_{\text{N},\vek{u}}$ and $\partial\Gamma_{\text{N},\omega}$ with their corresponding boundary conditions
\begin{gather}	
    \label{eq: boundary conditions u}
        \vek{u} = \hat{\vek{u}} \:\: \text{on} \:\: \partial\Gamma_{\text{D},\vek{u}}, \\
    \label{eq: boundary conditions p}
        \tilde{\vek{p}} = \hat{\tilde{\vek{p}}} \:\: \text{on} \:\: \partial\Gamma_{\text{N},\vek{u}}, \\
    \label{eq: boundary conditions rot}
        \omega_{\vek{t}} = \hat{\omega}_{\vek{t}} \:\: \text{on} \:\: \partial\Gamma_{\text{D},\omega}, \\
    \label{eq: boundary conditions m}
        m_{\vek{t}} = \hat{m}_{\vek{t}} \:\: \text{on} \:\: \partial\Gamma_{\text{N},\omega},
\end{gather}
where a hat $\hat{\bullet}$ indicates a prescribed quantity at the boundary. The values $\tilde{\vek{p}}$ and $m_{\vek{t}}$ represent the \emph{effective} boundary force and moment around the tangential direction, respectively. The components of the displacement vector $\vek{u}$ can also be prescribed individually, while the other components are then prescribed through the respective components of $\tilde{\vek{p}}$. Those components may either be defined w.r.t.~the global (Cartesian) coordinate system (in $\vek{x}\vek{y}\vek{z}$-directions) or some local boundary coordinate system (in $\vek{t}\vek{n}\vek{q}$-directions), see Fig.~\ref{fig: vector on surface}.

A boundary condition for the rotation in the conormal direction $\omega_{\vek{q}}$ does not exist, since it is automatically specified for a prescribed displacement $\vek{u}$. Consequently, the moment around the conormal direction $m_{\vek{q}}$ lacks a conjugated displacement quantity and, as a result, cannot be prescribed directly via a corresponding Neumann boundary condition. In the Kirchhoff--Love shell theory, this issue is addressed by establishing a coupling between $m_{\vek{q}}$ and the boundary force $\vek{p}$, which results in the effective boundary force $\tilde{\vek{p}}$.

As specified in \cite{Schoellhammer_2019a}, the Neumann boundary quantities are related to the moment tensor and the effective normal force tensor as
\begin{gather}
    \label{eq: boundary moment in t}
    	m_{\vek{t}} = \big(\mat{m}_\Gamma \cdot \vek{q}\big) \ScalProd \vek{q},\\
    \label{eq: effective boundary force}
    	\tilde{\vek{p}} = \tilde{p}_{\vek{t}} \cdot \vek{t} + \tilde{p}_{\vek{q}} \cdot \vek{q} + \tilde{p}_{\vek{n}} \cdot \vek{n},
\end{gather}
where
\begin{gather}
    \label{eq: effective boundary force in t}
    	\tilde{p}_{\vek{t}} = p_{\vek{t}} + \big(\mat{H} \cdot \vek{t}\big) \ScalProd \vek{t} \cdot m_{\vek{q}},\\
    \label{eq: effective boundary force in q}
    	\tilde{p}_{\vek{q}} = p_{\vek{q}} + \big(\mat{H} \cdot \vek{t}\big) \ScalProd \vek{q} \cdot m_{\vek{q}},\\
    \label{eq: effective boundary force in n}
    	\tilde{p}_{\vek{n}} = p_{\vek{n}} + \nabla_\Gamma m_{\vek{q}} \cdot \vek{t},
\end{gather}
with
$p_{\vek{t}} = \big(\mat{n}_\Gamma^\text{real} \cdot \vek{q}\big) \ScalProd \vek{t}$,
$p_{\vek{q}} = \big(\mat{n}_\Gamma^\text{real} \cdot \vek{q}\big) \ScalProd \vek{q}$,
$p_{\vek{n}} = \big(\mat{P} \cdot \text{div}_\Gamma \mat{m}_\Gamma\big) \ScalProd \vek{q}$,
and $m_{\vek{q}} = \big(\mat{m}_\Gamma \cdot \vek{q}\big) \ScalProd \vek{t}$.
The rotations at the boundary can also be expressed in terms of the displacements $\vek{u}$ by splitting the difference vector $\vek{w}$ into
$\omega_{\vek{t}} = \vek{w} \ScalProd \vek{q}$ and
$\omega_{\vek{q}} = \vek{w} \ScalProd \vek{t}$.

\begin{figure}
    \centering
    
    \includegraphics[width=0.5\textwidth]{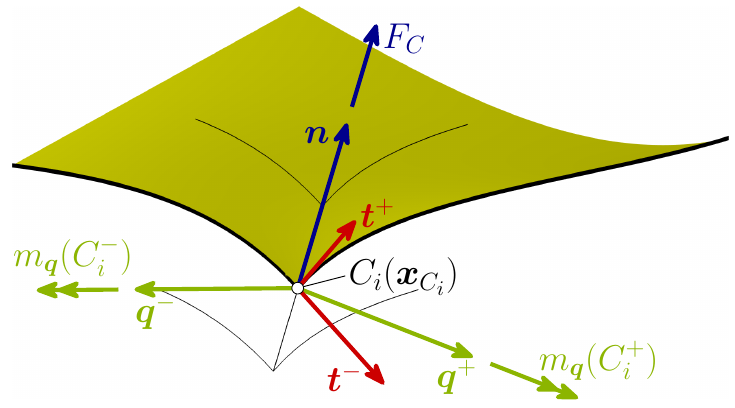}
    
    \caption{\label{fig: Kirchhoff force}Moments $m_{\vek{q}}(C_i^-)$ and $m_{\vek{q}}(C_i^+)$ around their conormal directions at the two sides of some arbitrary corner $C_i$, respectively, and the resulting Kirchhoff force.}
\end{figure}

Another characteristic of the Kirchhoff--Love model is the existence of Kirchhoff forces, also known as corner forces $F_{C}$ for non-smooth boundaries. We can interpret them as the conjugated stress quantity of the displacement  in the normal direction $u_{\vek{n}} = \vek{u} \ScalProd \vek{n}$ of a corner $C_i$, which in turn leads to the following corner boundary conditions
\begin{gather}
    \label{eq: corner conditions u}	
        u_{\vek{n}} = \hat{u}_{\vek{n}} \:\: \text{on} \:\: \vek{x}_{C_i,\text{D}}, \\
    \label{eq: corner conditions F}	
        F_{C} = \hat{F}_{C} \:\: \text{on} \:\: \vek{x}_{C_i,\text{N}},
\end{gather}
where $\vek{x}_{C_i,\text{D}}$ and $\vek{x}_{C_i,\text{N}}$ represent the position of a corner $C_i$ where either a pin support is present (Dirichlet boundary condition) or an external corner force (Neumann boundary condition) is applied. The index $i$ is defined as $i = \{ 1,\dots, n_{C, \text{N}}, n_{C, \text{N}}+1,\dots, n_{C, \text{N}}+n_{C, \text{D}} \}$, where Neumann and Dirichlet corners are sorted and $n_{C, \text{N}}$ and $n_{C, \text{D}}$ are representing the number of Neumann and Dirichlet corners of the shell, respectively. The Kirchhoff force is related to a jump in the moment $m_{\vek{q}}$ at some corner and is computed as
\begin{equation}\label{eq: Kirchhoff force}	
    F_{C} = [m_{\vek{q}}]_{C_i^-}^{C_i^+}=m_{\vek{q}}(C_i^+)-m_{\vek{q}}(C_i^-),
\end{equation}
where $C_i^+$ is the side of the corner to which the tangent vector $\vek{t}$ points and $C_i^-$ the other side, see Fig.~\ref{fig: Kirchhoff force}.

\section{Mixed-hybrid FEM for Kirchhoff--Love shells}\label{sec: A mixed-hybrid FEM for Kirchhoff--Love shells}
Based on the Hellinger--Reissner principle, the bending-related part of the governing equations is expressed in a mixed formulation and, therefore, the moment tensor $\mat{m}_\Gamma$ is included as a primary variable in addition to the displacement vector $\vek{u}$. As a result, the continuity requirements in the FEM analysis are reduced from $C^1$-continuity to $C^0$-continuity, thereby enabling the use of standard Lagrange elements. The boundary value problem of the Kirchhoff--Love shell model in \emph{strong} form may be stated as follows: Find $\mat{m}_\Gamma$ and $\vek{u}$ from suitable function spaces such that the following two field equations in the domain $\Gamma$ are fulfilled, see Eqs.~(\ref{eq: strong form equation 1}) and (\ref{eq: strong form equation 2}),
\begin{gather}
    \label{eq: strong form equation 1 repeat}
	-\vek{\varepsilon}_{\Gamma,\text{Bend}}(\mat{m}_\Gamma) + \vek{\varepsilon}_{\Gamma,\text{Bend}}(\vek{u}) = \mat{0},\\
    \label{eq: strong form equation 2 repeat}
    \vek{n} \cdot \text{div}_\Gamma \big(\mat{P} \cdot \text{div}_\Gamma \mat{m}_\Gamma\big)
    + \mat{H} \cdot \text{div}_\Gamma \mat{m}_\Gamma
    + \text{div}_\Gamma \big(\mat{H} \cdot \mat{m}_\Gamma\big)
    + \text{div}_\Gamma \tilde{\mat{n}}_\Gamma(\vek{u})
    = -\vek{f},
\end{gather}
subject to boundary conditions on $\partial\Gamma$ as in Eqs.~(\ref{eq: boundary conditions u}) to (\ref{eq: boundary conditions m}).

\subsection{Mixed weak form}\label{sec: Mixed weak form}
As for all analyses based on the FEM, the governing equations, herein, Eqs.~(\ref{eq: strong form equation 1 repeat}) and (\ref{eq: strong form equation 2 repeat}), need to be formulated in their weak forms. Different variants of the weak forms may be obtained based on how boundary conditions are enforced, e.g., through Lagrange multipliers, the Nitsche method, restricting function spaces, etc. This is of particular importance in the present context because the relevant boundary conditions for the moment $m_{\vek{t}}$ or rotation $\omega_{\vek{t}}$ on $\partial \Gamma_{\omega}$ are enforced through Lagrange multipliers, otherwise, these quantities are hardly accessible. In contrast, boundary conditions for either $\vek{u}$ or $\tilde{\vek{p}}$ on $\partial\Gamma_{\vek{u}}$ are enforced in the usual manner, i.e., by restricting the function space for $\vek{u}$ on $\partial\Gamma_{\text{D},\vek{u}}$ or evaluating the corresponding boundary integral for $\tilde{\vek{p}}$ on $\partial\Gamma_{\text{N},\vek{u}}$.

Hence, in the system of PDEs, the primary variables are the displacement vector $\vek{u}$, the moment tensor $\mat{m}_\Gamma$, and, as a Lagrange multiplier, the tangential boundary rotation $\omega_{\vek{t}}$. Related to the primary variables, we establish the test functions $\mat{V}_{\mat{m}_\Gamma}$ for Eq.~(\ref{eq: strong form equation 1 repeat}), $\vek{v}_{\vek{u}}$ for Eq.~(\ref{eq: strong form equation 2 repeat}), and $v_{\omega_{\vek{t}}}$ for the Lagrange multiplier. The corresponding function spaces are then defined as
\begin{eqnarray}
	\mathcal{S}_{\mat{m}_\Gamma} = \mathcal{V}_{\mat{m}_\Gamma} &=&
	\begin{Bmatrix}
		\mat{V} \in \mathcal{H}(\text{div}, \Gamma): \mat{V} = \mat{V}^\text{T}
	\end{Bmatrix} \label{eq: function space m}, \\
	\mathcal{S}_{\vek{u}} &=& 
	\begin{Bmatrix}
		\vek{v} \in [\mathcal{H}^1(\Gamma)]^3: \vek{v} = \hat{\vek{u}} \, \text{on} \, \partial\Gamma_{\text{D},\vek{u}}
	\end{Bmatrix} \label{eq: function space u}, \\
	\mathcal{V}_{\vek{u}} &=& 
	\begin{Bmatrix}
		\vek{v} \in [\mathcal{H}^1(\Gamma)]^3: \vek{v} = \vek{0} \, \text{on} \, \partial\Gamma_{\text{D},\vek{u}}
	\end{Bmatrix} \label{eq: function space vu}, \\
	\mathcal{S}_{\omega_{\vek{t}}} &=& 
	\begin{Bmatrix}
		v \in \mathcal{L}^2(\partial\Gamma): v = \hat{\omega}_{\vek{t}} \, \text{on} \, \partial\Gamma_{\text{D},\omega}
	\end{Bmatrix} \label{eq: function space omega}, \\
	\mathcal{V}_{\omega_{\vek{t}}} &=&
	\begin{Bmatrix}
		v \in \mathcal{L}^2(\partial\Gamma): v = 0 \, \text{on} \, \partial\Gamma_{\text{D},\omega}
	\end{Bmatrix} \label{eq: function space vomega},
\end{eqnarray}
where $\mathcal{L}^2$ is the Lebesgue space, $\mathcal{H}^1(\Gamma) = \{ v \in \mathcal{L}^2(\Gamma): \nabla_\Gamma v \in [\mathcal{L}^2(\Gamma)]^3 \}$ is the Sobolev space, and $\mathcal{H}(\text{div}, \Gamma) = \{ \mat{V} \in [\mathcal{L}^2(\Gamma)]^{3 \times 3}: \text{div}_\Gamma \mat{V} \in [\mathcal{L}^2(\Gamma)]^3 \}$. To derive the weak form, we multiply the PDEs of the strong form with their corresponding test functions and integrate over the domain $\Gamma$ for Eqs.~(\ref{eq: strong form equation 1 repeat})~and~(\ref{eq: strong form equation 2 repeat}) and over the Neumann boundary $\partial\Gamma_{\text{N},\omega}$ for Eq.~(\ref{eq: boundary conditions m}). Then, applying the integral theorems from Eqs.~(\ref{eq: integral theorem 1})~and~(\ref{eq: integral theorem 2}), this results in the weak formulation of the BVP of Kirchhoff-Love shells. The task is now to find $\mat{m}_\Gamma \in \mathcal{S}_{\mat{m}_\Gamma}$, $\vek{u} \in \mathcal{S}_{\vek{u}}$, and $\omega_{\vek{t}} \in \mathcal{S}_{\omega_{\vek{t}}}$ such that for all $(\mat{V}_{\mat{m}_\Gamma}, \vek{v}_{\vek{u}}, v_{\omega_{\vek{t}}}) \in \mathcal{V}_{\mat{m}_\Gamma} \times \mathcal{V}_{\vek{u}} \times \mathcal{V}_{\omega_{\vek{t}}}$ there holds
\begin{multline}\label{eq: weak form equation 1}
    \int_{\Gamma^h} \mat{V}_{\mat{m}_\Gamma} \FrobProd
    \big(
        -\vek{\varepsilon}_{\Gamma,\text{Bend}}(\mat{m}_\Gamma) + \mat{H} \cdot \nabla^\text{dir}_\Gamma \vek{u}  
    \big)
    +\text{div}_\Gamma\big(\mat{P} \cdot \mat{V}_{\mat{m}_\Gamma} \cdot \mat{P}\big) \ScalProd
    \big[
        (\nabla^\text{dir}_\Gamma \vek{u})^\text{T} \cdot \vek{n}
    \big]
    \, \text{d}\Gamma \\
    + \int_{\partial\Gamma} m_{\vek{q}}(\mat{V}_{\mat{m}_\Gamma}) \cdot \omega_{\vek{q}}(\vek{u})
    + m_{\vek{t}}(\mat{V}_{\mat{m}_\Gamma}) \cdot \omega_{\vek{t}} \, \text{d}\partial\Gamma = 0,
\end{multline}
\begin{multline}\label{eq: weak form equation 2}
    \int_{\Gamma}
    \big(
        \mat{H} \cdot \nabla^\text{dir}_\Gamma \vek{v}_{\vek{u}}
    \big) \FrobProd \mat{m}_\Gamma + (\nabla^\text{dir}_\Gamma \vek{v}_{\vek{u}})^\text{T} \cdot \vek{n} \ScalProd \text{div}_\Gamma \mat{m}_\Gamma
    + \nabla^\text{dir}_\Gamma \vek{v}_{\vek{u}} \FrobProd \tilde{\mat{n}}_\Gamma(\vek{u}) \, \text{d}\Gamma \\
    + \int_{\partial\Gamma} \omega_{\vek{q}}(\vek{v}_{\vek{u}}) \cdot m_{\vek{q}}(\mat{m}_\Gamma) \, \text{d}\partial\Gamma = \int_{\Gamma} \vek{v}_{\vek{u}} \ScalProd \vek{f} \, \text{d}\Gamma + \int_{\partial\Gamma_{\text{N},\vek{u}}}\!\!\!\!\!\!\! \vek{v}_{\vek{u}} \ScalProd \hat{\tilde{\vek{p}}} \, \text{d}\partial\Gamma + \sum_{i=1}^{n_{C,\text{N}}} \vek{v}_{\vek{u}}|_{C_i} \ScalProd \vek{n}|_{C_i} \cdot \hat{F}_{C},
\end{multline}
\begin{equation}\label{eq: weak form equation 3}
	\int_{\partial\Gamma_{\text{N},\omega}}\!\!\!\!\!\!\! v_{\omega_{\vek{t}}} \cdot m_{\vek{t}}(\mat{m}_\Gamma) \, \text{d}\partial\Gamma = \int_{\partial\Gamma_{\text{N},\omega}}\!\!\!\!\!\!\! v_{\omega_{\vek{t}}} \cdot \hat{m}_{\vek{t}} \, \text{d}\partial\Gamma.
\end{equation}
A detailed derivation may be found in Appendix~\ref{app: Derivation of the mixed weak form}.

Discretizing the weak form with suitable FE spaces as, e.g., spanned by classical $C^0$-continuous shape functions of Lagrange elements, leads to a linear system of equations, structured as
\begin{equation}\label{eq: system of equations mixed global}
	\begin{bmatrix}
		\mat{K}_{\mat{m}\mat{m}} & \mat{K}_{\mat{m}\vek{u}} & \mat{K}_{\mat{m}\omega} \\
		\mat{K}_{\vek{u}\mat{m}} & \mat{K}_{\vek{u}\vek{u}} & \mat{0} \\
		\mat{K}_{\omega\mat{m}} & \mat{0} & \mat{0}
	\end{bmatrix} \cdot
	\begin{bmatrix}
		\underline{\underline{\vek{m_\Gamma}}} \\
		\underline{\vek{u}} \\
		\vek{\omega_{\vek{t}}}
	\end{bmatrix} =
	\begin{bmatrix}
		\vek{0} \\
		\vek{b}_{\vek{u}} \\
		\vek{b}_{\omega} 
	\end{bmatrix}.
\end{equation}
with $[\underline{\underline{\vek{m_\Gamma}}}, \underline{\vek{u}}, \vek{\omega_{\vek{t}}}]^\text{T} = [\vek{m}_{11}, \vek{m}_{22}, \vek{m}_{33}, \vek{m}_{12}, \vek{m}_{13}, \vek{m}_{23}, \vek{u}, \vek{v}, \vek{w}, \vek{\omega_{\vek{t}}}]^\text{T}$ being the sought nodal components of the moment tensor, the displacement vector, and the Lagrange multiplier. $\mat{K}_{ij}$ and $\vek{b}_i$ are the entries of the stiffness matrix and the right-hand side, respectively, deduced from Eqs.~(\ref{eq: weak form equation 1})-(\ref{eq: weak form equation 3}).

\subsection{Discretization and hybridization}\label{sec: Discretization and hybridization}
Compared to the purely displacement-based formulation \cite{Schoellhammer_2019a}, the moment tensor $\mat{m}_\Gamma$ is calculated directly as a primary variable in this mixed approach and is not derived from the displacement vector $\vek{u}$ in a post-processing step. As seen later, this direct evaluation leads to a higher-order convergence of $\mat{m}_\Gamma$ but also to a significant increase in the number of DOFs.
Together with the fact that the stiffness matrix in Eq.~(\ref{eq: system of equations mixed global}) is also indefinite, the system of equations is more expensive to solve. Following \cite{Boffi_2013a}, a hybridization scheme is employed to address this problem.

\begin{figure}
	\centering
	
	\subfigure[discretized surface $\Gamma^h$]
	{\includegraphics[width=0.3\textwidth]{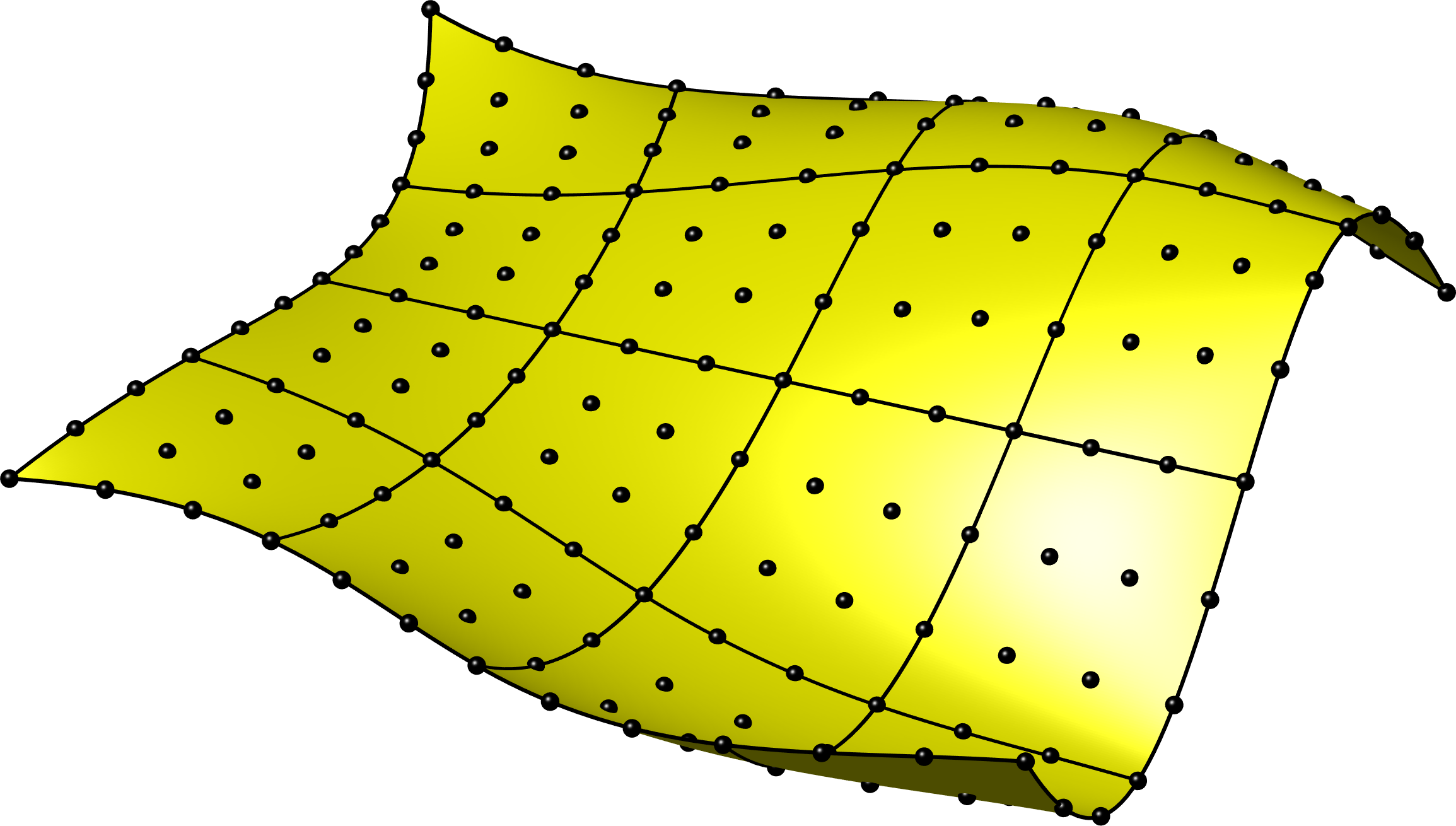}\label{fig: surface mesh}}\hfill
	\subfigure[discretized boundary $\partial\Gamma^h$]
	{\includegraphics[width=0.3\textwidth]{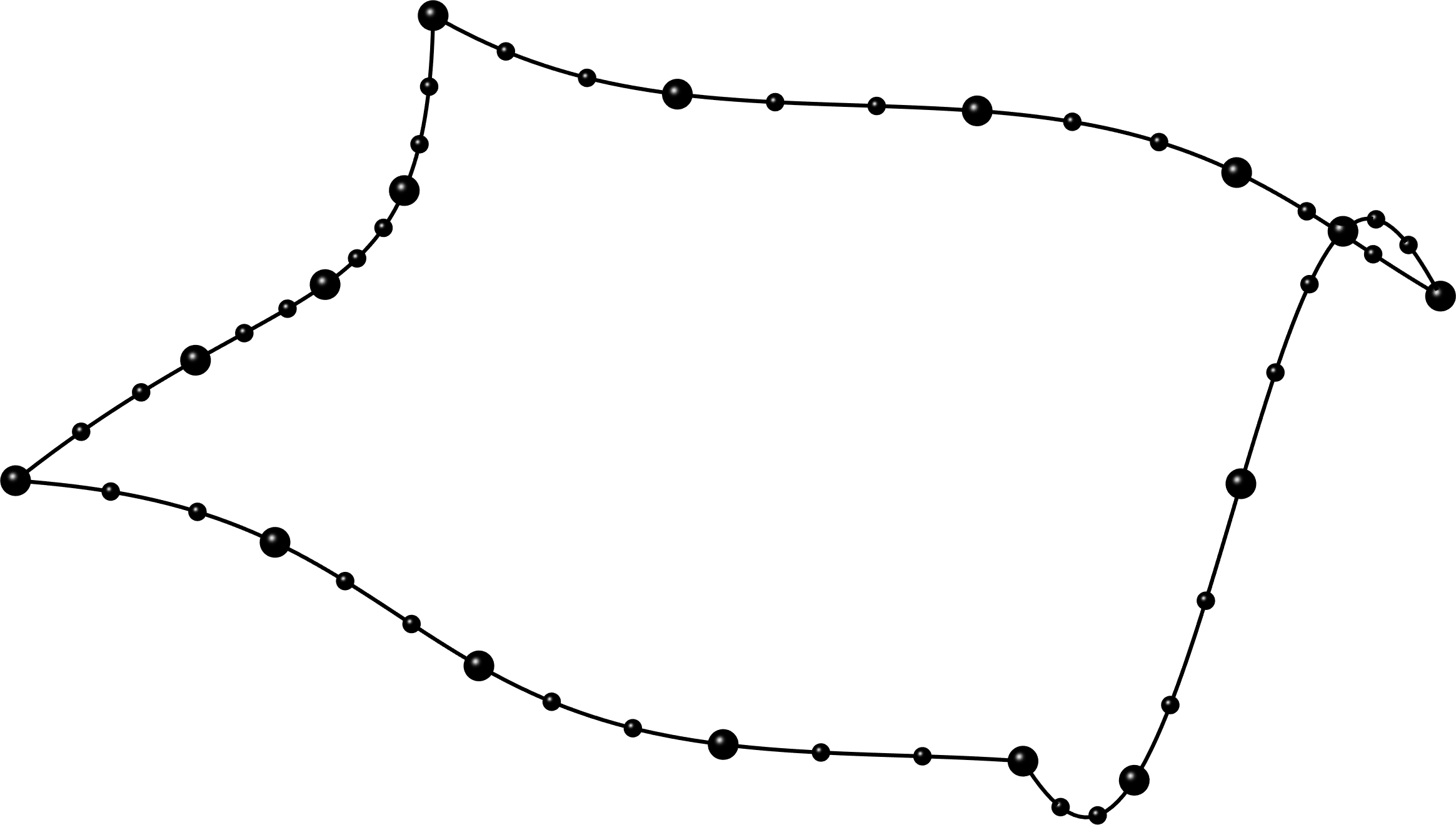}\label{fig: boundary mesh}}\hfill
	\subfigure[element interfaces $\Psi^h$]
	{\includegraphics[width=0.3\textwidth]{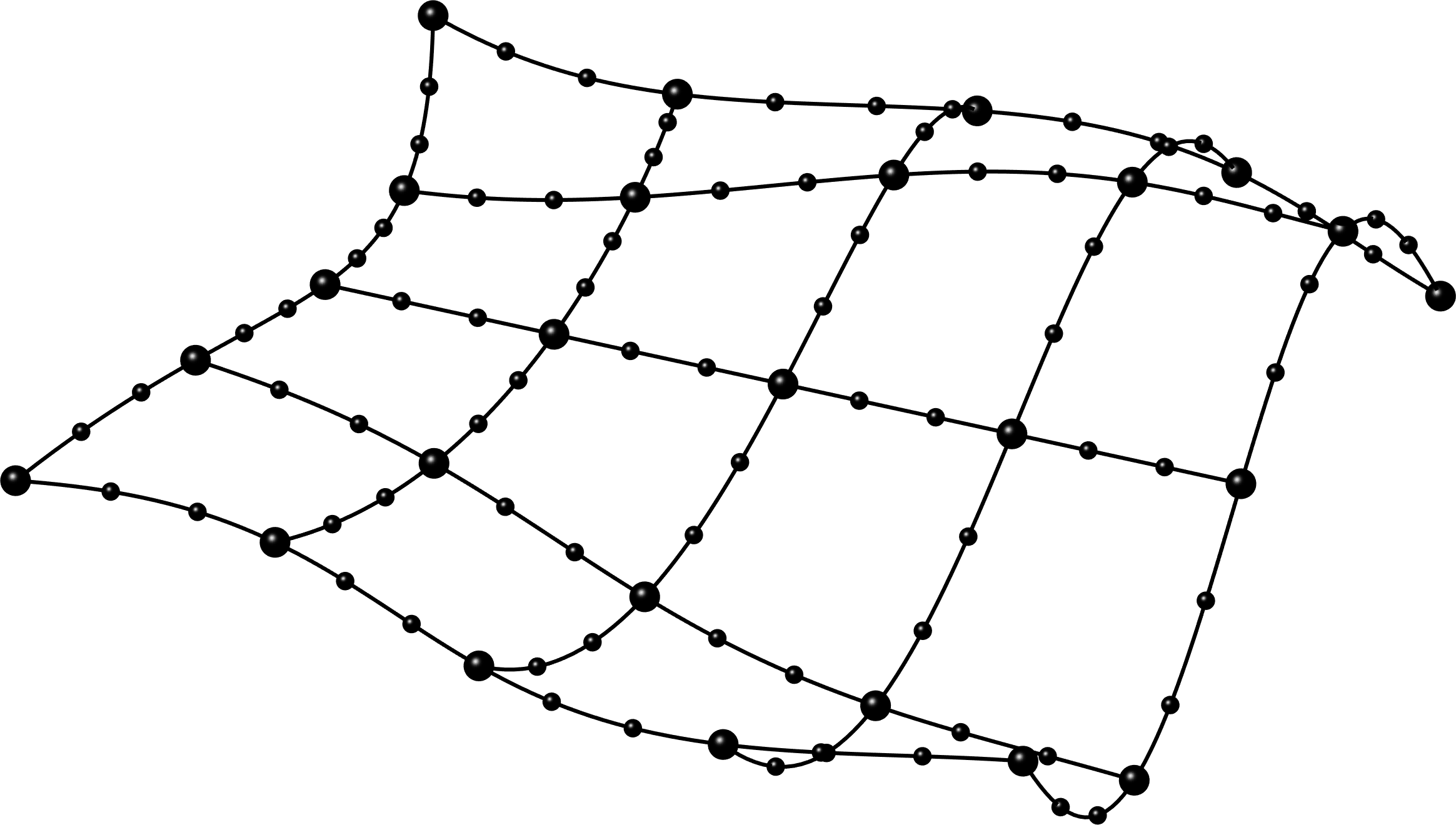}\label{fig: skeleton mesh}}
	
	\caption{\label{fig: meshes}For some surface $\Gamma$, (a) an example surface mesh composed by $4 \times 4$ cubic elements representing $\Gamma^h$ is shown, (b) the discretized boundary $\partial\Gamma^h$, and (c) the skeleton $\Psi^h$ composed by the element edges.}
\end{figure}

Before that, let there be a discretization of the surface $\Gamma$ and its boundary $\partial\Gamma$. Entities with the index $h$ indicate that these are discretized quantities. Subsequently, a conforming mesh $\Gamma^h$ and a conforming boundary mesh $\partial\Gamma^h$ are generated in the three-dimensional space. The surface $\Gamma^h$ is discretized by two-dimensional quadrilateral (or triangular) Lagrange elements so that its boundary $\partial\Gamma^h$ is discretized by one-dimensional Lagrange elements. $\Gamma^h$ and $\partial\Gamma^h$ are illustrated for an arbitrary surface in Fig.~\ref{fig: surface mesh} and Fig.~\ref{fig: boundary mesh}, respectively.

For the hybridization, the continuity between elements referring to the field of the moment tensor $\mat{m}_\Gamma$ is broken and subsequently reinforced weakly by a hybridization variable $\omega_{\vek{t}}$. In this context, the Lagrange multiplier $\omega_{\vek{t}}$ is applied to the set of edges of all elements $\Psi^h$, rather than solely to the boundary $\partial\Gamma^h$ as for the purely mixed case from above. $\Psi^h$ can be considered as the skeleton of $\Gamma^h$, consisting of one-dimensional elements in the three-dimensional space, as depicted in Fig.~\ref{fig: skeleton mesh}. In summary, the globally continuous finite element space is replaced by a local space for the field of $\mat{m}_\Gamma$. 

\begin{figure}
	\centering
	
	\subfigure[general situation]
	{\includegraphics[width=0.5\textwidth]{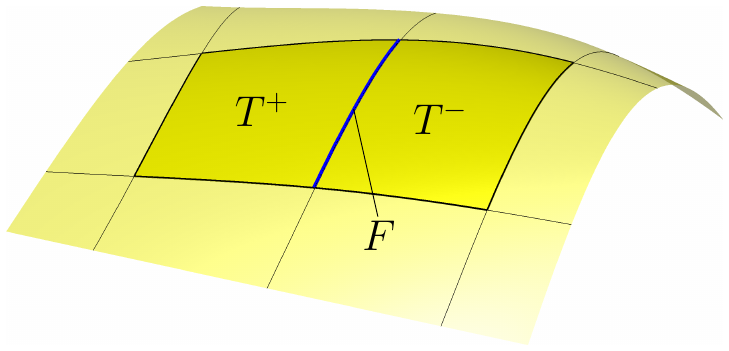}\label{fig: general position}}\hfill
	\subfigure[orientation of vectors]
	{\includegraphics[width=0.5\textwidth]{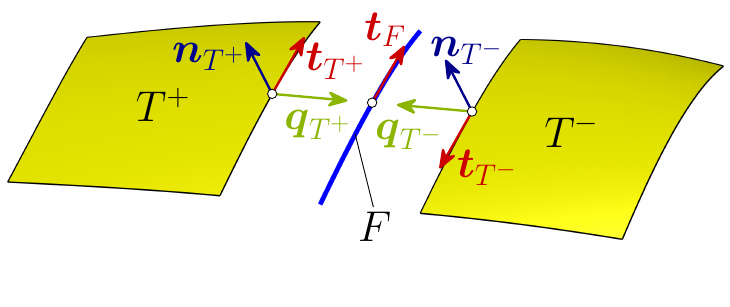}\label{fig: direction vectors}}
	
	\caption{\label{fig: edge orientation}Neighboring surface elements $T^+$ and $T^-$ and the edge element $F$ located between them.}
\end{figure}

Classic Lagrange elements with interpolating shape functions are used here, however, the methodology straightforwardly extends to hierarchic shape functions. The nodal coordinates in the surface mesh are labeled as $\vek{x}_i$ with $i = 1,\dots, n_p$, where $n_p$ is the number of nodes in the mesh. Note that for discontinuous fields, nodes are only associated locally with a single element, leading to multiple nodes with identical coordinates on shared element edges, a typical situation in DG-FEM. The nodal basis functions $N_i^p(\vek{x})$ of a certain order $p$, used as test and trial functions, are obtained in the usual manner from maps of one- or two-dimensional reference elements to the physical elements. Finite element spaces for each field are defined as
\begin{eqnarray}
    \mathcal{Q}_{\Gamma,\mat{m}_\Gamma}^h &:=&
    \begin{Bmatrix}
        v^h \in C^{-1}(\Gamma^h): v^h = \sum_{i=1}^{n_{p_{\mat{m}_\Gamma}}} N_i^{p_{\mat{m}_\Gamma}}(\vek{x}) \cdot \hat{v}_i \, \text{with} \, \hat{v}_i \in \mathbb{R}
    \end{Bmatrix}
        \subset \mathcal{L}^2(\Gamma^h) \label{eq: finite element space m},\\
        \mathcal{Q}_{\Gamma,\vek{u}}^h &:=&
    \begin{Bmatrix}
        v^h \in C^{0}(\Gamma^h): v^h = \sum_{i=1}^{n_{p_{\vek{u}}}} N_i^{p_{\vek{u}}}(\vek{x}) \cdot \hat{v}_i \, \text{with} \, \hat{v}_i \in \mathbb{R}
    \end{Bmatrix}
        \subset \mathcal{H}^1(\Gamma^h) \label{eq: finite element space u},\\
    \mathcal{Q}_{\Psi,\omega_{\vek{t}}}^h &:=&
    \begin{Bmatrix}
        v^h \in C^{-1}(\Psi^h): v^h = \sum_{i=1}^{n_{p_{\omega_{\vek{t}}}}} N_i^{p_{\omega_{\vek{t}}}}(\vek{x}) \cdot \hat{v}_i \, \text{with} \, \hat{v}_i \in \mathbb{R}
    \end{Bmatrix}
        \subset \mathcal{L}^2(\Psi^h) \label{eq: finite element space omega}.
\end{eqnarray}
Based on Eqs.~(\ref{eq: finite element space m})-(\ref{eq: finite element space omega}), the discrete test and trial function spaces are
\begin{eqnarray}
	\mathcal{S}_{\mat{m}_\Gamma}^h = \mathcal{V}_{\mat{m}_\Gamma}^h &=&
	\begin{Bmatrix}
		\mat{V}^h \in \big[\mathcal{Q}_{\Gamma,\mat{m}_\Gamma}^h\big]^{3 \times 3}: \mat{V}^h = {\mat{V}^h}^{\text{T}}
	\end{Bmatrix} \label{eq: h function space m}, \\
	\mathcal{S}_{\vek{u}}^h &=& 
	\begin{Bmatrix}
		\vek{v}^h \in \big[\mathcal{Q}_{\Gamma,\vek{u}}^h\big]^3: \vek{v}^h = \hat{\vek{u}} \:\: \text{on} \:\: \partial\Gamma_{\text{D},\vek{u}}^h
	\end{Bmatrix} \label{eq: h function space u}, \\
	\mathcal{V}_{\vek{u}}^h &=& 
	\begin{Bmatrix}
		\vek{v}^h \in \big[\mathcal{Q}_{\Gamma,\vek{u}}^h\big]^3: \vek{v}^h = \vek{0} \:\: \text{on} \:\: \partial\Gamma_{\text{D},\vek{u}}^h
	\end{Bmatrix} \label{eq: h function space vu}, \\
	\mathcal{S}_{\omega_{\vek{t}}}^h &=& 
	\begin{Bmatrix}
		v^h \in \mathcal{Q}_{\Psi,\omega_{\vek{t}}}^h: v^h = \hat{\omega}_{\vek{t}} \:\: \text{on} \:\: \partial\Gamma_{\text{D},\omega}^h
	\end{Bmatrix} \label{eq: h function space omega}, \\
	\mathcal{V}_{\omega_{\vek{t}}}^h &=&
	\begin{Bmatrix}
		v^h \in \mathcal{Q}_{\Psi,\omega_{\vek{t}}}^h: v^h = 0 \:\: \text{on} \:\: \partial\Gamma_{\text{D},\omega}^h
	\end{Bmatrix} \label{eq: h function space vomega}.
\end{eqnarray}
In order to weakly enforce (i) boundary conditions for $m_{\vek{t}}$ on $\partial \Gamma_{N,\omega}$ and (ii) the continuity of $m_{\vek{t}}$ on all element interfaces (those element edges that are shared by two surface elements), it is necessary to evaluate the conormal vector $\vek{q}$ and the tangential vector $\vek{t}$ on  $\Psi^h$. Note that $\vek{q}$ and $\vek{t}$ feature opposite orientations on the edges from two neighboring elements. Let us designate these elements $T^+$ and $T^-$ with the edge element $F$ between them. $F$ is oriented such that its tangent vector aligns with the one from the corresponding edge of element $T^+$ and is opposite to the one of $T^-$, i.e., $\vek{t}_{F} = \vek{t}_{T^+} = -\vek{t}_{T^-}$ see Fig.~\ref{fig: edge orientation} for details. In order to relate quantities on the edge elements $F$, i.e., $\omega_{\vek{t}}^h$ and $v_{\omega_{\vek{t}}}^h$, with quantities on the surface elements $T$, i.e., $\mat{m}_\Gamma^h$ and $\mat{V}_{\mat{m}_\Gamma}^h$, we need the jump operator defined as
\begin{equation}\label{eq: jump operator}
    \jumpl f \jumpr = 
    \begin{cases}
        f|_{T^+} - f|_{T^-}& \text{on} \:\: \Psi^h \setminus \partial\Gamma^h, \\
        f|_{T}&  \text{on} \:\: \Psi^h \cap \partial\Gamma^h.
    \end{cases}
\end{equation}

The discretized and hybridized variational problem is defined as follows. Find $\mat{m}_\Gamma^h \in \mathcal{S}_{\mat{m}_\Gamma}^h$, $\vek{u}^h \in \mathcal{S}_{\vek{u}}^h$, and $\omega_{\vek{t}}^h \in \mathcal{S}_{\omega_{\vek{t}}}^h$ such that for all test functions $(\mat{V}_{\mat{m}_\Gamma}^h, \vek{v}_{\vek{u}}^h, v_{\omega_{\vek{t}}}^h) \in \mathcal{V}_{\mat{m}_\Gamma}^h \times \mathcal{V}_{\vek{u}}^h \times \mathcal{V}_{\omega_{\vek{t}}}^h$, there holds
\begin{multline}\label{eq: hyb weak form equation 1}
    \int_{\Gamma^h} \mat{V}_{\mat{m}_\Gamma}^h \FrobProd
    \big(
        -\vek{\varepsilon}_{\Gamma,\text{Bend}}(\mat{m}_\Gamma^h) + \mat{H} \cdot \nabla^\text{dir}_\Gamma \vek{u}^h  
    \big) \, \text{d}\Gamma
    + \sum_{T \in \Gamma^h} \int_T \text{div}_\Gamma\big(\mat{P} \cdot \mat{V}_{\mat{m}_\Gamma}^h \cdot \mat{P}\big) \ScalProd
    \big[
        (\nabla^\text{dir}_\Gamma \vek{u}^h)^\text{T} \cdot \vek{n}
    \big]
    \, \text{d}\Gamma \\
    + \int_{\Psi^h} m_{\vek{q}}(\mat{V}_{\mat{m}_\Gamma}^h) \cdot \omega_{\vek{q}}(\vek{u}^h)
    + \jumpl m_{\vek{t}}(\mat{V}_{\mat{m}_\Gamma}^h) \jumpr \cdot \omega_{\vek{t}}^h \, \text{d}\Psi = 0,
\end{multline}
\begin{multline}\label{eq: hyb weak form equation 2}
    \int_{\Gamma^h}
    \big(
        \mat{H} \cdot \nabla^\text{dir}_\Gamma \vek{v}_{\vek{u}}^h
    \big) \FrobProd \mat{m}_\Gamma^h 
    + \nabla^\text{dir}_\Gamma \vek{v}_{\vek{u}}^h \FrobProd \tilde{\mat{n}}_\Gamma(\vek{u}^h) \, \text{d}\Gamma
    + \sum_{T \in \Gamma^h} \int_T (\nabla^\text{dir}_\Gamma \vek{v}_{\vek{u}}^h)^\text{T} \cdot \vek{n} \ScalProd \text{div}_\Gamma \mat{m}_\Gamma^h \, \text{d}\Gamma \\
    + \int_{\Psi^h} \omega_{\vek{q}}(\vek{v}_{\vek{u}}^h) \cdot m_{\vek{q}}(\mat{m}_\Gamma^h) \, \text{d}\Psi = \int_{\Gamma^h} \vek{v}_{\vek{u}}^h \ScalProd \vek{f}^h \, \text{d}\Gamma + \int_{\partial\Gamma_{\text{N},\vek{u}}^h}\!\!\!\!\!\!\! \vek{v}_{\vek{u}}^h \ScalProd \hat{\tilde{\vek{p}}}^h \, \text{d}\partial\Gamma + \sum_{i=1}^{n_{C,\text{N}}} \vek{v}_{\vek{u}}^h|_{C_i^h} \ScalProd \vek{n}|_{C_i^h} \cdot \hat{F}_{C}^h,
\end{multline}
\begin{equation}\label{eq: hyb weak form equation 3}
	\int_{\Psi^h} v_{\omega_{\vek{t}}}^h \cdot \jumpl m_{\vek{t}}(\mat{m}_\Gamma^h) \jumpr \, \text{d}\Psi = \int_{\partial\Gamma_{\text{N},\omega}^h}\!\!\!\!\!\!\! v_{\omega_{\vek{t}}}^h \cdot \hat{m}_{\vek{t}}^h \, \text{d}\partial\Gamma.
\end{equation}
Note that for geometric quantities such as $\vek{n}$ and differential operators such as $\nabla_\Gamma$, when referring to the discretized setup, we still do \emph{not} write $\nabla_\Gamma^h$ or $\vek{n}^h$, that is, the additional index $h$ is avoided for brevity there.

Deduced from Eqs.~(\ref{eq: hyb weak form equation 1})-(\ref{eq: hyb weak form equation 3}), the local stiffness matrix $\mat{K}^\text{el}$ and the local right hand side $\vek{b}^\text{el}$ is determined as
\begin{equation}\label{eq: stiffness matrix and right hand side}
    \mat{K}^\text{el} =
    \begin{bmatrix}
        \mat{K}_{\mat{m}\mat{m}}^\text{el} & \mat{K}_{\mat{m}\vek{u}}^\text{el} & \mat{K}_{\mat{m}\omega}^\text{el} \\
        \mat{K}_{\vek{u}\mat{m}}^\text{el} & \mat{K}_{\vek{u}\vek{u}}^\text{el} & \mat{0} \\
        \mat{K}_{\omega\mat{m}}^\text{el} & \mat{0} & \mat{0}
    \end{bmatrix}, \quad
    \vek{b}^\text{el} =
    \begin{bmatrix}
        \vek{0} \\
        \vek{b}_{\vek{u}}^\text{el} \\
        \vek{b}_{\omega}^\text{el} 
    \end{bmatrix}.
\end{equation}
The resulting definitions of the individual entries of $\mat{K}^\text{el}$ and $\vek{b}^\text{el}$ are given in Appendix~\ref{app: Entries of the local stiffness matrix}. The field of the moment tensor $\mat{m}_\Gamma$, as it is entirely local, may now be eliminated element-wise by static condensation via the Schur complement, as usual in mixed-hybrid FEMs \cite{Arnold_1985a, Cockburn_2004a, Boffi_2013a},
\begin{equation}\label{eq: static condensation}
    \tilde{\mat{K}}^\text{el} =
    \begin{bmatrix}
        \mat{K}_{\vek{u}\vek{u}}^\text{el} & \mat{0} \\
        \mat{0} & \mat{0}
    \end{bmatrix} -
    \begin{bmatrix}
        \mat{K}_{\vek{u}\mat{m}}^\text{el} \\
        \mat{K}_{\omega\mat{m}}^\text{el}
    \end{bmatrix} \cdot
    \begin{bmatrix}
        \mat{K}_{\mat{m}\mat{m}}^\text{el}	
    \end{bmatrix}^{-1} \cdot
    \begin{bmatrix}
        \mat{K}_{\mat{m}\vek{u}}^\text{el} & \mat{K}_{\mat{m}\omega}^\text{el}
    \end{bmatrix}, \quad
    \tilde{\vek{b}}^\text{el} =
    \begin{bmatrix}
        \vek{b}_{\vek{u}}^\text{el} \\
        \vek{b}_{\omega}^\text{el} 
    \end{bmatrix},
\end{equation}
with $\tilde{\mat{K}}^\text{el}$ and $\tilde{\vek{b}}^\text{el}$ being the local, condensed versions of the stiffness matrix and the right-hand side, respectively. The assembly of the condensed element entries results in the global, condensed stiffness matrix $\tilde{\mat{K}}$ and right-hand side $\tilde{\vek{b}}$ representing a linear system of equations where only the displacement fields $\vek{u}$ and $\omega_{\vek{t}}$ remain as primary variables
\begin{equation}\label{eq:condensed system of equations}
    \tilde{\mat{K}} \cdot
    \begin{bmatrix}
        \underline{\vek{u}} \\
        \vek{\omega_{\vek{t}}}
    \end{bmatrix}
    = \tilde{\vek{b}}.
\end{equation}
The stiffness matrix is significantly smaller and positive-definite after the hybridization and condensation. These properties confer a considerable advantage regarding the efficient solvability of the system of equations. Provided that the Dirichlet boundary conditions for $\vek{u}$ and $\omega_{\vek{t}}$ have been considered (herein by strongly prescribing the nodal values), the resulting linear system of equations can be solved accordingly.

In a post-processing step, the previously condensed field of the moment tensor $\mat{m}_\Gamma$ can be recovered element-wise by
\begin{equation}\label{eq: post-processing m}
	\underline{\underline{\vek{m_\Gamma^\text{el}}}} =
	\begin{bmatrix}
		\mat{K}_{\mat{m}\mat{m}}^\text{el}	
	\end{bmatrix}^{-1} \cdot
	\begin{bmatrix}
		\mat{K}_{\mat{m}\vek{u}}^\text{el} & \mat{K}_{\mat{m}\omega}^\text{el} \\
	\end{bmatrix} \cdot
	\begin{bmatrix}
		\underline{\vek{u}}^\text{el} \\
		\vek{\omega_{\vek{t}}^\text{el}}
	\end{bmatrix}.
\end{equation}
Concerning higher-order convergence, the moment tensor $\mat{m}_\Gamma$ in the hybridized formulation obtained by Eq.~(\ref{eq: post-processing m}) is, except for some minor rounding errors, equivalent to the one of the simple mixed formulation without static condensation.

\section{Numerical results}\label{sec: Numerical results}

Various test cases concerning the performance of the mixed-hybrid Kirchhoff--Love shell model are presented. Some of the test cases are (variations of) classical benchmarks; however, new test cases allowing for smooth solutions and, thereby, enabling higher-order convergence rates are also proposed. Therefore, different error measures are introduced next. For classical benchmark tests, often only point-wise displacements are available as reference solutions. Other useful error measures may be based on the $\mathcal{L}^2$-error, residual errors, and the stored energy error and enable systematic convergence studies in the context of the $hp$-FEM. In the present context based on a mixed formulation, it is noteworthy that the element orders of the components of $\vek{u}$ and $\mat{m}_{\Gamma}$ and the Lagrange multiplier $\omega_{\vek{t}}$ are equal, hence, $p = p_{\mat{m}_\Gamma} = p_{\vek{u}} = p_{\omega_{\vek{t}}}$.

For the evaluation of the $\mathcal{L}^2$-error, the analytic solution for some test cases must be known and can then be compared to the approximated FEM-solution. The relative $\mathcal{L}^2$-error is
\begin{equation}\label{eq:L2-error}
    \varepsilon_{\mathcal{L}^2,\text{rel}}^2(f) = \frac{\int_{\Gamma^h} (f^{\text{ex}} - f^h)^2 \, \text{d}\Gamma}{\int_{\Gamma^h} (f^{\text{ex}})^2 \, \text{d}\Gamma},
\end{equation}
where the superscript "ex" and $h$ indicate the analytical and approximated solution  of a certain quantity $f$, respectively.

For the calculation of the residual errors, the approximate solutions $\mat{m}_\Gamma^h$ and $\vek{u}^h$ are inserted into the strong form, see Eqs.~(\ref{eq: strong form equation 1 repeat}) and (\ref{eq: strong form equation 2 repeat}), and residuals are computed. The residuals are zero for the analytical solution, however, for the approximate solution, an error remains. An element-wise integration over all elements results in the corresponding error measure:
\begin{equation}\label{eq:ResidualError1}
    \begin{gathered}
        \varepsilon_{\text{res},1}^2 = \sum_{T \in \Gamma^h} \int_T \mathfrak{r}_1(\mat{m}_\Gamma^h,\vek{u}^h) \FrobProd \mathfrak{r}_1(\mat{m}_\Gamma^h,\vek{u}^h) \, \text{d}\Gamma,\\
        \text{with} \quad \mathfrak{r}_1(\mat{m}_\Gamma^h,\vek{u}^h) = -\vek{\varepsilon}_{\Gamma,\text{Bend}}(\mat{m}_\Gamma^h) + \vek{\varepsilon}_{\Gamma,\text{Bend}}(\vek{u}^h),
    \end{gathered}  
\end{equation}
\begin{equation}\label{eq:ResidualError2}
    \begin{gathered}
        \varepsilon_{\text{res},2}^2 = \sum_{T \in \Gamma^h} \int_T \mathfrak{r}_2(\mat{m}_\Gamma^h,\vek{u}^h) \ScalProd \mathfrak{r}_2(\mat{m}_\Gamma^h,\vek{u}^h) \, \text{d}\Gamma,\\
        \text{with} \quad \mathfrak{r}_2(\mat{m}_\Gamma^h,\vek{u}^h) = \text{div}_\Gamma \mat{n}_\Gamma^\text{real}(\mat{m}_\Gamma^h,\vek{u}^h) + \vek{n} \cdot \text{div}_\Gamma \big(\mat{P} \cdot \text{div}_\Gamma \mat{m}_\Gamma^h\big) + \mat{H} \cdot \text{div}_\Gamma \mat{m}_\Gamma^h + \vek{f}^h.
    \end{gathered}  
\end{equation}
For the second residual, a \emph{relative} error can be achieved for non-zero load vectors $\vek{f}^h$ by dividing Eq.~(\ref{eq:ResidualError2}) through $\sum_{T \in \Gamma^h} \int_T {\vek{f}^h}^2 \, \text{d}\Gamma$. With second-order derivatives of $\vek{u}^h$ and $\mat{m}_\Gamma^h$ required for evaluating the residuals, a convergence rate of $\mathcal{O}(p-1)$ is expected in both these error measures. The residual error based on Eq.~\ref{eq:ResidualError2} has also been analyzed for the displacement-based formulation in \cite{Schoellhammer_2019a}. There, the moment tensor was obtained from higher-order derivatives of $\vek{u}^h$ and, hence, only a convergence of $\mathcal{O}(p-3)$ was achieved.

To assess the quality of the enforcement of boundary conditions, another residual error based on the force equilibrium on the Neumann boundary $\partial\Gamma_{\text{N},\vek{u}}$ is defined as
\begin{equation}\label{eq:ResidualErrorBoundary}
    \begin{gathered}
        \varepsilon_{\text{res},\text{bound}}^2 = \sum_{F \in \partial\Gamma_{\text{N},\vek{u}}^h} \int_F \mathfrak{r}_{\text{bound}}(\mat{m}_\Gamma^h,\vek{u}^h) \ScalProd \mathfrak{r}_{\text{bound}}(\mat{m}_\Gamma^h,\vek{u}^h) \, \text{d}\partial\Gamma,\\
        \text{with} \quad \mathfrak{r}_{\text{bound}}(\mat{m}_\Gamma^h,\vek{u}^h) = \tilde{\vek{p}}(\mat{m}_\Gamma^h,\vek{u}^h) - \hat{\tilde{\vek{p}}}^h.
    \end{gathered}
\end{equation}
A relative error can be evaluated by dividing through $\sum_{F \in \partial\Gamma_{\text{N},\vek{u}}^h} \int_F {\hat{\tilde{\vek{p}}}^h}^2 \, \text{d}\partial\Gamma$ for non-zero line load vectors $\hat{\tilde{\vek{p}}}^h$.

The relative stored energy error is computed as
\begin{equation}\label{eq:EnergyError}
    \varepsilon_{\mathfrak{e}} = \frac{|\mathfrak{e}_{\text{ref}} - \mathfrak{e}(\mat{m}_\Gamma^h,\vek{u}^h)|}{\mathfrak{e}_{\text{ref}}},
\end{equation}
where the approximated stored elastic energy is defined as
\begin{equation}\label{eq:StoredEnergy}
    \mathfrak{e}(\mat{m}_\Gamma^h,\vek{u}^h) = \frac{1}{2} \int_{\Gamma^h} \vek{\varepsilon}_{\Gamma,\text{Memb}}(\vek{u}^h) \FrobProd \tilde{\mat{n}}_\Gamma(\vek{u}^h) + \vek{\varepsilon}_{\Gamma,\text{Bend}}(\mat{m}_\Gamma^h) \FrobProd \mat{m}_\Gamma^h \, \text{d}\Gamma.
\end{equation}
The reference energy $\mathfrak{e}_{\text{ref}}$ may be determined with an overkill solution using extremely fine meshes and higher-order elements. The error of the stored energy, evaluated by the square of the stress resultants, is expected to converge with a rate of $\mathcal{O}(2p)$, see, e.g., \cite[p.~229]{Zienkiewicz_2013a}. This type of energy error shall not be mistaken by the classical energy error norm as described in \cite[p.~494]{Zienkiewicz_2013a}.

\subsection{Scordelis--Lo roof}\label{sec: Scordelis--Lo roof}
\begin{figure}
	\centering
	
	\subfigure[geometry]
	{\includegraphics[width=0.5\textwidth]{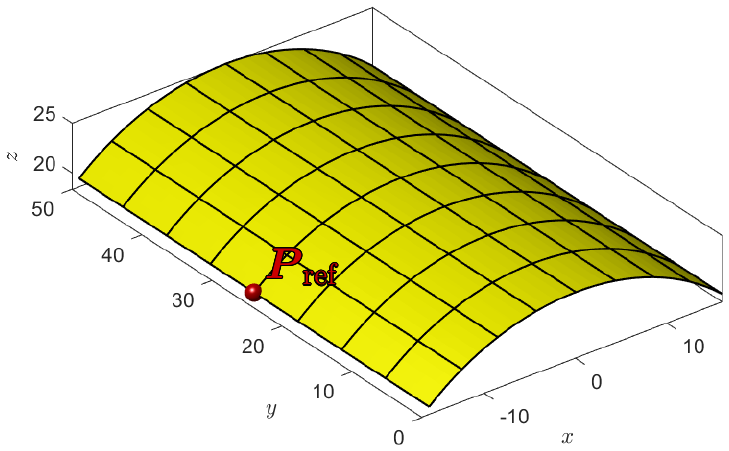}\label{fig:SLRGeometry}}\hfill
	\subfigure[displacements]
	{\includegraphics[width=0.5\textwidth]{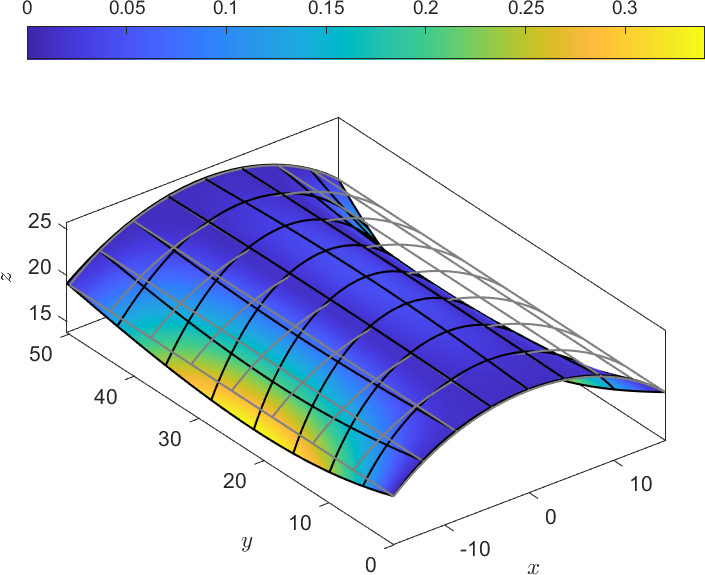}\label{fig:SLRDisplacement}}
	
	\caption{\label{fig:ScordelisLoRoof}(a) Geometry of the Scordelis--Lo roof. The red dot indicates the reference point $\vek{P}_{\text{ref}}$ for the vertical benchmark displacement $w_i$. (b) Scaled deformed configuration with the Euclidean norm of the displacement $\|\vek{u}\|$ as a color plot. The gray mesh lines depict the undeformed configuration.}
\end{figure}

As a part of the well-known \textit{shell obstacle course} in \cite{Belytschko_1985a}, the Scordelis--Lo roof is a frequently used benchmark test to verify shell models in various publications, e.g., \cite{Kiendl_2009a, Echter_2013a, Schoellhammer_2019a, Rafetseder_2019a}. The geometric setup and the resulting displacements of the Scordelis--Lo roof are depicted in \ref{fig:SLRGeometry} and \ref{fig:SLRDisplacement}, respectively. For the convergence study, the vertical displacement $w_i$ of the reference point $\vek{P}_{\text{ref}}$ is compared to a reference solution of $w_{\text{ref}} = 0.3006$ evaluated in \cite{Kiendl_2009a}, which is also the recommended solution of \cite{Krysl_2023a}. Additional required information is summarized in Tab.~\ref{tab:ScordelisLoRoof}.

\begin{table}
    \centering
    \begin{tabular}{|r|c|}
        \toprule
        \multirow{2}{*}{Geometry:}& cylindrical shell: length in $y$-direction $L_y = 50$,\\
        &radius $R = 25$, central angle $\theta = 80^{\circ}$\\ \hline
        Thickness:& $t = 0.25$\\ \hline
        Material:& $E = 4.32 \cdot 10^{8}$, $\nu = 0.0$\\ \hline
        Load:& $\vek{f} = [0,0,-90]^\text{T}$\\ \hline
        BCs:& rigid diaphragm at $y = 0$ and $y = 50$, free everywhere else\\ \hline
        Ref. point:& $\vek{P}_{\text{ref}} = [\pm R \cos(50^{\circ}), 25, R \sin(50^{\circ})]^\text{T}$\\ \hline
        Ref. solution \cite{Kiendl_2009a}:& $w_{\text{ref}} = -0.3006$\\
         \bottomrule
    \end{tabular}
    \caption{Collection of data for the Scordelis--Lo roof.}
    \label{tab:ScordelisLoRoof}
\end{table}

\begin{figure}
	\centering

	{\includegraphics[width=0.5\textwidth]{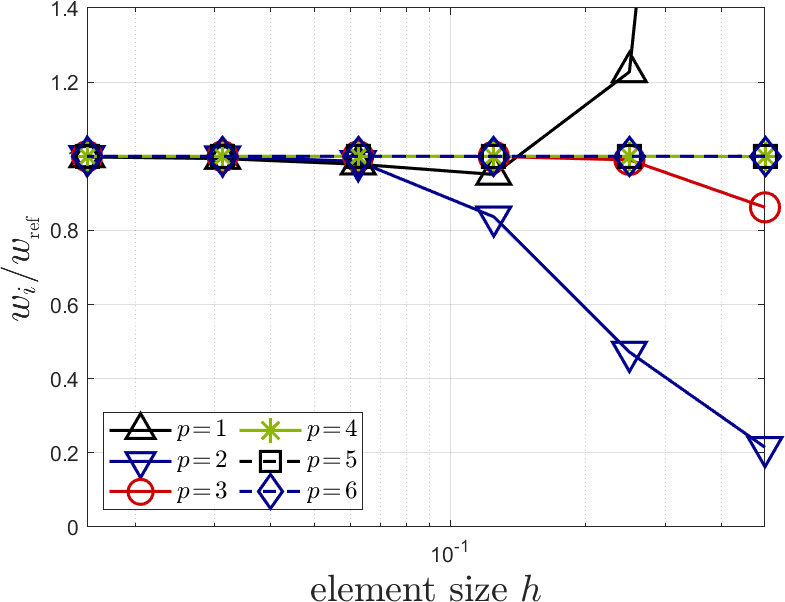}}
	
	\caption{\label{fig:SLRError}Normalized convergence of the reference displacement $w_{\text{ref}} = -0.3006$ for the Scordelis--Lo roof.}
\end{figure}

The convergence analysis in Fig.~\ref{fig:SLRError} clearly shows that the approximations converge well to the reference solution. Higher-order elements are able to achieve the expected reference displacement even on rather coarse meshes. The exact solution converges to $w_i = -0.30059246$, which is in perfect agreement to the results of \cite{Coox_2017a}. However, it is also noted that this classical test case does not allow for \emph{optimal} convergence rates when using higher-order elements due to the presence of singularities in the stresses and strains.

\subsection{Partly clamped hyperbolic paraboloid}\label{sec: Partly clamped hyperbolic paraboloid}

\begin{figure}
	\centering
	
	\subfigure[geometry]
	{\includegraphics[width=0.5\textwidth]{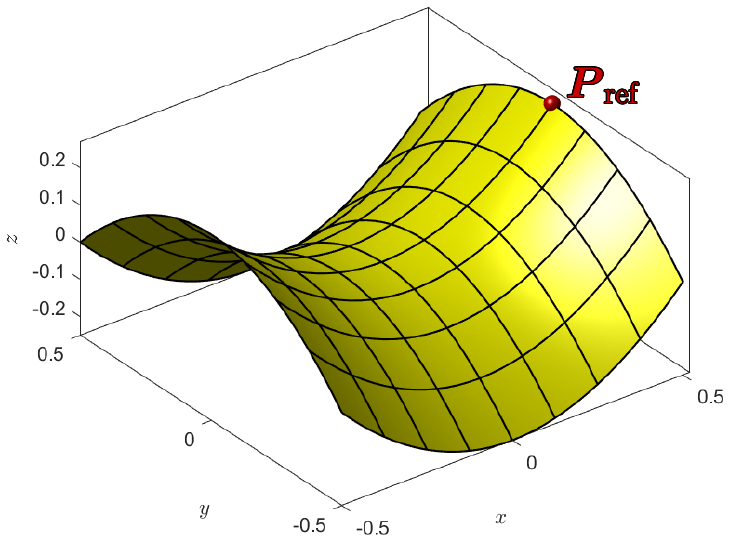}\label{fig:PCHPGeometry}}\hfill
	\subfigure[displacements]
	{\includegraphics[width=0.5\textwidth]{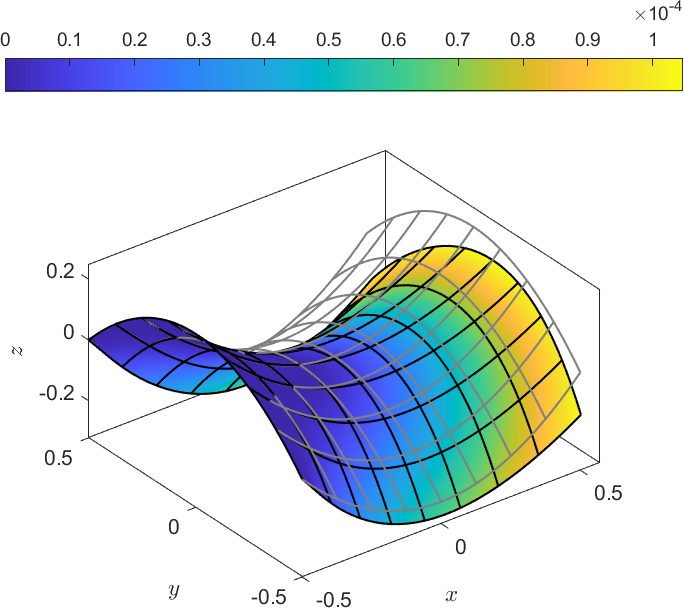}\label{fig:PCHPDisplacement}}
	
	\caption{\label{fig:PartlyClampedHyperbolicParaboloid}(a) Geometry of the partly clamped hyperbolic paraboloid. The red dot indicates the reference point $\vek{P}_{\text{ref}}$ for the vertical benchmark displacement $w_i$. (b) Scaled deformed configuration with the Euclidean norm of the displacement $\|\vek{u}\|$ as a color plot for the case $t = 0.01$. The gray mesh lines depict the undeformed configuration.}
\end{figure}

Compared to the Scordelis--Lo roof, the partly clamped hyperbolic paraboloid introduced in \cite{Chapelle_1998a} represents a more complex test case due to its double curved geometry as illustrated in Fig.~\ref{fig:PCHPGeometry}. Fig.~\ref{fig:PCHPDisplacement} depicts the displacements and Tab.~\ref{tab:PartlyClampedHyperbolicParaboloid} summarizes the system data of the test case. As for the Scordelis--Lo roof, the hyperbolic paraboloid has been used in various publications, e.g., \cite{Bathe_2000a, Rafetseder_2019a, Balobanov_2019a}, to verify shell models using a pointwise vertical reference displacement. As proposed in \cite{Rafetseder_2019a}, we examine three different values for the thickness $t$ to investigate the effect of locking.

\begin{table}
    \centering
    \begin{tabular}{|r|c|c|c|}
        \toprule
        & Case~1& Case~2& Case~3\\
        \midrule
        Geometry:& \multicolumn{3}{c|}{$\hat{x},\hat{y} \in [-0.5,0.5] \Rightarrow \vek{x}(\hat{\vek{x}}) = \left[ \begin{array}{c}  \hat{x} \\ \hat{y} \\ \hat{x}^2 - \hat{y}^2 \end{array}\right]$}\\ \hline
        Thickness:& $t = 0.01$& $t = 0.001$& $t = 0.0001$\\ \hline
        Material:& \multicolumn{3}{c|}{$E = 2.0 \cdot 10^{11}$, $\nu = 0.3$}\\ \hline
        Load:& \multicolumn{3}{c|}{$\vek{f} = [0,0,-8000 \cdot t]^\text{T}$}\\ \hline
        BCs:& \multicolumn{3}{c|}{clamped support at $x = -0.5$, free everywhere else}\\ \hline
        Ref. point:& \multicolumn{3}{c|}{$\vek{P}_{\text{ref}} = [0.5, 0, 0.25]^\text{T}$}\\ \hline
        Ref. solution \cite{Rafetseder_2019a}:& $w_{\text{ref}} = -9.3327 \! \cdot \! 10^{-5}$& $w_{\text{ref}} = -6.3955 \! \cdot \! 10^{-3}$& $w_{\text{ref}} = -5.2948 \! \cdot \! 10^{-1}$\\
        \bottomrule
    \end{tabular}
    \caption{Collection of data for the partly clamped hyperbolic paraboloid.}
    \label{tab:PartlyClampedHyperbolicParaboloid}
\end{table}

\begin{figure}
	\centering
	
	\subfigure[Case~1]
	{\includegraphics[width=0.33\textwidth]{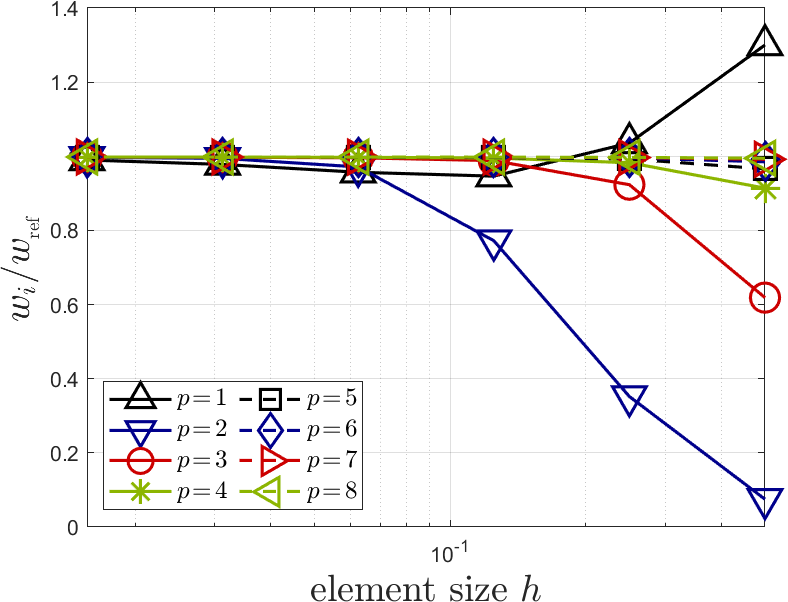}\label{fig:PCHPError1}}\hfill
	\subfigure[Case~2]
	{\includegraphics[width=0.33\textwidth]{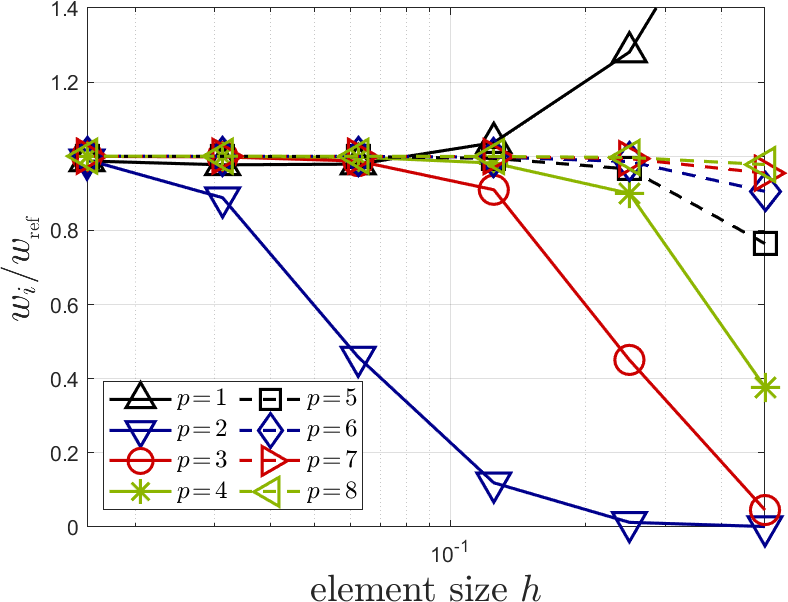}\label{fig:PCHPError2}}\hfill
	\subfigure[Case~3]
	{\includegraphics[width=0.33\textwidth]{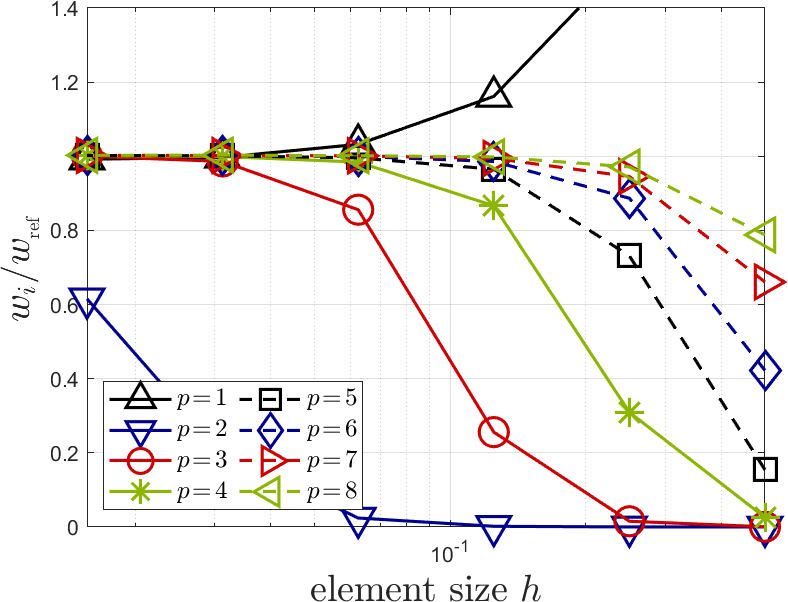}\label{fig:PCHPError3}}
	
	\caption{\label{fig:PCHPError}Normalized convergence of the reference displacement $w_{\text{ref}}$ for the partly clamped hyperbolic paraboloid for different thicknesses $t$. (a) Case~1 with $t = 0.01$, $w_{\text{ref}} = 9.3327 \cdot 10^{-5}$, (b)  Case~2 with $t = 0.001$, $w_{\text{ref}} = 6.3955 \cdot 10^{-3}$, (c) Case~3 with $t = 0.0001$, $w_{\text{ref}} = 5.2948 \cdot 10^{-1}$.}
\end{figure}

As shown in Fig.~\ref{fig:PCHPError}, the numerical results converge to the reference solutions with increasing mesh resolution; higher-order elements generally converge better. Yet this classical test case does not allow for optimal convergence rates in the above mentioned error measures for lack of smoothness in the resulting mechanical fields. Fig.~\ref{fig:PCHPError1} depicts the convergence study of case~1 with the highest thickness of $t = 0.01$. For this case, all element orders converge to the reference displacement at some point. However, mild locking behaviors are observed for lower element orders and coarser meshes. These locking phenomena intensify with decreasing thickness for case~2 and even more for case~3 as seen in Figs.~\ref{fig:PCHPError2} and ~\ref{fig:PCHPError3}. This behavior is in agreement with the results of \cite{Rafetseder_2019a} since the presented model does not address a locking-free formulation.

\subsection{Extruded arc}\label{sec: Sliced arc}

\begin{figure}
	\centering
	
	\subfigure[geometry]
	{\includegraphics[width=0.5\textwidth]{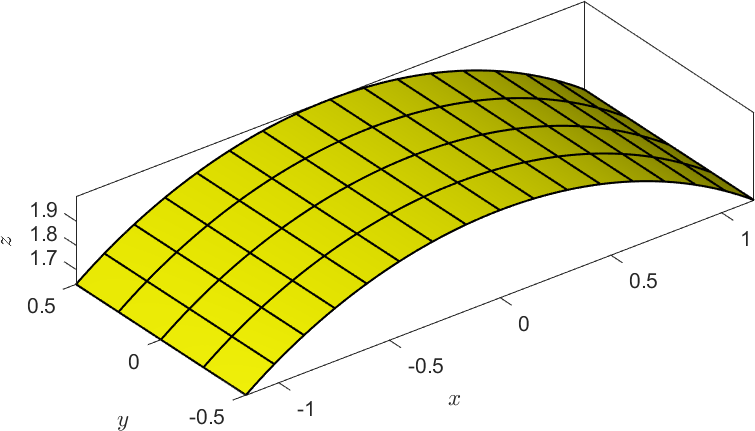}\label{fig:SAGeometry}}\hfill
	\subfigure[displacements]
	{\includegraphics[width=0.5\textwidth]{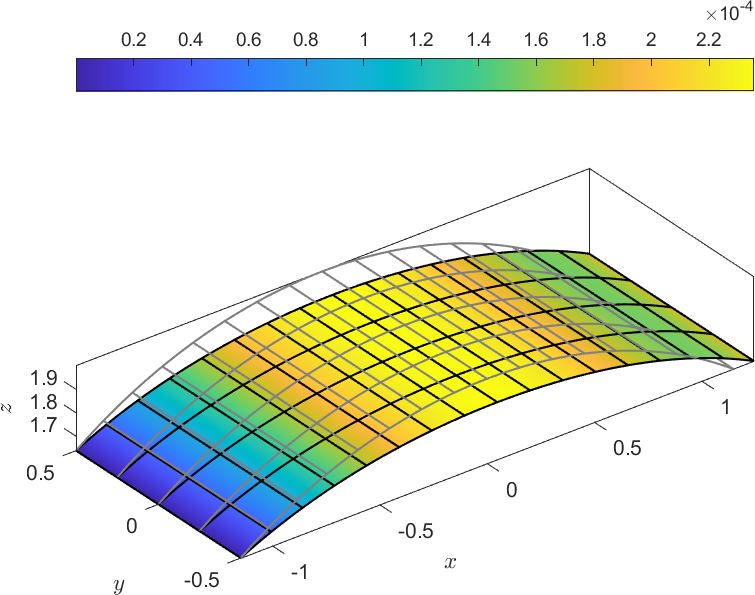}\label{fig:SADisplacement}}
	
	\caption{\label{fig:SlicedArc}(a) Geometry of the extruded arc. (b) Scaled deformed configuration with the Euclidean norm of the displacement $\|\vek{u}\|$ as a color plot. The gray mesh lines depict the undeformed configuration.}
\end{figure}

The previous test cases only enabled a verification based on point-wise reference displacements. The next test case features the geometry of an extruded arc where, for the present boundary conditions and load, an analytic (and smooth) solution is available and given in Appendix~\ref{app: Analytical solution of the sliced arc}. This enables a more systematic convergence study, confirming higher-order convergence rates in the $\mathcal{L}^2$-errors. The geometry and the displacements of the test case are visualized in Figs.~\ref{fig:SAGeometry} and \ref{fig:SADisplacement}, respectively, and details are summarized in Tab.~\ref{tab:SlicedArc}. This test case is the extruded version of a popular curved beam test case, e.g., from \cite{Kaiser_2023a}. Therefore, Poisson's ratio is set to $\nu = 0$ to avoid any lateral response. As a result, only one non-zero eigenvalue of the physical normal force tensor $\mat{n}_{\Gamma}^{\text{real}}$ and the moment tensor $\mat{m}_{\Gamma}$ remains, which corresponds directly to the analytical solutions of the normal force $N$ and the bending moment $M$ in the curved beam model. An analytical solution also exists for the displacement $\vek{u}$ and the Euclidean norm of the transverse shear force vector $\|\vek{q}\| = \| \mat{P} \cdot \text{div}_\Gamma \mat{m}_\Gamma \|$, which corresponds directly to the shear force $Q$ in the curved beam model. The analytical beam solution for displacements and internal forces $M$, $Q$, and $N$ can be found in Appendix~\ref{app: Analytical solution of the sliced arc}.
\begin{table}
    \centering
    \begin{tabular}{|r|c|}
        \toprule
        \multirow{2}{*}{Geometry:}& cylindrical shell: length in $y$-direction $L_y = 1$,\\
        &radius $R = 2$, central angle $\theta = 70^{\circ}$\\ \hline
        Thickness:& $t = 0.1$\\ \hline
        Material:& $E = 2.1 \cdot 10^{8}$, $\nu = 0.0$\\ \hline
        Load:& $\vek{f} = [0,0,-10]^\text{T}$\\ \hline
        \multirow{3}{*}{BCs:}& Navier support at $x = -R \cos(55^{\circ})$, \\
        & no vertical displacement at $x = R \cos(55^{\circ})$, \\
        & free everywhere else\\
         \bottomrule
    \end{tabular}
    \caption{Collection of data for the extruded arc.}
    \label{tab:SlicedArc}
\end{table}

\begin{figure}
	\centering
	
	\subfigure[convergence in $\varepsilon_{\mathcal{L}^2,\text{rel}}(\vek{u})$]
	{\includegraphics[width=0.33\textwidth]{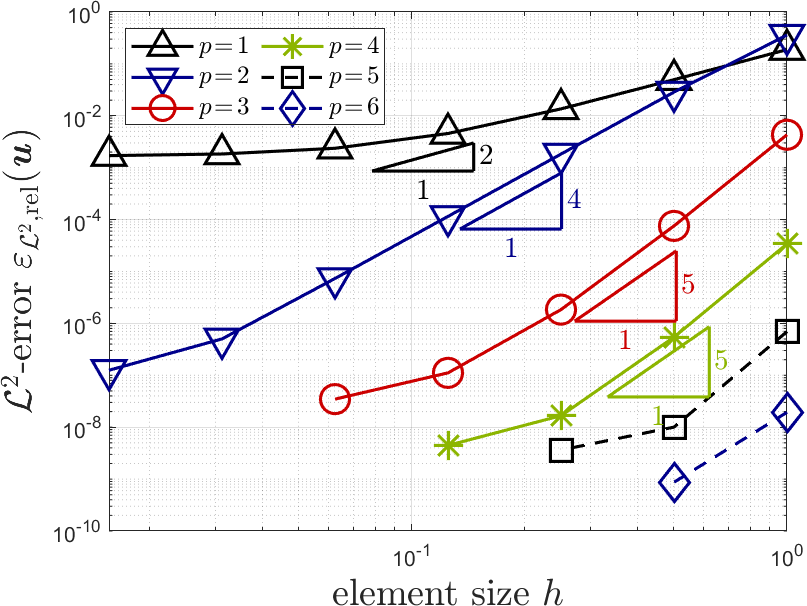}\label{fig:SAErrorU}}
	\subfigure[convergence in $\varepsilon_{\mathcal{L}^2,\text{rel}}(\mat{n}_{\Gamma}^{\text{real}})$]
	{\includegraphics[width=0.33\textwidth]{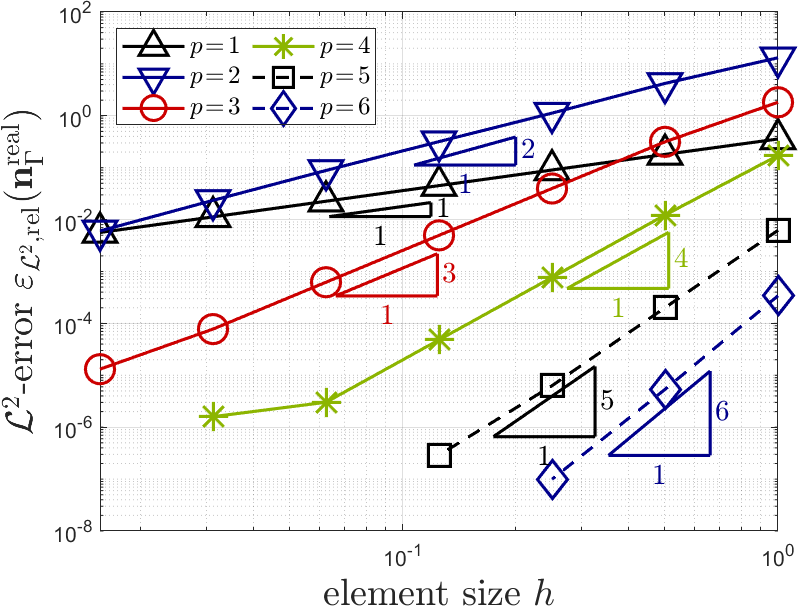}\label{fig:SAErrorN}}\hfill \\
	\subfigure[convergence in $\varepsilon_{\mathcal{L}^2,\text{rel}}(\mat{m}_{\Gamma})$]
	{\includegraphics[width=0.33\textwidth]{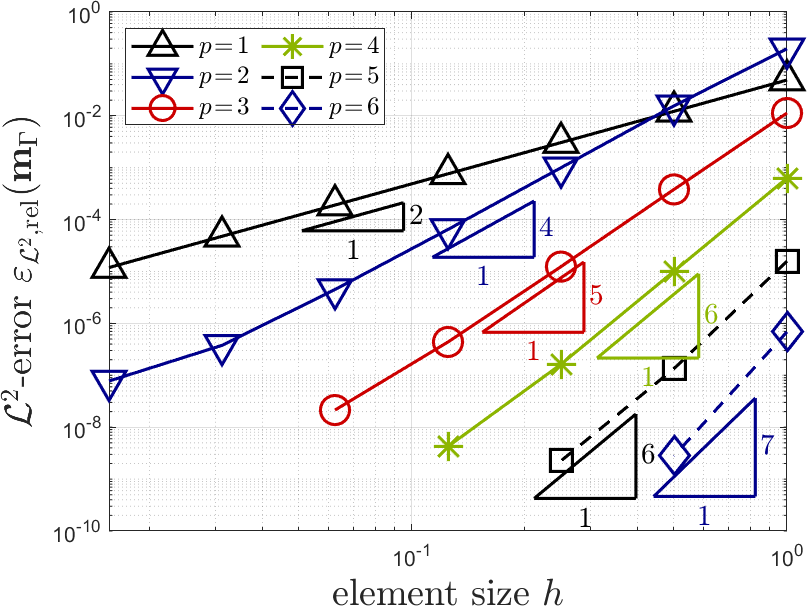}\label{fig:SAErrorM}}
    \subfigure[convergence in $\varepsilon_{\mathcal{L}^2,\text{rel}}(\vek{q})$]
    {\includegraphics[width=0.33\textwidth]{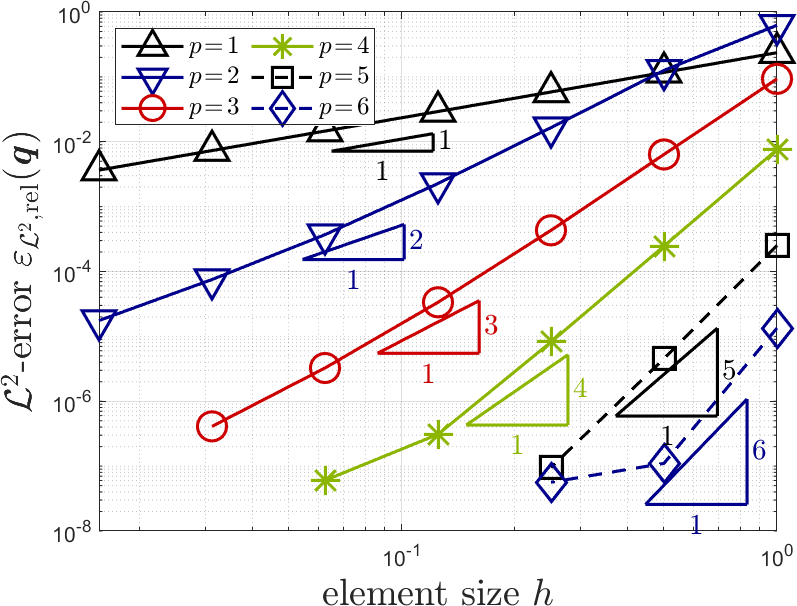}\label{fig:SAErrorQ}}
	
	\caption{\label{fig:SAError}Convergence results for the extruded arc. Convergence rates in the $\mathcal{L}^2$-error for (a) the displacement $\vek{u}$, (b) the only non-zero eigenvalue of the physical normal force tensor $\mat{n}_{\Gamma}^{\text{real}}$, (c) the only non-zero eigenvalue of the moment tensor $\mat{m}_{\Gamma}$, and (d) the Euclidean norm of the transverse shear force $\vek{q}$.}
 \end{figure}

Convergence studies for the relative $\mathcal{L}^2$-error, as defined in Eq.~(\ref{eq:L2-error}), are shown in Fig.~\ref{fig:SAError}. The convergence studies in the primal variables $\vek{u}$ and $\mat{m}_\Gamma$ show optimal convergence rates of at least $\mathcal{O}(p+1)$, see Figs.~\ref{fig:SAErrorU} and \ref{fig:SAErrorM}. For the derived quantities $\mat{n}_{\Gamma}^{\text{real}}$ and $\vek{q}$, the expected convergence rate of $\mathcal{O}(p)$ is also achieved as depicted in Figs.~\ref{fig:SAErrorN} and \ref{fig:SAErrorQ}. In summary, optimal convergence rates are obtained for all quantities, confirming the accuracy of the proposed model.

\subsection{Hemispherical shell}\label{sec: Hemispherical shell}

\begin{figure}
	\centering
	
	\subfigure[geometry]
	{\includegraphics[width=0.33\textwidth]{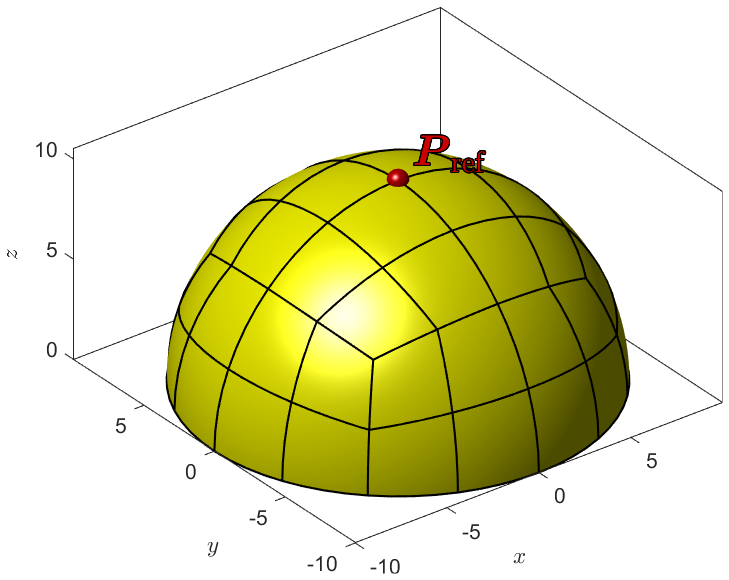}\label{fig:HSGeometry}}\hfill
	\subfigure[displacements, clamped]
	{\includegraphics[width=0.33\textwidth]{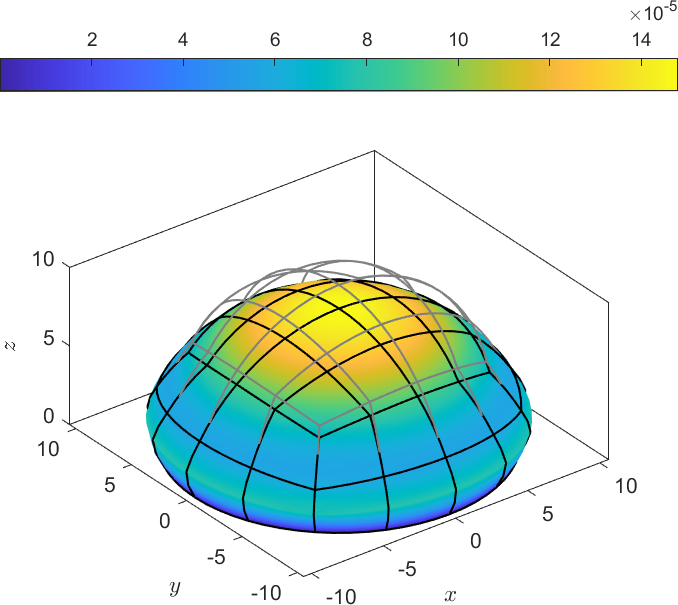}\label{fig:HSCDisplacement}}
    \subfigure[displacements, Navier]
	{\includegraphics[width=0.33\textwidth]{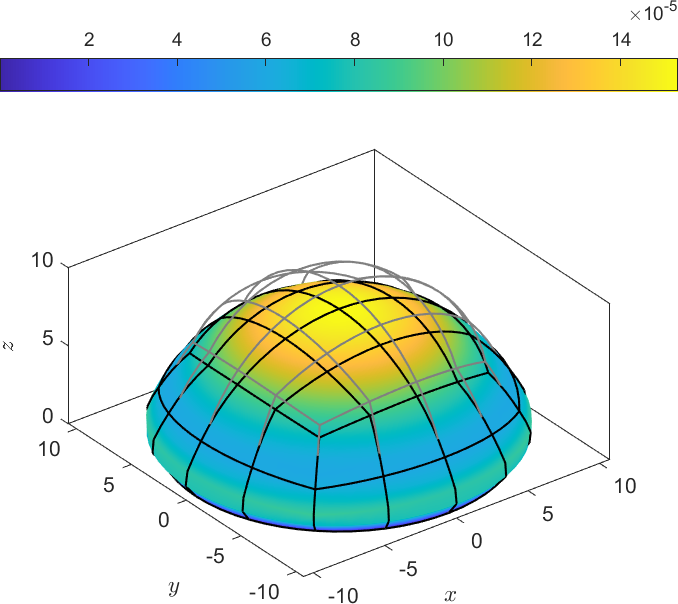}\label{fig:HSNDisplacement}}
	
	\caption{\label{fig:HemisphericalShell}(a) Geometry of the hemispherical shell. The red dot indicates the reference point $\vek{P}_{\text{ref}}$ for the vertical benchmark displacement $w_i$. Scaled deformed configuration with the Euclidean norm of the displacement $\|\vek{u}\|$ as a color plot for (b) clamped and (c) Navier support. The gray mesh lines depict the undeformed configuration.}
\end{figure}

Hemispherical shells are a common type of benchmark test in the field of shell mechanics \cite{Belytschko_1985a, Simo_1989b, Kiendl_2010a, Kaiser_2024a}. The hemispherical test case proposed herein, see Fig.~\ref{fig:HSGeometry}, aims to verify high-order convergence rates using the residual and stored energy errors described in Section~\ref{sec: Numerical results}. A collection of data to replicate this test case, including the resulting maximum vertical displacements and reference energies for comparison are summarized in Tab.~\ref{tab:HemisphericalShell}. We distinguish between full clamped support and full Navier support on the boundary. Figs.~\ref{fig:HSCDisplacement} and \ref{fig:HSNDisplacement} show the resulting displacements for both cases, respectively. As expected, the displacements for the clamped case are (slightly) smaller than for the Navier-supported case. However, because for this test case the load-bearing behavior is rather membrane-dominated, the differences are small.

\begin{table}
    \centering
    \begin{tabular}{|r|c|c|}
        \toprule
        & Case~1& Case~2\\
        \midrule
        Geometry:& \multicolumn{2}{c|}{hemisphere: radius $R = 10$}\\ \hline
        Thickness:& \multicolumn{2}{c|}{$t = 0.1$}\\ \hline
        Material:& \multicolumn{2}{c|}{$E = 3.0 \cdot 10^{7}$, $\nu = 0.3$}\\ \hline
        Load:& \multicolumn{2}{c|}{$\vek{f} = [0,0,-25 t]^\text{T}$}\\ \hline
        BCs:& clamped support& Navier support\\ \hline
        Ref. point:& \multicolumn{2}{c|}{$\vek{P}_{\text{ref}} = [0, 0, R ]^\text{T}$}\\ \hline
        Ref. solution:& $w_{\text{ref}} = -1.48203237 \cdot 10^{-4}$& $w_{\text{ref}} = -1.52964593 \cdot 10^{-4}$\\ \hline
        Ref. energy:& $\mathfrak{e}_{\text{ref}} = 4.717240184 \cdot 10^{-2}$& $\mathfrak{e}_{\text{ref}} = 5.039873241 \cdot 10^{-2}$\\   
        \bottomrule
    \end{tabular}
    \caption{Collection of data for the hemispherical shell.}
    \label{tab:HemisphericalShell}
\end{table}

\begin{figure}
	\centering
	
	\subfigure[convergence in $\varepsilon_{\text{res},1}$]
	{\includegraphics[width=0.33\textwidth]{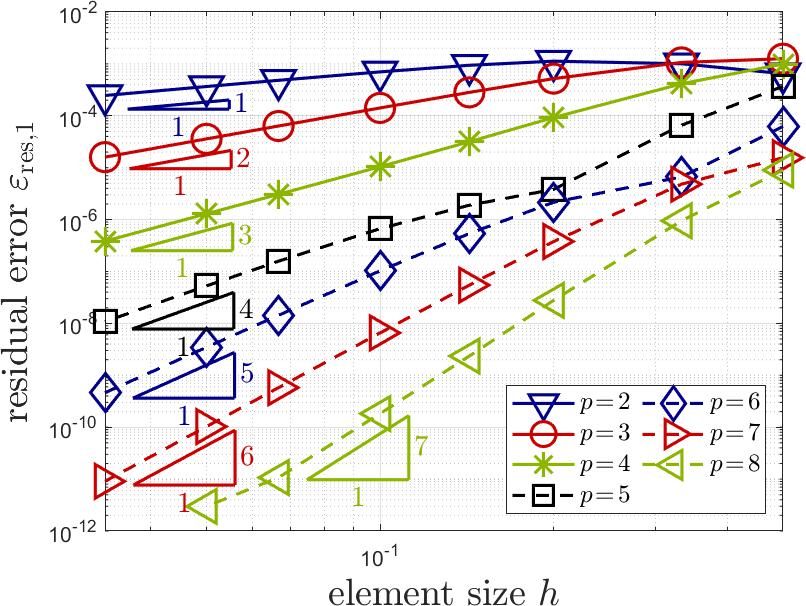}\label{fig:HSCErrorRes1}}\hfill
	\subfigure[convergence in $\varepsilon_{\text{res},2}$]
	{\includegraphics[width=0.33\textwidth]{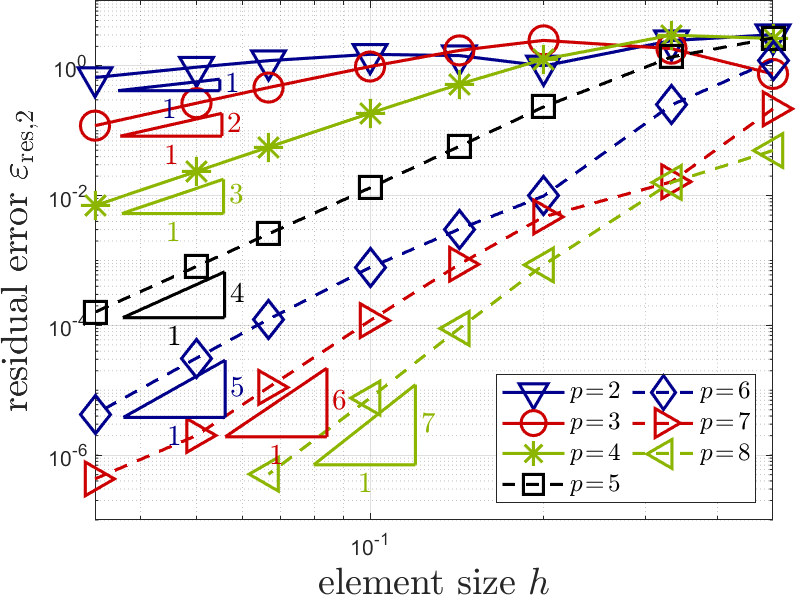}\label{fig:HSCErrorRes2}}\hfill
	\subfigure[convergence in $\varepsilon_{\mathfrak{e}}$]
	{\includegraphics[width=0.33\textwidth]{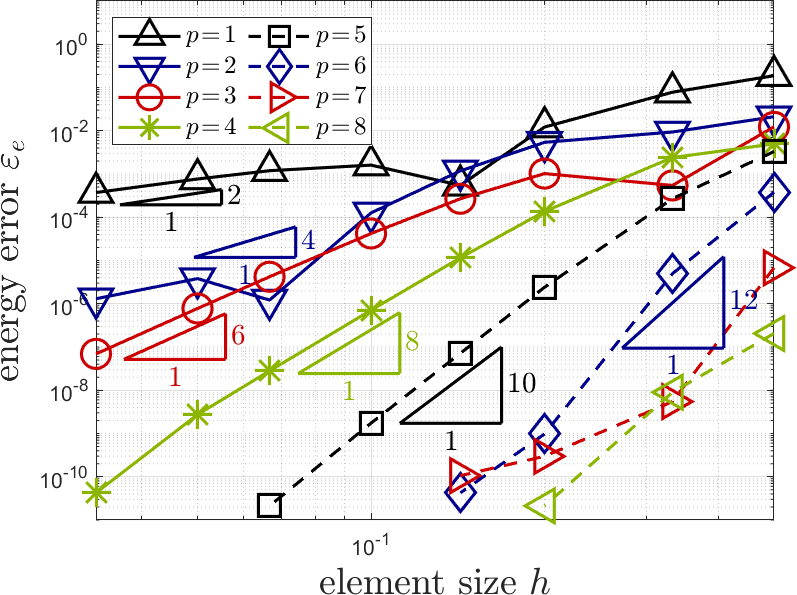}\label{fig:HSCErrorEnergy}}
	
	\caption{\label{fig:HSCError}Convergence studies for the hemispherical shell with clamped support. Convergence rates in (a) the absolute residual error $\varepsilon_{\text{res},1}$, (b) the relative residual error $\varepsilon_{\text{res},2}$, and (c) the relative energy error $\varepsilon_{\mathfrak{e}}$.}
\end{figure}

\begin{figure}
	\centering
	
	\subfigure[convergence in $\varepsilon_{\text{res},1}$]
	{\includegraphics[width=0.33\textwidth]{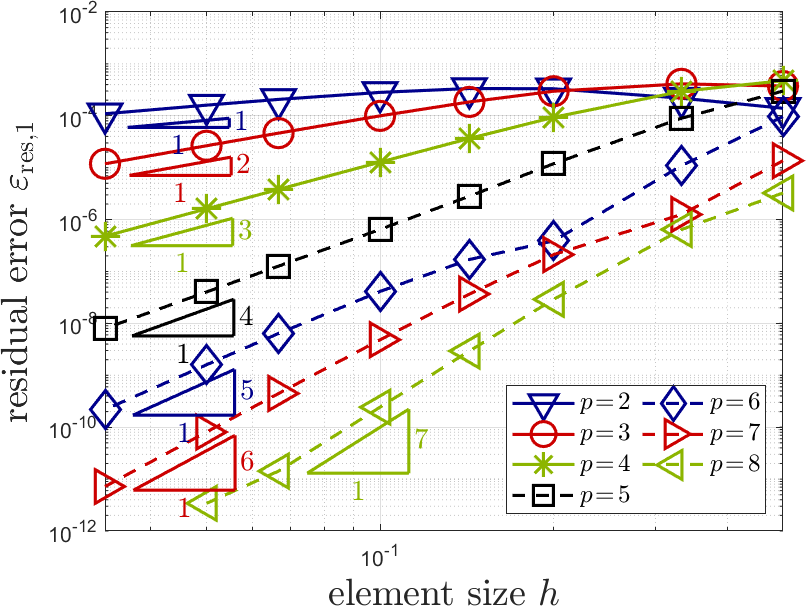}}\label{fig:HSNErrorRes1}\hfill
	\subfigure[convergence in $\varepsilon_{\text{res},2}$]
	{\includegraphics[width=0.33\textwidth]{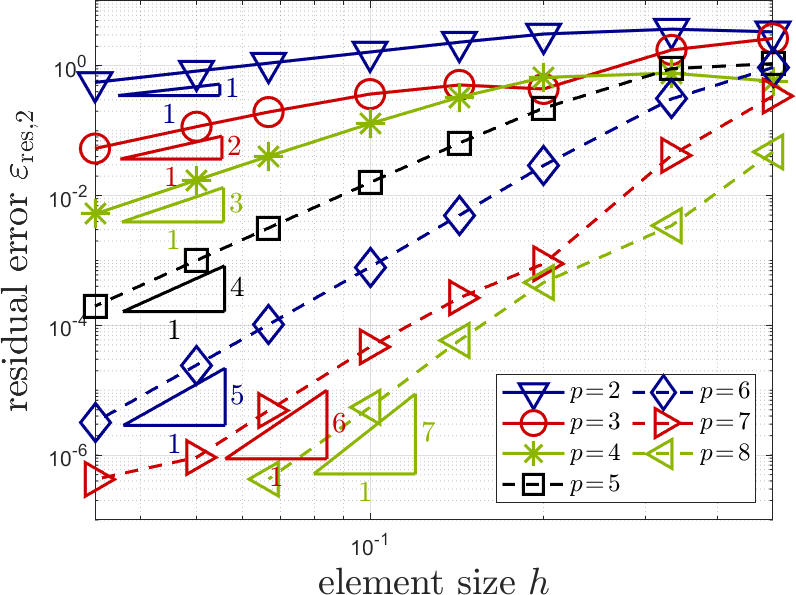}\label{fig:HSNErrorRes2}}\hfill
	\subfigure[convergence in $\varepsilon_{\mathfrak{e}}$]
	{\includegraphics[width=0.33\textwidth]{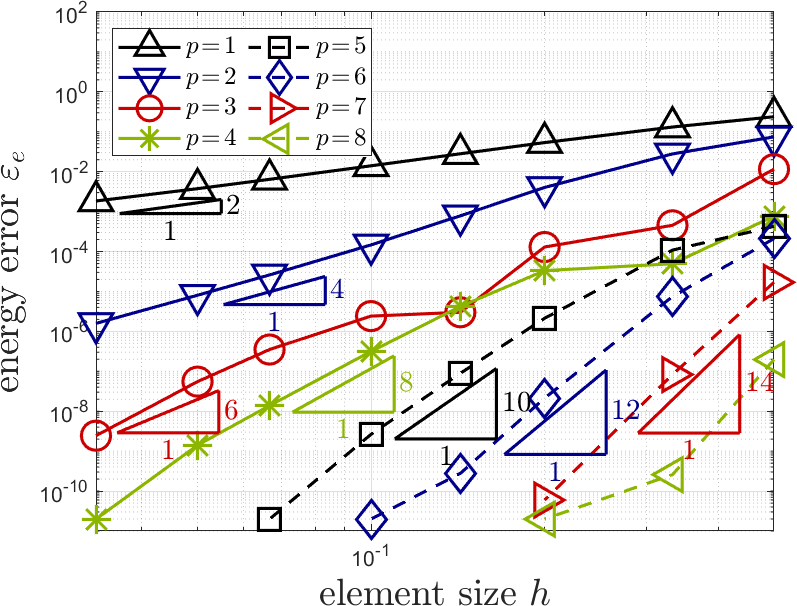}\label{fig:HSNErrorEnergy}}
	
	\caption{\label{fig:fig:HSNError}Convergence studies for the hemispherical shell with Navier support. Convergence rates in (a) the absolute residual error $\varepsilon_{\text{res},1}$, (b) the relative residual error $\varepsilon_{\text{res},2}$, and (c) the relative energy error $\varepsilon_{\mathfrak{e}}$.}
\end{figure}

In Figs.~\ref{fig:HSCErrorRes1} and \ref{fig:HSCErrorRes2}, the convergence studies for the clamped case in the residual errors, as defined in Eqs.~(\ref{eq:ResidualError1}) and (\ref{eq:ResidualError2}), are displayed. The convergence curves flatten for coarser meshes with larger element sizes $h$, marking a pre-asymptotic regime. For finer meshes, convergence rates achieve the expected optimal order of $\mathcal{O}(p-1)$ as second-order derivatives are required to compute these residual errors. The convergence study in the stored energy error, see Eq.~(\ref{eq:EnergyError}), as depicted in Fig.~\ref{fig:HSCErrorEnergy}, confirms the expected convergence rate of $\mathcal{O}(2p)$. Qualitatively the same results are seen in the convergence results for Navier rather than clamped supports, see Fig.~\ref{fig:fig:HSNError}, so that all conclusions carry over to that case as well.

\subsection{Flower-shaped shell}\label{sec: flower-shaped shell}

\begin{figure}
	\centering
	
	\subfigure[geometry]
	{\includegraphics[width=0.5\textwidth]{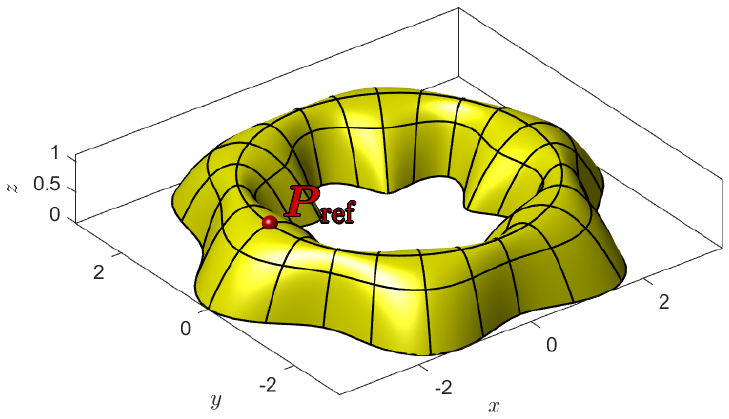}\label{fig:FSSGeometry}}\hfill
	\subfigure[displacements]
	{\includegraphics[width=0.5\textwidth]{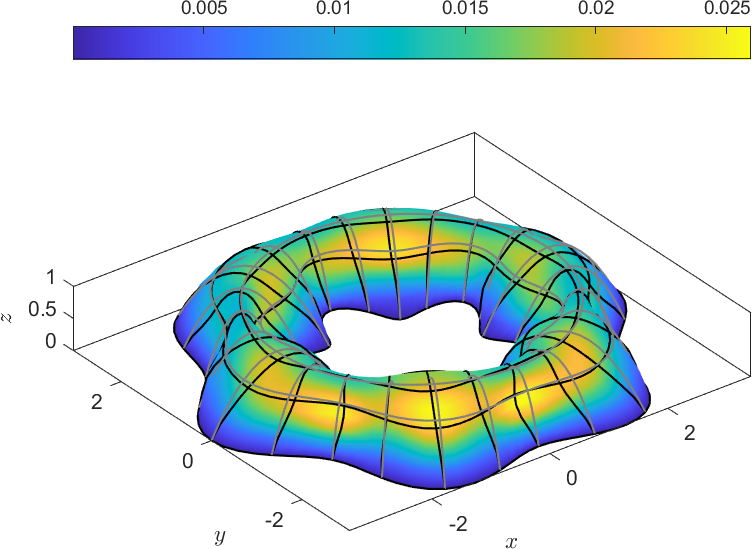}\label{fig:FSSDisplacement}}
	
	\caption{\label{fig:FlowerShapedShell}(a) Geometry of the flower-shaped shell. The red dot indicates the reference point $\vek{P}_{\text{ref}}$ for the vertical benchmark displacement $w_i$. (b) Scaled deformed configuration with the Euclidean norm of the displacement $\|\vek{u}\|$ as a color plot. The gray mesh lines depict the undeformed configuration.}
\end{figure}

\begin{figure}
	\centering
	
	{\includegraphics[width=0.65\textwidth]{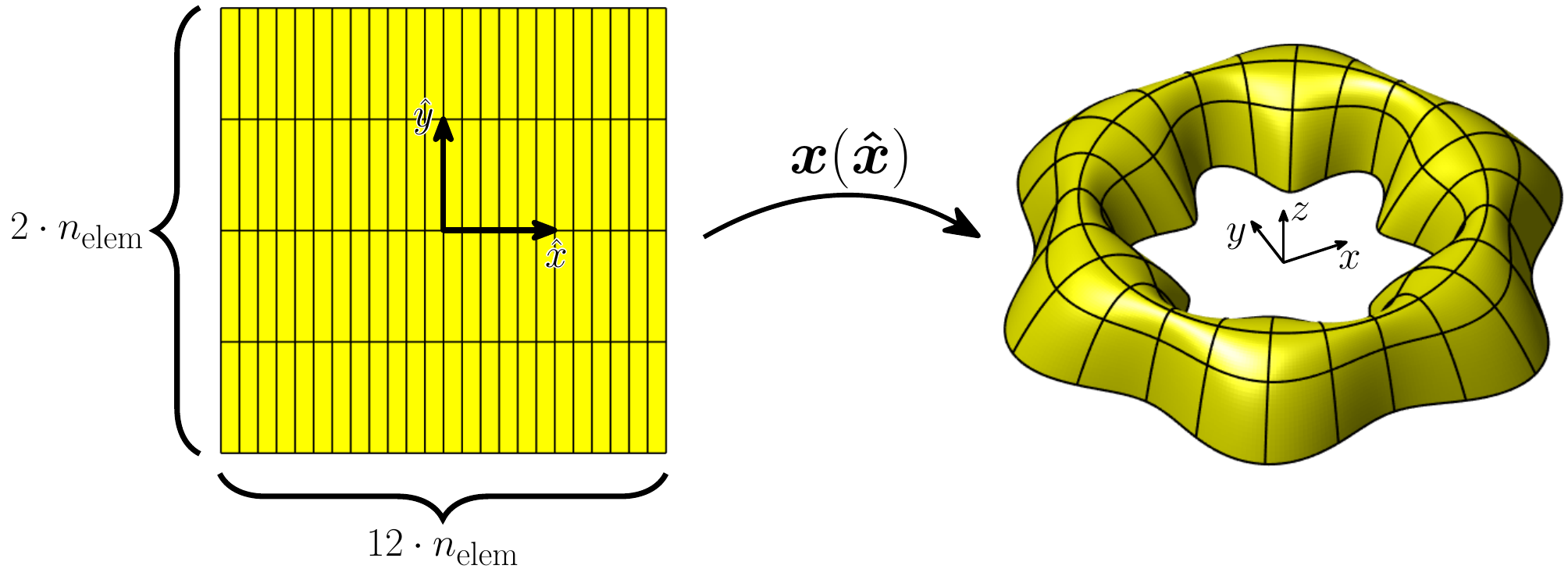}}
	
	\caption{\label{fig:FSSMapping}Map of the flower-shaped shell with a mesh resolution of $n_{\text{elem}}=2$. The function $\vek{x}(\hat{\vek{x}})$ maps the reference square $\hat{x},\hat{y}\in[-1,1]$ (left) to the desired geometry of the test case (right).}
\end{figure}

The flower-shaped shell follows \cite{Schoellhammer_2019a}, representing a more complex geometry as shown in Fig.~\ref{fig:FSSGeometry}. The function $\vek{x}(\hat{\vek{x}})$ given in Tab.~\ref{tab:FlowerShapedShell} maps the reference square $\hat{x},\hat{y}\in[-1,1]$ to the desired geometry of this test case, see Fig.~\ref{fig:FSSMapping}. Further details and reference quantities of the test case are summarized in Tab.~\ref{tab:FlowerShapedShell}.

\begin{table}
    \centering
    \begin{tabular}{|r|c|}
        \toprule
        \multirow{2}{*}{Geometry:}& $\hat{x},\hat{y} \in [-1,1] \Rightarrow \vek{x}(\hat{\vek{x}}) = \left[ \begin{array}{c}  \left[ 2.3 - \hat{y} \cdot (0.8+0.3 \cos(6 \theta)) \right] \cos(\theta) \\ \left[ 2.3 - \hat{y} \cdot (0.8+0.3 \cos(6 \theta)) \right] \sin(\theta) \\ 1 - \hat{y}^2 \end{array}\right]$\\
        & with $\theta = \pi \cdot (\hat{x} + 1)$\\ \hline
        Thickness:& $t = 0.1$\\ \hline
        Material:& $E = 1.0 \cdot 10^{4}$, $\nu = 0.3$\\ \hline
        Load:& $\vek{f} = [1,2,-10]^\text{T}$\\ \hline
        BCs:& clamped support\\ \hline
        Ref. point:& $\vek{P}_{\text{ref}} = [-2.3,0,1]^\text{T}$\\ \hline
        Ref. solution:& $w_{\text{ref}} = -1.48331874 \cdot 10^{-2}$\\ \hline
        Ref. energy:& $\mathfrak{e}_{\text{ref}} = 1.763595793$\\
        \bottomrule
    \end{tabular}
    \caption{Collection of data for the flower-shaped shell.}
    \label{tab:FlowerShapedShell}
\end{table}

\begin{figure}
	\centering
	
	\subfigure[convergence in $\varepsilon_{\text{res},1}$]
	{\includegraphics[width=0.33\textwidth]{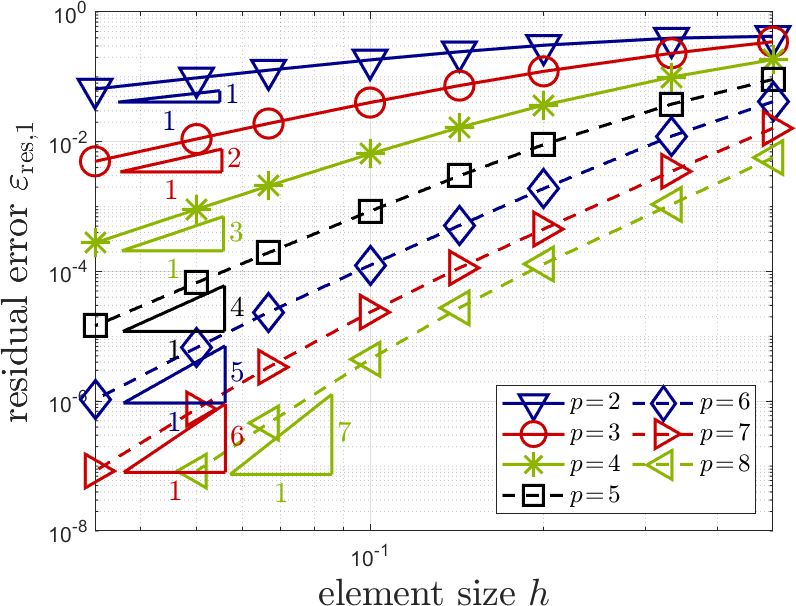}}\label{fig:FSSErrorRes1}\hfill
	\subfigure[convergence in $\varepsilon_{\text{res},2}$]
	{\includegraphics[width=0.33\textwidth]{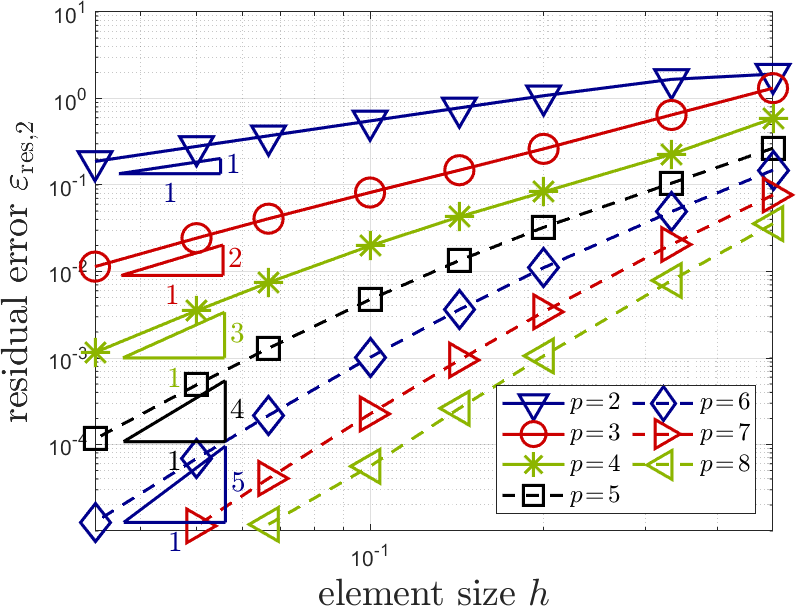}\label{fig:FSSErrorRes2}}\hfill
	\subfigure[convergence in $\varepsilon_{\mathfrak{e}}$]
	{\includegraphics[width=0.33\textwidth]{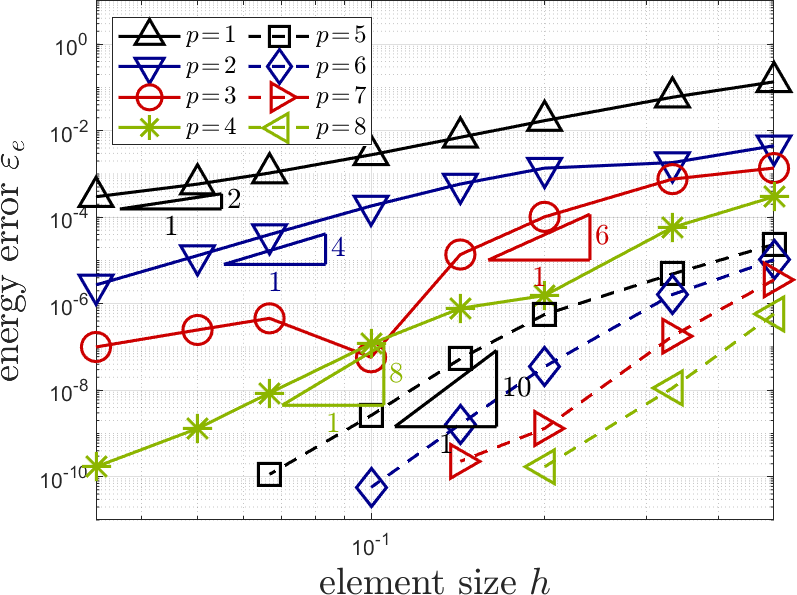}\label{fig:FSSErrorEnergy}}
	
	\caption{\label{fig:FSSError}Convergence studies for the flower-shaped shell. Convergence rates in (a) the absolute residual error $\varepsilon_{\text{res},1}$, (b) the relative residual error $\varepsilon_{\text{res},2}$, and (c) the relative energy error $\varepsilon_{\mathfrak{e}}$.}
\end{figure}

Comparing the displacements of our model, displayed in Fig.~\ref{fig:FSSDisplacement}, with the displacement plot illustrated in \cite{Schoellhammer_2019a}, excellent agreement is found. Also, the stored energy converges as expected to the reference energy given there, see also Tab.~\ref{tab:FlowerShapedShell}. Fig.~\ref{fig:FSSError} shows the convergence results in the residual and stored energy errors. As before in Section~\ref{sec: Hemispherical shell}, the convergence rates in all error measures achieve the expected rates, confirming optimal results also for this example.

\subsection{Ring-shaped shell with distinct boundary conditions}\label{sec: Shell with two separated boundaries}

\begin{figure}
	\centering

	\subfigure[geometry]
	{\includegraphics[width=0.33\textwidth]{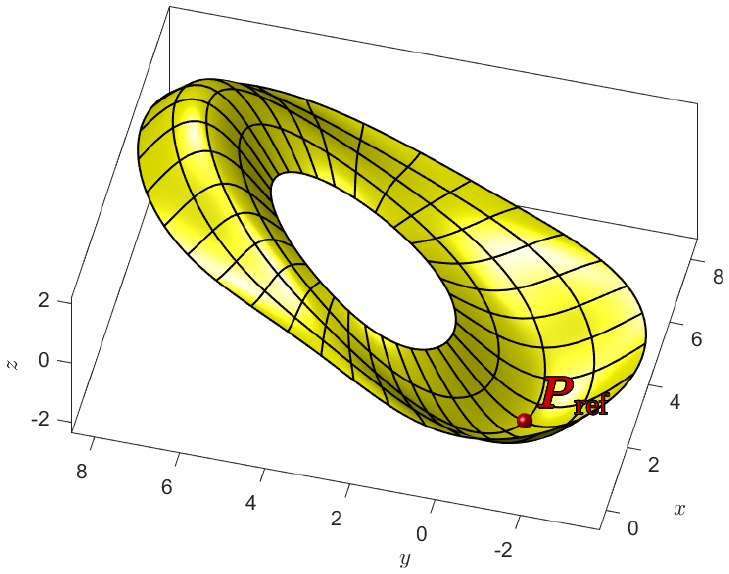}\label{fig:SWTSBGeometry}}
	\subfigure[diplacments, case~1]
	{\includegraphics[width=0.33\textwidth]{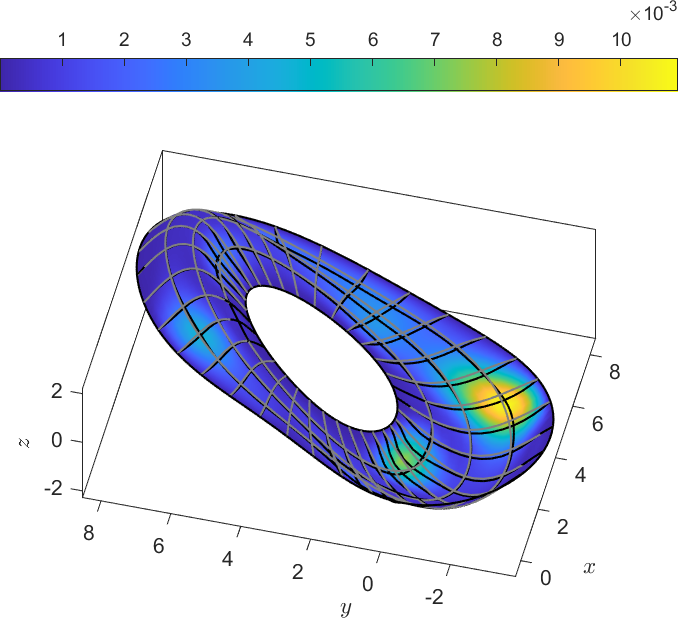}\label{fig:SWTSBClampedDisplacement}}\hfill \\
    \subfigure[diplacments, case~2]
    {\includegraphics[width=0.33\textwidth]{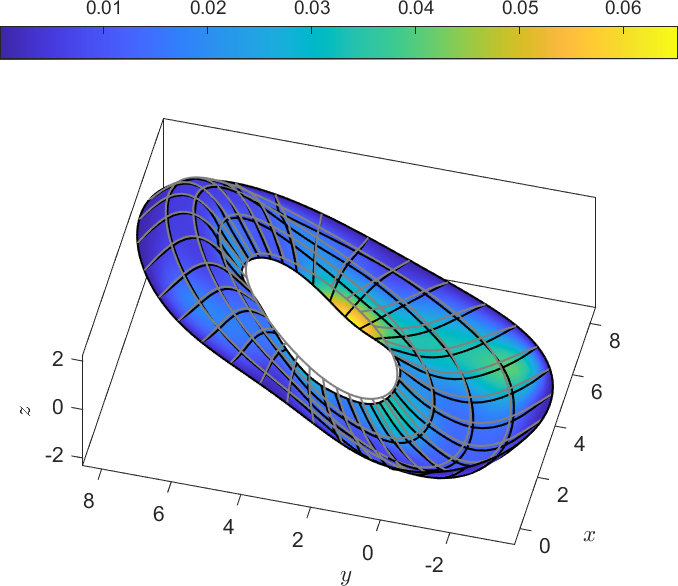}\label{fig:SWTSBFreeHomDisplacement}}
    \subfigure[diplacments, case~3]
    {\includegraphics[width=0.33\textwidth]{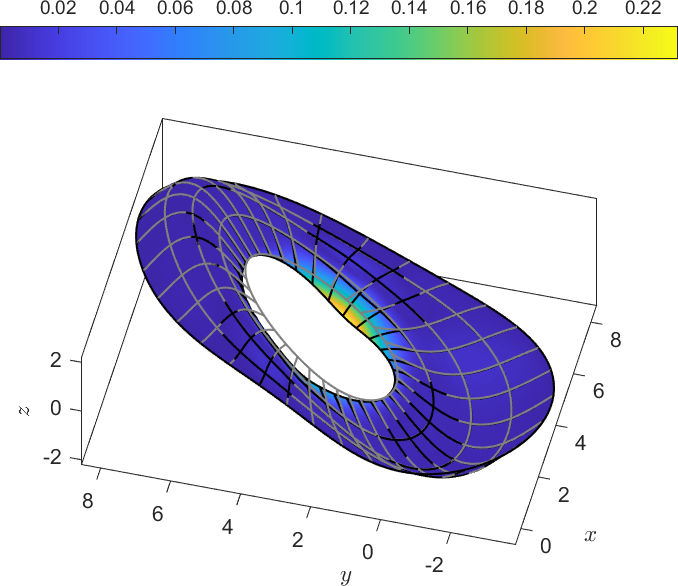}\label{fig:SWTSBFreeInhomDisplacement}}
 
	\caption{\label{fig:ShellWithTwoSeparatedBoundaries}(a) Geometry of the ring-shaped shell. The red dot indicates the reference point $\vek{P}_{\text{ref}}$ for the vertical benchmark displacement $w_i$. Scaled deformed configuration with the Euclidean norm of the displacement $\|\vek{u}\|$ as a color plot for (b) case~1, (c) case~2, and (d) case~3. The gray mesh lines depict the undeformed configuration.}
\end{figure}

\begin{figure}
	\centering
	
	{\includegraphics[width=1.0\textwidth]{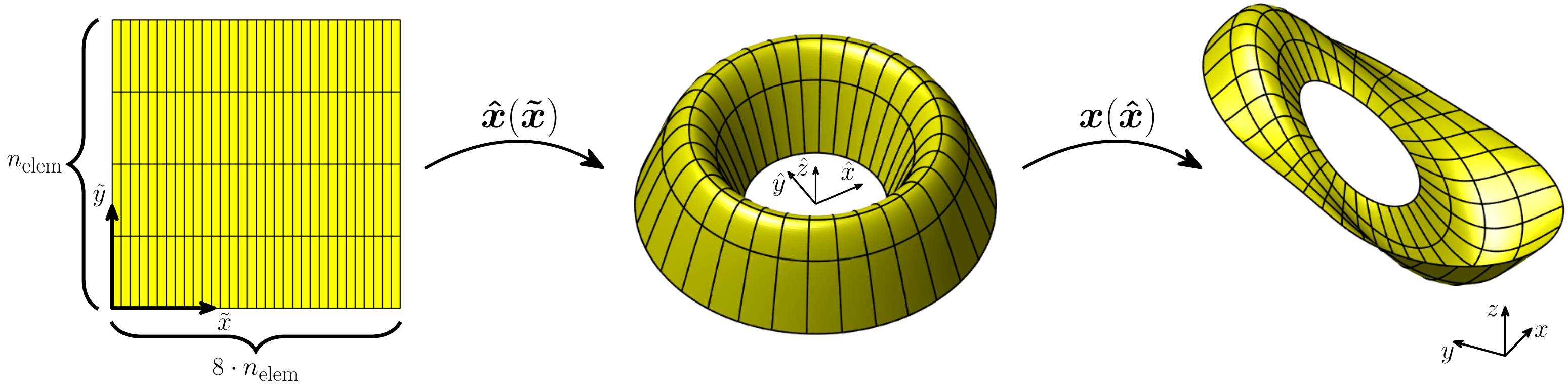}}
	
	\caption{\label{fig:SWTSBMapping}Map of the ring-shaped shell with a mesh resolution of $n_{\text{elem}}=4$. A first mapping $\hat{\vek{x}}(\tilde{\vek{x}})$ transforms the reference square $\tilde{x},\tilde{y}\in[0,1]$ (left) into the intermediate geometry (center) and a second map $\vek{x}(\hat{\vek{x}})$ further transforms it to the desired geometry of the test case (right).}
\end{figure}

The final test case demonstrates higher-order convergence rates for shells with different boundary conditions on individual boundary sections. For this purpose, an arbitrary ring-shaped geometry with two separate boundaries, as shown in Fig.~\ref{fig:SWTSBGeometry}, is proposed. Along the inner and outer boundaries, two different boundary conditions can be prescribed respectively, avoiding a point where different boundary conditions adjoin and, therefore, preventing singularities. Three cases with different combinations of boundary conditions are presented with Navier support at the outer boundary for all cases. For case~1, the whole inner boundary is clamped. For the other cases, a free boundary is prescribed with homogeneous Neumann boundary conditions ($\hat{\tilde{\vek{p}}} = \vek{0},m_{\vek{t}} = 0$) for case~2 and inhomogeneous Neumann boundary conditions ($\hat{\tilde{\vek{p}}} \neq \vek{0},m_{\vek{t}} \neq 0$) for case~3. Displacement plots of each case are presented in Figs.~\ref{fig:SWTSBClampedDisplacement}-\ref{fig:SWTSBFreeInhomDisplacement}.
Again, the desired geometry of the undeformed shell is obtained through maps, however, in a two-step precedure here, as depicted in Fig.~\ref{fig:SWTSBMapping}. The first map $\hat{\vek{x}}(\tilde{\vek{x}})$ transforms a reference square to a ring with a sinusoidal arc in the radial direction and a second map $\vek{x}(\hat{\vek{x}})$ further transforms this to the desired geometry of this example. Further details and resulting reference quantities are summarized in Tab.~\ref{tab:ShellWithTwoSeparatedBoundaries}.

\begin{table}
    \centering
    \begin{tabular}{|r|c|c|c|}
        \toprule
        & Case~1& Case~2& Case~3\\
        \midrule
        \multirow{3}{*}{Geometry:}& \multicolumn{3}{c|}{$\tilde{x},\tilde{y} \in [0,1] \Rightarrow \hat{\vek{x}}(\tilde{\vek{x}}) = \left[ \begin{array}{c} (0.4 + 0.6 \tilde{y}) \cos(2\pi \cdot \tilde{x})\\ (0.4 + 0.6 \tilde{y}) \sin(2\pi \cdot \tilde{x})\\ 0.6 \sin(\pi \cdot \tilde{y}) \end{array}\right]$}\\
        & \multicolumn{3}{c|}{$\Rightarrow \vek{x}(\hat{\vek{x}}) = \left[ \begin{array}{c} 4 - 4 \hat{y} + 2 \sin(2 \hat{x} + 0.5 \hat{y})\\ 2 + 6 \hat{x} + \hat{y} + \cos(\hat{x} + 1.5 \hat{y})\\ (1 + 2.5 \hat{z}) \sin(1.5 \hat{x}) \cos(2 \hat{y}) \end{array}\right]$}\\ \hline
        Thickness:& \multicolumn{3}{c|}{$t = 0.05$}\\ \hline
        Material:& \multicolumn{3}{c|}{$E = 3.0 \cdot 10^{7}$, $\nu = 0.2$}\\ \hline
        Load:& \multicolumn{2}{c|}{$\vek{f} = [40,60,-100]^\text{T}$}& $\vek{f} = [0,0,0]^\text{T}$\\ \hline
        Line force:& -& $\hat{\tilde{\vek{p}}} = [0,0,0]^\text{T}$& $\hat{\tilde{\vek{p}}} = [40,60,-100]^\text{T}$\\ \hline
        Line moment:& -& $\hat{m}_{\vek{t}} = 0$& $\hat{m}_{\vek{t}} = 100$\\ \hline
        Outer BCs:& Navier support& Navier support& Navier support\\ \hline
        Inner BCs:& clamped support& free& free\\ \hline
        Ref. point:& \multicolumn{3}{c|}{$\vek{P}_{\text{ref}} = \vek{x}(\hat{\vek{x}} = [-0.7, 0, 0.6]^\text{T})$}\\ \hline
        Ref. solution:& $w_{\text{ref}} = \text{-}8.9292835 \! \cdot \! 10^{\text{-}4}$& $w_{\text{ref}} = \text{-}8.6651573 \! \cdot \! 10^{\text{-}3}$& $w_{\text{ref}} = \text{-}1.4199220 \! \cdot \! 10^{\text{-}3}$\\ \hline
        Ref. energy:& $\mathfrak{e}_{\text{ref}} = 6.25385804$& $\mathfrak{e}_{\text{ref}} = 44.4257042$& $\mathfrak{e}_{\text{ref}} = 236.455536$\\
        \bottomrule
    \end{tabular}
    \caption{Collection of data for the ring-shaped shell.}
    \label{tab:ShellWithTwoSeparatedBoundaries}
\end{table}

\begin{figure}
	\centering
	
	\subfigure[convergence in $\varepsilon_{\text{res},1}$]
	{\includegraphics[width=0.328\textwidth]{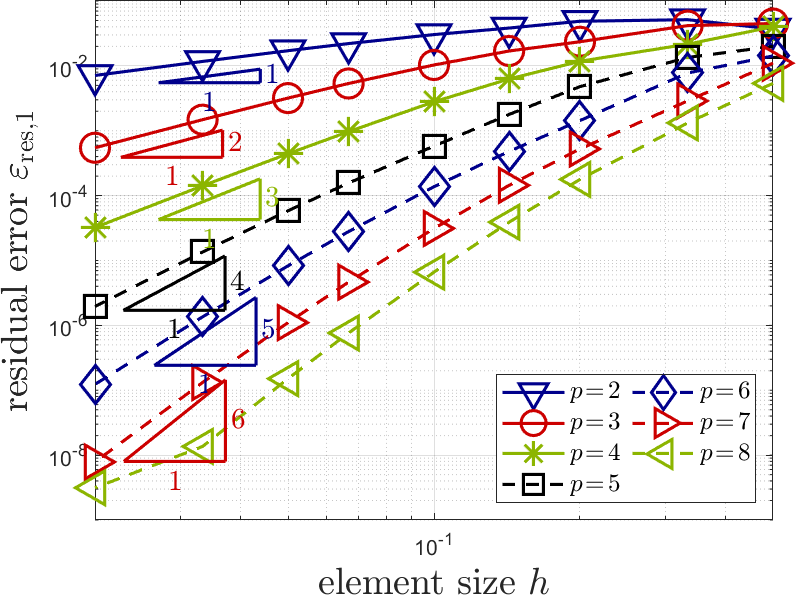}\label{fig:SWTSBClampedErrorRes1}}
	\subfigure[convergence in $\varepsilon_{\text{res},2}$]
	{\includegraphics[width=0.328\textwidth]{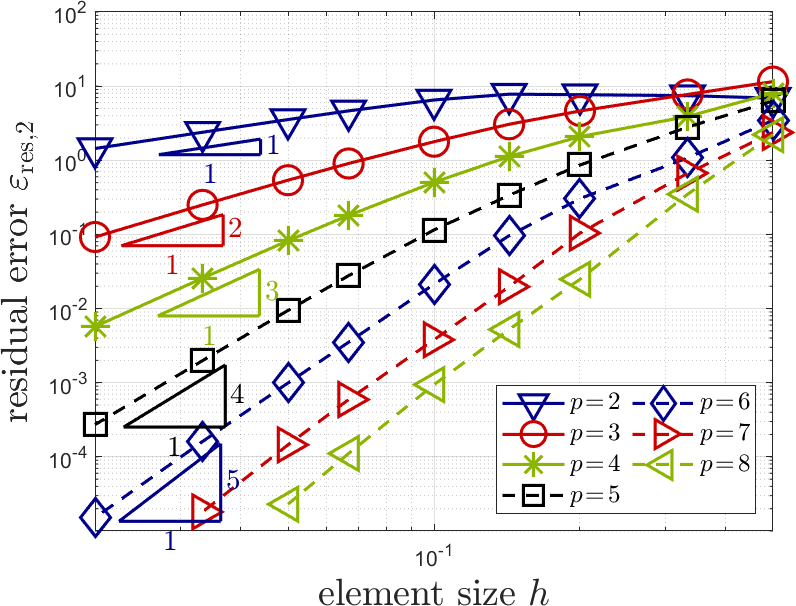}\label{fig:SWTSBClampedErrorRes2}}
	\subfigure[convergence in $\varepsilon_{\mathfrak{e}}$]
	{\includegraphics[width=0.328\textwidth]{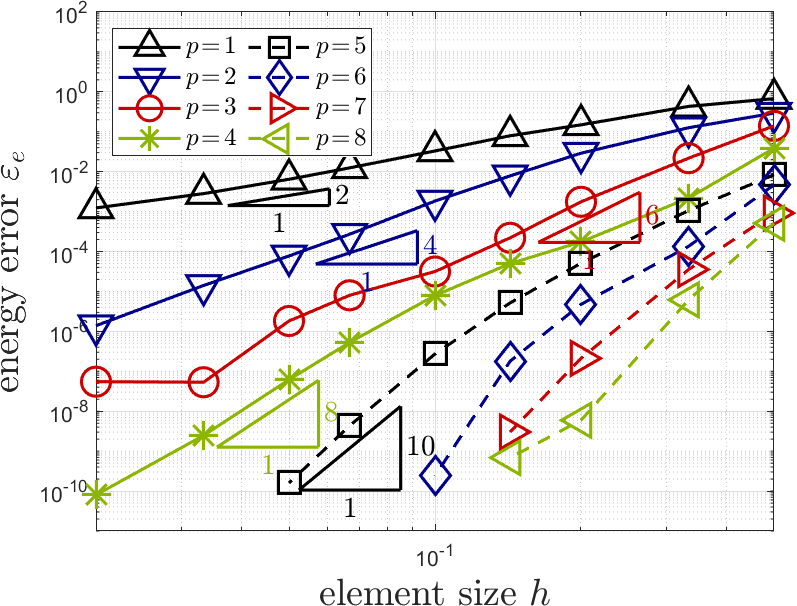}\label{fig:SWTSBClampedErrorEnergy}}
    
	\caption{\label{fig:SWTSBClampedError}Convergence studies for the ring-shaped shell for case 1 (Navier outer, clamped inner boundary). Convergence rates in (a) the absolute residual error $\varepsilon_{\text{res},1}$, (b) the relative residual error $\varepsilon_{\text{res},2}$, and (c) the relative energy error $\varepsilon_{\mathfrak{e}}$.}

\end{figure}

\begin{figure}
	\centering
	
	\subfigure[convergence in $\varepsilon_{\text{res},1}$]
	{\includegraphics[width=0.328\textwidth]{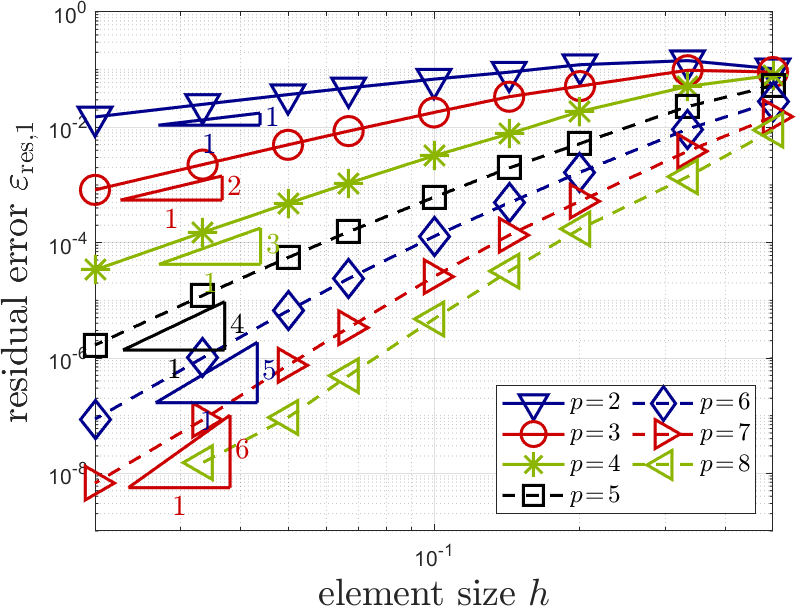}\label{fig:SWTSBFreeHomErrorRes1}}
	\subfigure[convergence in $\varepsilon_{\text{res},2}$]
	{\includegraphics[width=0.328\textwidth]{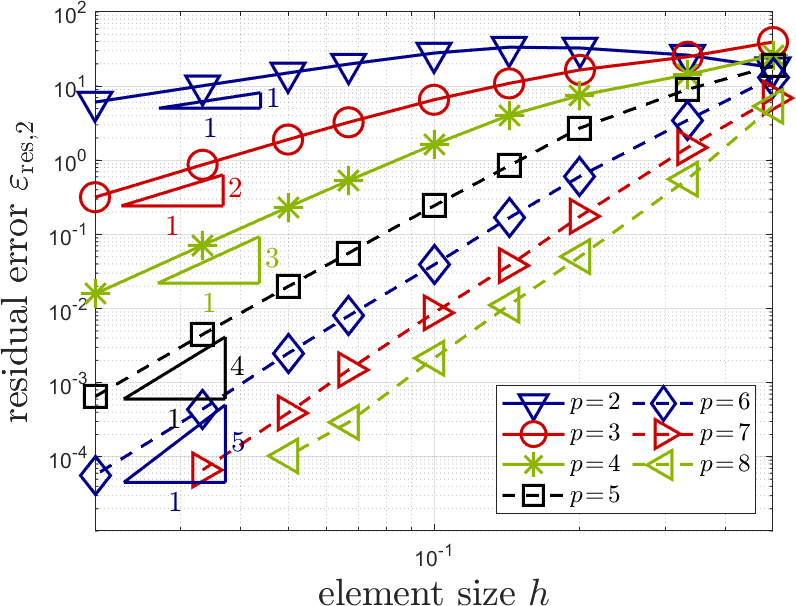}\label{fig:SWTSBFreeHomErrorRes2}}
        \subfigure[convergence in $\varepsilon_{\mathfrak{e}}$]
	{\includegraphics[width=0.328\textwidth]{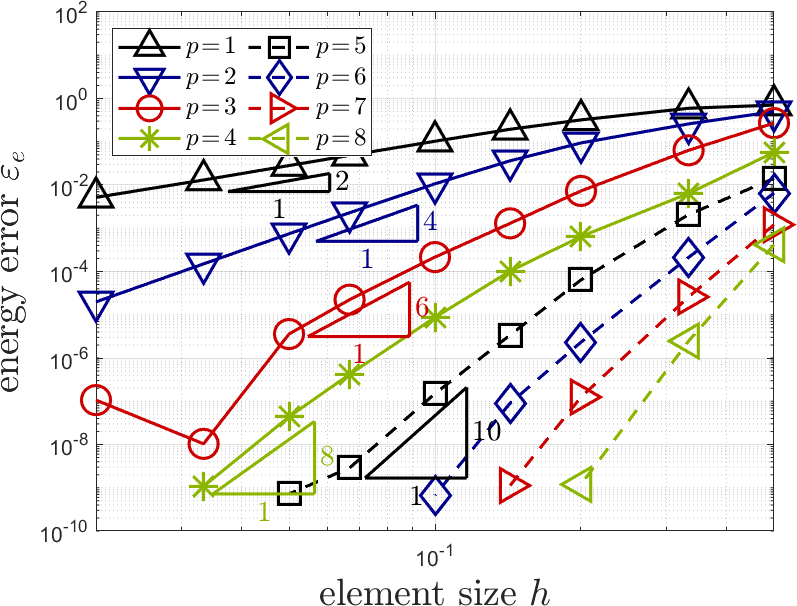}\label{fig:SWTSBFreeHomErrorEnergy}}
	
	\caption{\label{fig:SWTSBFreeHomError}Convergence studies for the ring-shaped shell for case 2 (Navier outer, free homogeneous inner boundary). Convergence rates in (a) the absolute residual error $\varepsilon_{\text{res},1}$, (b) the relative residual error $\varepsilon_{\text{res},2}$, and (c) the relative energy error $\varepsilon_{\mathfrak{e}}$.}
\end{figure}

\begin{figure}
	\centering
	
	\subfigure[convergence in $\varepsilon_{\text{res},1}$]
	{\includegraphics[width=0.33\textwidth]{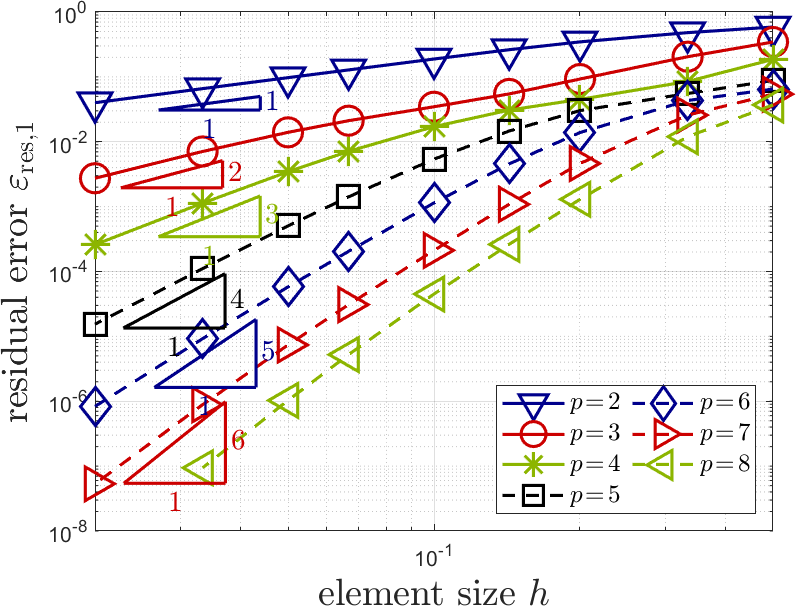}\label{fig:SWTSBFreeInhomErrorRes1}}\hfill
	\subfigure[convergence in $\varepsilon_{\text{res},2}$]
	{\includegraphics[width=0.33\textwidth]{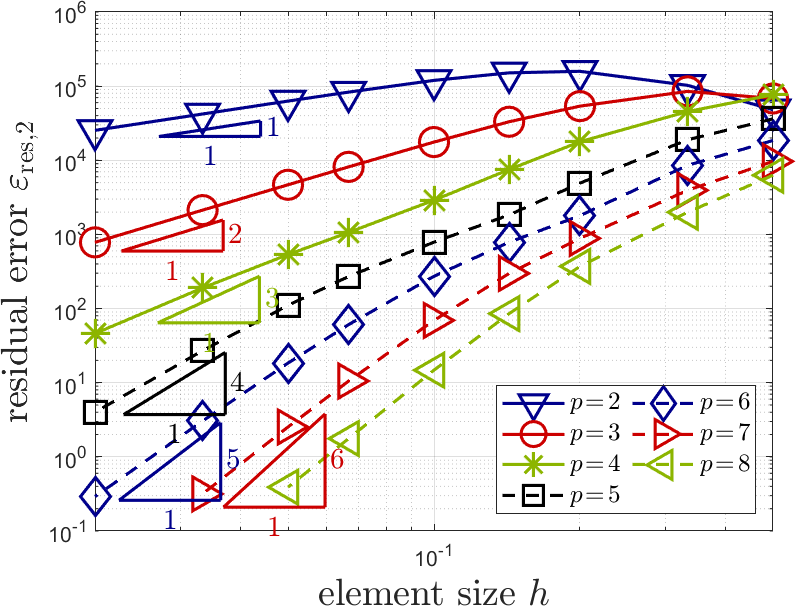}\label{fig:SWTSBFreeInhomErrorRes2}}
    \subfigure[convergence in $\varepsilon_{\text{res},\text{bound}}$]
	{\includegraphics[width=0.33\textwidth]{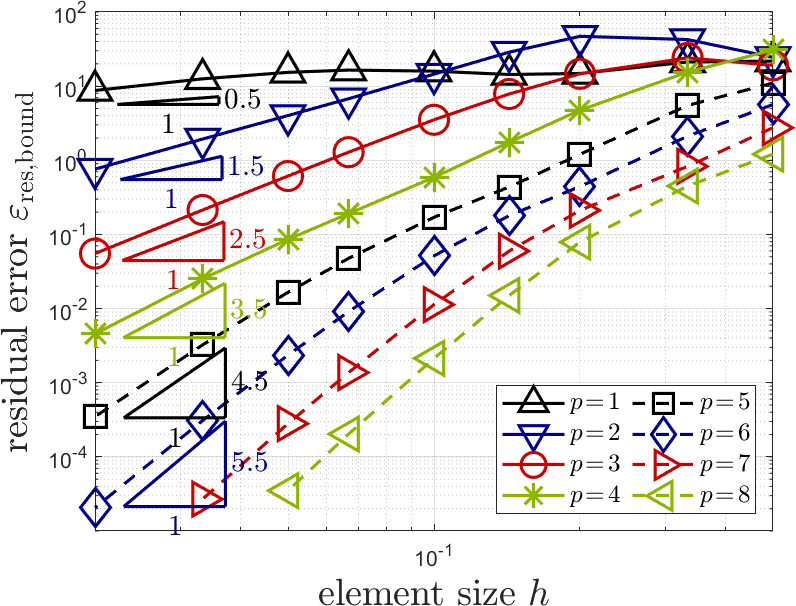}\label{fig:SWTSBFreeInhomErrorResBoundForce}}
	
	\caption{\label{fig:SWTSBFreeInhomError}Convergence studies for the ring-shaped shell for case 3 (Navier outer, free inhomogeneous inner boundary). Convergence rates in (a) the absolute residual error $\varepsilon_{\text{res},1}$, (b) the absolute residual error $\varepsilon_{\text{res},2}$, (c) the relative residual error for the force equilibrium at the free boundary $\varepsilon_{\text{res},\text{bound}}$.}
\end{figure}

Analogously to the three previous test cases, convergence studies in the residual errors and the stored energy error are displayed, see Figs.~\ref{fig:SWTSBClampedError}-\ref{fig:SWTSBFreeInhomError}. Generally, it can be stated that optimal convergence rates are obtained similarly to the test cases in Sections~\ref{sec: Hemispherical shell} and \ref{sec: flower-shaped shell}. However, due to the complex geometry, the pre-asymptotic regime is more pronounced than in the other test cases. Additionally, the convergence in the residual error based on the force equilibrium is evaluated on the inner boundary for case~3, cf.~Eq.~\ref{eq:ResidualErrorBoundary}. A convergence rate of $\mathcal{O}(p-0.5)$ can be observed in this error measure, evaluated along the corresponding boundary section $\partial\Gamma_{\text{N},\vek{u}}$.

\section{Conclusions and outlook}\label{sec: Conclusions and outlook}

A novel mixed-hybrid FEM for the linear Kirchhoff--Love shell is formulated. A mixed ansatz is proposed where components of the moment tensor serve as additional primary variables next to the displacements, thus lowering the continuity requirements for the employed shape functions. Consequently, using $C^0$-continuous shape functions as furnished by classical Lagrange and hierarchical elements are enabled in the context of $hp$-FEM analysis. As this approach leads to additional DOFs and an indefinite stiffness matrix, a hybridization is suggested. The continuity of the moment tensor is broken across element interfaces and reinforced weakly by a Lagrange multiplier living only on the element edges. Then, the DOFs related to the moment tensor can be condensed element-wise, leaving only displacements and rotations along element edges as unknowns.

The mixed-hybrid weak form was consistently derived from the strong form, formulated coordinate-free based on the TDC. All mechanically relevant Neumann and Dirichlet boundary conditions for the Kirchhoff--Love shells and the hybridization terms are consistently deduced during this process. Although mixed formulations enable individual ansatz orders for the individual fields involved, the current formulation achieves optimal convergence rates even for the most convenient choice of the same order for the moments and displacements (and Lagrange multiplier), thus resulting in a usual isoparametric FEM. Hence, there is no need for any special FEM spaces (such as Taylor--Hood or Raviart--Thomas elements, FEM-variants based on the de~Rham complex, etc.), which enables a straightforward implementation in standard FEM programs.

For the numerical examples, some classical test cases are replicated and new test cases are proposed, featuring smooth solutions and thus enabling higher-order convergence rates in the numerical analyses. Different error measures are considered to demonstrate optimal higher-order convergence rates for the presented method. In particular, $\mathcal{L}^2$-errors, residual errors, and stored energy errors are investigated. For the new test cases, reference global scalar values such as selected displacements and stored elastic energies are provided, offering the opportunity to verify future shell implementations (without necessarily implementing the other, more advanced error measures). We observe optimal convergence rates of $\mathcal{O}(p+1)$ not only for the displacements but also for the components of the moment tensor, as they are also primary variables in the present formulation. This observation is an additional advantage compared to displacement-based formulations, where the moment tensor is obtained in a post-processing step involving differentiation.

Future research shall focus on extending this Kirchhoff--Love shell formulation to other FEM technologies. As the governing equations are formulated in terms of the TDC, they also apply to \emph{implicit} geometry descriptions based on level sets, enabling the use of fictitious domain methods such as the Trace FEM and the Bulk Trace FEM \cite{Dziuk_2013a, Olshanskii_2017a, Fries_2020a, Schoellhammer_2020a}, following previous works in \cite{Fries_2023a, Kaiser_2024a, Kaiser_2024b}. As only $C^0$-continuity is required, appropriate FEM spaces are also more straightforward to construct for those methods. Another field of application is seen in IGA in the context of general multi-patch geometries of shells, addressing the $C^0$-continuity across the NURBS-patches.

%% file: Acknowledgments.tex
The authors would like to thank Günther Of for valuable discussions and insightful comments related to this work.

%% file: Appendix.tex
\section{Derivation of the mixed weak form}\label{app: Derivation of the mixed weak form}
The mixed weak form is derived from the strong form of the governing equations in the usual manner: (i) multiplication of the strong form with test functions, (ii) integration over the domain, and (iii) application of integral theorems to shift derivatives, thus, obtaining a weak form that is suitable for FEM analyses using $C^0$-continuous shape functions. In the following, this is subsequently done for Eqs.~(\ref{eq: strong form equation 1 repeat}), (\ref{eq: strong form equation 2 repeat}), and the weak enforcement of continuity of $m_{\vek{t}}$ across elements.

Starting with equation, Eq.~(\ref{eq: strong form equation 1 repeat}), after multiplication with test functions $\mat{V}_{\mat{m}_\Gamma}$ and integration over the domain $\Gamma$, we obtain
\begin{equation*}
    \int_{\Gamma} -\mat{V}_{\mat{m}_\Gamma} \FrobProd \vek{\varepsilon}_{\Gamma,\text{Bend}}(\mat{m}_\Gamma) + \mat{V}_{\mat{m}_\Gamma} \FrobProd \vek{\varepsilon}_{\Gamma,\text{Bend}}(\vek{u}) \, \text{d}\Gamma = 0,
\end{equation*}
where $\FrobProd$ refers to the Frobenius inner product. With $\vek{\varepsilon}_{\Gamma,\text{Bend}}(\vek{u})$ from Eq.~(\ref{eq: strain tensor bending}) and some reformulations, we get
\begin{equation}
    \label{AppendixEqA}
    \int_{\Gamma} -\mat{V}_{\mat{m}_\Gamma} \FrobProd \vek{\varepsilon}_{\Gamma,\text{Bend}}(\mat{m}_\Gamma) + \mat{V}_{\mat{m}_\Gamma} \FrobProd \big( \mat{H} \cdot \nabla^\text{dir}_\Gamma \vek{u} \big) -\underbrace{\big(\mat{P} \cdot \mat{V}_{\mat{m}_\Gamma} \cdot \mat{P}\big) \FrobProd \nabla^\text{dir}_\Gamma \big( (\nabla^\text{dir}_\Gamma\vek{u})^\text{T} \cdot \vek{n}\big)}_{(\text{A})} \, \text{d}\Gamma = 0.
\end{equation}
The integral theorem from Eq.~(\ref{eq: integral theorem 2}) is applied on term (A), resulting in
\begin{equation}
    \label{AppendixEqB}
    \begin{split}
    &-\int_{\Gamma} \big(\mat{P} \cdot \mat{V}_{\mat{m}_\Gamma} \cdot \mat{P}\big) \FrobProd \nabla^\text{dir}_\Gamma  \big( (\nabla^\text{dir}_\Gamma\vek{u})^\text{T} \cdot \vek{n} \big) \, \text{d}\Gamma
    \\=& \int_{\Gamma} \text{div}_\Gamma \big(\mat{P} \cdot \mat{V}_{\mat{m}_\Gamma} \cdot \mat{P}\big) \ScalProd \big[ (\nabla^\text{dir}_\Gamma\vek{u})^\text{T} \cdot \vek{n} \big] \, \text{d}\Gamma - \int_{\partial\Gamma} \big(\mat{V}_{\mat{m}_\Gamma} \cdot \vek{q}\big) \ScalProd \big[ (\nabla^\text{dir}_\Gamma\vek{u}) ^\text{T} \cdot \vek{n} \big] \, \text{d}\partial\Gamma.
    \end{split}
\end{equation}
The boundary term is modified as
\begin{equation}
    \label{AppendixEqC}
    \begin{split}
        &- \int_{\partial\Gamma} \big(\mat{V}_{\mat{m}_\Gamma} \cdot \vek{q}\big) \ScalProd \big[ (\nabla^\text{dir}_\Gamma\vek{u}) ^\text{T} \cdot \vek{n} \big] \, \text{d}\partial\Gamma
        \\=& \int_{\partial\Gamma} \big( \mat{V}_{\mat{m}_\Gamma} \cdot \vek{q}  \big) \ScalProd \big( \omega_{\vek{q}} \cdot \vek{t} + \omega_{\vek{t}} \cdot \vek{q} \big) \, \text{d}\partial\Gamma
        \\=& \int_{\partial\Gamma} m_{\vek{q}}(\mat{V}_{\mat{m}_\Gamma}) \cdot \omega_{\vek{q}}(\vek{u})
    		+ m_{\vek{t}}(\mat{V}_{\mat{m}_\Gamma}) \cdot \omega_{\vek{t}} \, \text{d}\partial\Gamma.
    \end{split}
\end{equation}
and, inserting this into Eq.~(\ref{AppendixEqB}), then (\ref{AppendixEqA}), finally gives the weak form of Eq.~(\ref{eq: weak form equation 1}).

Continuing with the second equation, Eq.~(\ref{eq: strong form equation 2 repeat}) is multiplied with the test function $\vek{v}_{\vek{u}}$ and is integrated over the domain $\Gamma$,
\begin{multline*}
    -\int_{\Gamma}
    \underbrace{\vek{v}_{\vek{u}} \ScalProd \vek{n} \cdot \text{div}_\Gamma \big(\mat{P} \cdot \text{div}_\Gamma \mat{m}_\Gamma\big)}_{(\text{B})}
    + \vek{v}_{\vek{u}} \ScalProd \big( \mat{H} \cdot \text{div}_\Gamma \mat{m}_\Gamma \big)
    + \underbrace{\vek{v}_{\vek{u}} \ScalProd \text{div}_\Gamma \big(\mat{H} \cdot \mat{m}_\Gamma\big)}_{(\text{C})}
    \\ + \underbrace{\vek{v}_{\vek{u}} \ScalProd \text{div}_\Gamma \tilde{\mat{n}}_\Gamma(\vek{u})}_{(\text{D})} \, \text{d}\Gamma
    = \int_{\Gamma} \vek{v}_{\vek{u}} \ScalProd \vek{f} \, \text{d}\Gamma.
\end{multline*}
After applying the integral theorems Eq.~(\ref{eq: integral theorem 1}) on term (B), and Eq.~(\ref{eq: integral theorem 2}) on terms (C) and (D), we obtain
\begin{equation*}
    \begin{split}
        &-\int_{\Gamma} \vek{v}_{\vek{u}} \ScalProd \vek{n} \cdot \text{div}_\Gamma \big(\mat{P} \cdot \text{div}_\Gamma \mat{m}_\Gamma\big) \, \text{d}\partial\Gamma
        \\=& \int_{\Gamma} \nabla_\Gamma \big( \vek{v}_{\vek{u}} \ScalProd \vek{n} \big) \ScalProd \text{div}_\Gamma \mat{m}_\Gamma \, \text{d}\Gamma 
        - \int_{\partial\Gamma} \big( \vek{v}_{\vek{u}} \ScalProd \vek{n} \big) \cdot \big(\mat{P} \cdot \text{div}_\Gamma \mat{m}_\Gamma\big) \ScalProd \vek{q} \, \text{d}\partial\Gamma,
    \end{split}
\end{equation*}
\begin{equation*}
    -\int_{\Gamma} \vek{v}_{\vek{u}} \ScalProd \text{div}_\Gamma \big(\mat{H} \cdot \mat{m}_\Gamma\big) \, \text{d}\Gamma
    = \int_{\Gamma} \nabla^\text{dir}_\Gamma\vek{v}_{\vek{u}} \FrobProd \big(\mat{H} \cdot \mat{m}_\Gamma\big)  \, \text{d}\Gamma 
    - \int_{\partial\Gamma} \vek{v}_{\vek{u}} \ScalProd \big(\mat{H} \cdot \mat{m}_\Gamma \cdot \vek{q} \big) \, \text{d}\partial\Gamma,
\end{equation*}
\begin{equation*}
    -\int_{\Gamma} \vek{v}_{\vek{u}} \ScalProd \text{div}_\Gamma \tilde{\mat{n}}_\Gamma(\vek{u}) \, \text{d}\Gamma
    = \int_{\Gamma} \nabla^\text{dir}_\Gamma\vek{v}_{\vek{u}} \FrobProd \tilde{\mat{n}}_\Gamma(\vek{u}) \, \text{d}\Gamma 
    - \int_{\partial\Gamma} \vek{v}_{\vek{u}} \ScalProd \big( \tilde{\mat{n}}_\Gamma(\vek{u}) \cdot \vek{q} \big) \, \text{d}\partial\Gamma.
\end{equation*}
The boundary terms are reformulated in terms of the effective boundary forces given by Eqs.~(\ref{eq: effective boundary force in t})-(\ref{eq: effective boundary force in n})
\begin{equation*}
    \begin{split}
            &- \int_{\partial\Gamma} \big( \vek{v}_{\vek{u}} \ScalProd \vek{n} \big) \cdot \big(\mat{P} \cdot \text{div}_\Gamma \mat{m}_\Gamma\big) \ScalProd \vek{q}
            + \vek{v}_{\vek{u}} \ScalProd \big(\mat{H} \cdot \mat{m}_\Gamma \cdot \vek{q} \big)
            + \vek{v}_{\vek{u}} \ScalProd \big( \tilde{\mat{n}}_\Gamma(\vek{u}) \cdot \vek{q} \big)
            \, \text{d}\partial\Gamma
            \\ =& - \int_{\partial\Gamma} \vek{v}_{\vek{u}} \ScalProd \big( p_{\vek{t}} \cdot \vek{t} + p_{\vek{q}} \cdot \vek{q} + p_{\vek{n}} \cdot \vek{n} \big) \, \text{d}\partial\Gamma
            \\ =& - \int_{\partial\Gamma} \vek{v}_{\vek{u}} \ScalProd \big[ \big( \tilde{p}_{\vek{t}} - \mat{H} \cdot \vek{t} \ScalProd \vek{t} \cdot m_{\vek{q}} \big) \cdot \vek{t} + \big( \tilde{p}_{\vek{q}} - \mat{H} \cdot \vek{t} \ScalProd \vek{q} \cdot m_{\vek{q}} \big) \cdot \vek{q} + \big( \tilde{p}_{\vek{n}} - \nabla_{\Gamma} m_{\vek{q}} \ScalProd \vek{t} \big) \cdot \vek{n} \big] \, \text{d}\partial\Gamma
            \\ =& \int_{\partial\Gamma} - \vek{v}_{\vek{u}} \ScalProd \tilde{\vek{p}} + \vek{v}_{\vek{u}} \ScalProd \big( \mat{H} \cdot \vek{t} \cdot m_{\vek{q}} \big) + \underbrace{\big( \vek{v}_{\vek{u}}  \ScalProd \vek{n} \big) \cdot \big( \nabla_{\Gamma} m_{\vek{q}} \ScalProd \vek{t} \big)}_{\text{(E)}} \, \text{d}\partial\Gamma.
    \end{split}
\end{equation*}
Integration by parts of the term (E) gives
\begin{equation*}
    \int_{\partial\Gamma} \big( \vek{v}_{\vek{u}}  \ScalProd \vek{n} \big) \cdot \big( \nabla_{\Gamma} m_{\vek{q}} \ScalProd \vek{t} \big) \, \text{d}\partial\Gamma = - \int_{\partial\Gamma} \nabla_{\Gamma} \big(\vek{v}_{\vek{u}}  \ScalProd \vek{n} \big) \ScalProd \vek{t} \cdot  m_{\vek{q}} \, \text{d}\partial\Gamma + \sum_{i=1}^{n_{C}} \vek{v}_{\vek{u}}|_{C_i} \ScalProd \vek{n}|_{C_i} \cdot [m_{\vek{q}}]_{C_i^+}^{C_i^-}.
\end{equation*}
Some further reformulations and a simplification due to the restriction of the function space, i.e., $\vek{v}_{\vek{u}} = \vek{0} \, \text{on} \, \partial\Gamma_{\text{D},\vek{u}}$, are carried out
\begin{equation*}
    \begin{split}
        & \int_{\partial\Gamma} - \vek{v}_{\vek{u}} \ScalProd \tilde{\vek{p}} + \vek{v}_{\vek{u}} \ScalProd \big( \mat{H} \cdot \vek{t} \cdot m_{\vek{q}} \big) - \nabla_{\Gamma} \big(\vek{v}_{\vek{u}}  \ScalProd \vek{n} \big) \ScalProd \vek{t} \cdot  m_{\vek{q}} \, \text{d}\partial\Gamma + \sum_{i=1}^{n_{C}} \vek{v}_{\vek{u}}|_{C_i} \ScalProd \vek{n}|_{C_i} \cdot [m_{\vek{q}}]_{C_i^+}^{C_i^-}
        \\=& \int_{\partial\Gamma} - \vek{v}_{\vek{u}} \ScalProd \tilde{\vek{p}} - (\nabla^\text{dir}_\Gamma \vek{v}_{\vek{u}})^\text{T} \cdot \vek{n} \ScalProd \vek{t} \cdot m_{\vek{q}} \, \text{d}\partial\Gamma - \sum_{i=1}^{n_{C}} \vek{v}_{\vek{u}}|_{C_i} \ScalProd \vek{n}|_{C_i} \cdot F_{C}
        \\=& \int_{\partial\Gamma} \omega_{\vek{q}}(\vek{v}_{\vek{u}}) \cdot m_{\vek{q}}(\mat{m}_\Gamma) \, \text{d}\partial\Gamma - \int_{\partial\Gamma_{\text{N},\vek{u}}}\!\!\!\!\!\!\! \vek{v}_{\vek{u}} \ScalProd \hat{\tilde{\vek{p}}} \, \text{d}\partial\Gamma - \sum_{i=1}^{n_{C,\text{N}}} \vek{v}_{\vek{u}}|_{C_i} \ScalProd \vek{n}|_{C_i} \cdot \hat{F}_{C}.
    \end{split}
\end{equation*}
These boundary terms are inserted back into the original equation to give the desired weak form of Eq.~(\ref{eq: weak form equation 2}).

For the last equation of the weak from, referred to in Eq.~(\ref{eq: weak form equation 3}), Eq.~(\ref{eq: boundary conditions m}) is multiplied with the test function $v_{\omega_{\vek{t}}}$ and is integrated over the Neumann boundary $\partial\Gamma_{\text{N},\omega}$ to obtain
\begin{equation*}
    \int_{\partial\Gamma_{\text{N},\omega}}\!\!\!\!\!\!\! v_{\omega_{\vek{t}}} \cdot m_{\vek{t}}(\mat{m}_\Gamma) \, \text{d}\partial\Gamma =  \int_{\partial\Gamma_{\text{N},\omega}}\!\!\!\!\!\!\! v_{\omega_{\vek{t}}} \cdot \hat{m}_{\vek{t}} \, \text{d}\partial\Gamma.
\end{equation*}

\section{Entries of the local stiffness matrix and right hand side}\label{app: Entries of the local stiffness matrix}
After evaluating the discretized and hybridized weak form from Eqs.~(\ref{eq: hyb weak form equation 1})-(\ref{eq: hyb weak form equation 3}) under consideration of the discrete test and trial spaces of Eqs.~(\ref{eq: h function space m})-(\ref{eq: h function space vomega}), a system of equations in the form of $\mat{K} \cdot [\underline{\underline{\vek{m_\Gamma}}}, \underline{\vek{u}}, \vek{\omega_{\vek{t}}}]^\text{T} = \vek{b}$ results, with $\mat{K}$ as the global stiffness matrix and $\vek{b}$ as the global right hand side. For the sake of accessibility and replicability, the individual entries of the stiffness matrix and right hand side are given in this appendix. Due to the subsequent static condensation on an element-wise, hence, local level, the element versions of the stiffness matrix $\mat{K}^\text{el}$ and right hand side $\vek{b}^\text{el}$ are presented. Related to Eq.~\ref{eq: stiffness matrix and right hand side}, the individual entries are computed as
\allowdisplaybreaks
\begin{align*}
    \mat{K}_{\mat{m}_{ij}\mat{m}_{kl}}^\text{el}=&
    \big(1 - \frac{\delta_{ij}}{2}\big) \cdot \frac{24}{E \cdot t^3} \int_T \vek{N}_{\mat{m}_\Gamma} \cdot \vek{N}_{\mat{m}_\Gamma}^\text{T} \cdot \big[\delta_{kl} \cdot P_{ij} \cdot \nu - \delta_{ik} \cdot \delta_{jl} \cdot \big(1 + \nu\big)\big] \, \text{d}\Gamma,\\
    \begin{split}
        \mat{K}_{\mat{m}_{ij}\vek{u}_{k}}^\text{el}=&
        \big(1 - \frac{\delta_{ij}}{2}\big) \cdot \big[\int_T \vek{N}_{\mat{m}_\Gamma} \cdot \big(\partial^\Gamma_i\vek{N}_{\vek{u}}^\text{T} \cdot H_{jk} + \partial^\Gamma_j\vek{N}_{\vek{u}}^\text{T} \cdot H_{ik}\big)\\&
        + n_{k} \sum_{l,m=1}^3 \partial^\Gamma_l\vek{N}_{\mat{m}_\Gamma} \cdot \partial^\Gamma_m\vek{N}_{\vek{u}}^\text{T} \cdot (P_{im} \cdot P_{jl} + P_{jm} \cdot P_{il}\big)\\&
        + \vek{N}_{\mat{m}_\Gamma} \cdot \sum_{l,m=1}^3 \partial^\Gamma_l\vek{N}_{\vek{u}}^\text{T} \cdot \big(P_{im} \cdot \partial^\Gamma_m P_{jl} + P_{jl} \cdot \partial^\Gamma_m P_{im} + P_{jm} \cdot \partial^\Gamma_m P_{il} + P_{il} \cdot \partial^\Gamma_m P_{jm}\big) \, \text{d}\Gamma\\&
        - \int_{\partial T} \vek{N}_{\mat{m}_\Gamma} \cdot n_{k} \cdot \big(t_{i} \cdot q_{j} + t_{j} \cdot q_{i}\big) \cdot \sum_{l=1}^3 \partial^\Gamma_l\vek{N}_{\vek{u}}^\text{T} \cdot t_{l} \, \text{d}\Psi \big],
    \end{split}\\
    \mat{K}_{\mat{m}_{ij}\omega}^\text{el}=&
    \big(2 - \delta_{ij}\big) \int_{\partial T} \vek{N}_{\mat{m}_\Gamma} \cdot \vek{N}_{\omega_{\vek{t}}}^\text{T} \cdot q_{i} \cdot q_{j} \, \text{d}\Psi,\\
    \begin{split}
        \mat{K}_{\vek{u}_{i}\mat{m}_{jk}}^\text{el}=&
        \big(1 - \frac{\delta_{ij}}{2}\big) \cdot \big[\int_T \big(\partial^\Gamma_j\vek{N}_{\vek{u}} \cdot H_{ik} + \partial^\Gamma_k\vek{N}_{\vek{u}}^\text{T} \cdot H_{ij}\big) \cdot \vek{N}_{\mat{m}_\Gamma}^\text{T}\\&
        + n_{i} \big(\partial^\Gamma_j\vek{N}_{\vek{u}} \cdot \partial^\Gamma_k\vek{N}_{\mat{m}_\Gamma}^\text{T} + \partial^\Gamma_k\vek{N}_{\vek{u}} \cdot \partial^\Gamma_j\vek{N}_{\mat{m}_\Gamma}^\text{T}\big) \, \text{d}\Gamma\\&
        - \int_{\partial T} \big(\sum_{l=1}^3 \partial^\Gamma_l\vek{N}_{\vek{u}} \cdot t_{l}\big) \cdot \vek{N}_{\mat{m}_\Gamma}^\text{T} \cdot n_{i} \cdot \big(t_{j} \cdot q_{k} + t_{k} \cdot q_{j}\big) \, \text{d}\Psi \big],
    \end{split}\\
    \begin{split}
        \mat{K}_{\vek{u}_{i}\vek{u}_{j}}^\text{el}=&
        \frac{E \cdot t}{2 - 2 \nu^2} \int_T \sum_{k=1}^3 P_{ik} \cdot \big[\big(\partial^\Gamma_j\vek{N}_{\vek{u}} \cdot \partial^\Gamma_k\vek{N}_{\vek{u}}^\text{T} + \delta_{jk} \sum_{l=1}^3\partial^\Gamma_l\vek{N}_{\vek{u}} \cdot \partial^\Gamma_l\vek{N}_{\vek{u}}^\text{T}\big) \cdot \big(1 - \nu\big)\\&
        + 2 \partial^\Gamma_k\vek{N}_{\vek{u}} \cdot \partial^\Gamma_j\vek{N}_{\vek{u}}^\text{T} \cdot \nu \big] \, \text{d}\Gamma,
    \end{split}\\
    \mat{K}_{\vek{u}_{i}\omega}^\text{el}=&
    \mat{0},\\
    \mat{K}_{\omega\mat{m}_{ij}}^\text{el}=&
    \big(2 - \delta_{ij}\big) \int_{\partial T} \vek{N}_{\omega_{\vek{t}}} \cdot \vek{N}_{\mat{m}_\Gamma}^\text{T} \cdot q_{i} \cdot q_{j} \, \text{d}\Psi,\\
    \mat{K}_{\omega\vek{u}_{i}}^\text{el}=&
    \mat{0},\\
    \mat{K}_{\omega\omega}^\text{el}=&
    \mat{0},\\
    \vek{b}_{\mat{m}_{ij}}=&
    \vek{0},\\
    \vek{b}_{\vek{u}_{i}}=&
    \int_T \vek{N}_{\vek{u}} \cdot f_{i}^h \, \text{d}\Gamma
    + \int_{\partial T_{\text{N},\vek{u}}}\!\!\!\!\!\!\! \vek{N}_{\vek{u}} \cdot \hat{\tilde{p}}_{i}^h \, \text{d}\partial\Gamma + \vek{N}_{\vek{u}}|_{C_{\partial T}} \cdot n_{i}|_{C_{\partial T}} \cdot \hat{F}_{C}^h,\\
    \vek{b}_{\omega}=&
    \int_{\partial T_{\text{N},\omega}}\!\!\!\!\!\!\! \vek{N}_{\omega} \cdot \hat{m}_t^h \, \text{d}\partial\Gamma,
\end{align*}
where $T$ is the element under consideration and $\partial T$ the corresponding edges, $\delta_{ij}$ is the Kronecker delta, $P_{ij}$, $H_{ij}$, $n_{i}$, $t_{i}$, and $q_{i}$ are certain entries of $\mat{P}$, $\mat{H}$, $\vek{n}$, $\vek{t}$, and $\vek{q}$ respectively and $\vek{N}_{\mat{m}_\Gamma}$, $\vek{N}_{\vek{u}}$, and $\vek{N}_{\omega}$ are the sets of the local basis functions of the three primary fields. For the Neumann boundary conditions, we have $\partial T_{\text{N},\vek{u}} = \partial T \cap \partial\Gamma_{\text{N},\vek{u}}$, $\partial T_{\text{N},\omega} = \partial T \cap \partial\Gamma_{\text{N},\omega}$, and $C_{\partial T} = C_{i} \, \text{ if } \, \vek{x}_{C_i,\text{N}} \in \partial T$ where a Neumann boundary and corner coexist on an element edge. Note that the jump operator terms, concerning the entries $\mat{K}_{\mat{m}_{ij}\omega}^\text{el}$ and $\mat{K}_{\omega\mat{m}_{ij}}^\text{el}$, are considered in this element-local formulation by assigning the proper signs to the corresponding shape functions.

\section{Analytical solution of the extruded arc}\label{app: Analytical solution of the sliced arc}
\begin{figure}
    \centering
    
    \subfigure[setup]
    {\includegraphics[width=0.4\textwidth]{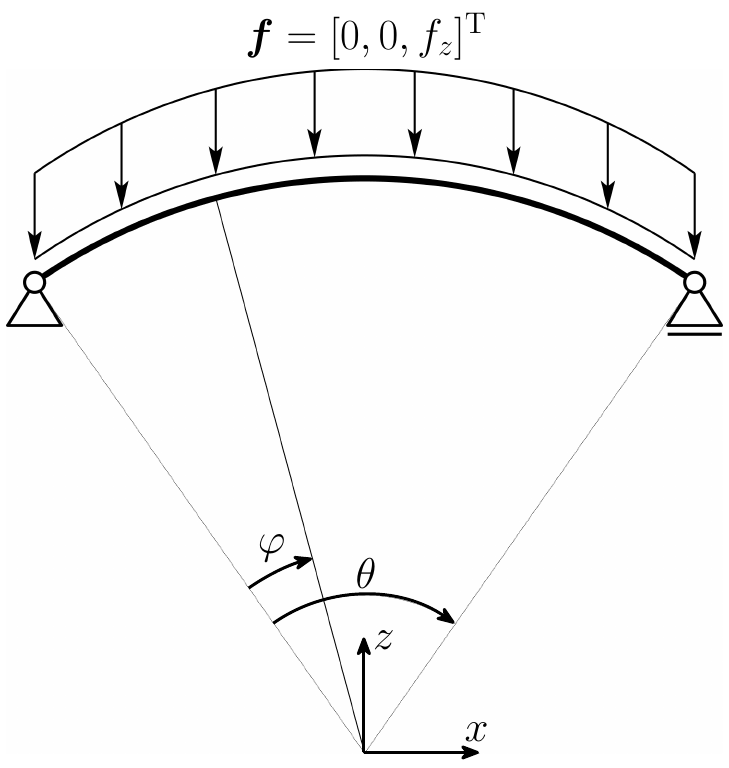}\label{fig: sliced arc 2d view}}\qquad
    \subfigure[free body diagram]
    {\includegraphics[width=0.4\textwidth]{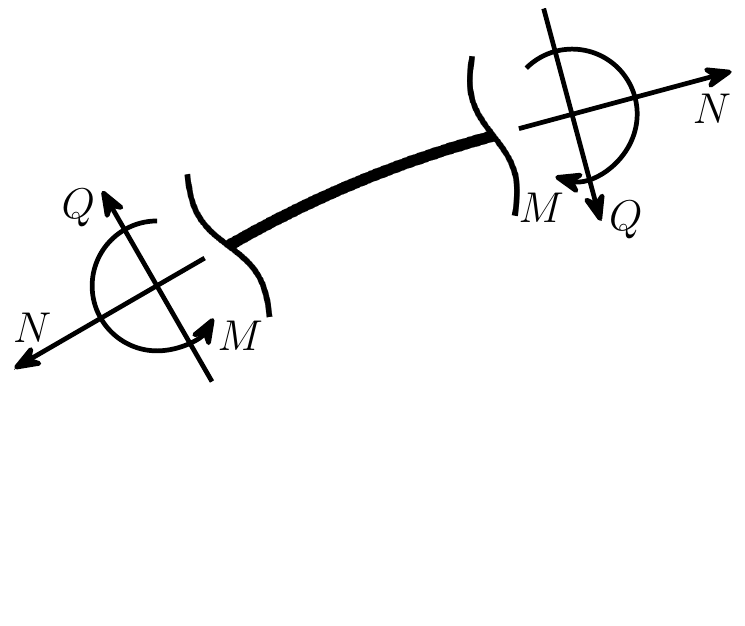}\label{fig: sliced arc free body diagram}}
    
    \caption{\label{fig: sliced arc situation}Visualization of the beam case related to the extruded arc in Section \ref{sec: Sliced arc}, (a) the general setup of the test case and (b) a free body diagram highlighting the inner force quantities.}
\end{figure}

The test case in Section \ref{sec: Sliced arc} is based on an extrusion of the beam test case shown in Fig.~\ref{fig: sliced arc situation}. The analytic solutions of the force and displacement quantities of the curved beam are given next:
\begin{equation*}
N(\varphi) = R \cdot L_y \cdot f_z \sin(\psi - \varphi) \cdot (\psi - \varphi),
\end{equation*}
\begin{equation*}
Q(\varphi) = R \cdot L_y \cdot f_z \cos(\psi - \varphi) \cdot (\psi - \varphi),
\end{equation*}
\begin{equation*}
M(\varphi) = R^2 \cdot L_y \cdot f_z \cdot \big(- \cos(\psi - \varphi) + \cos(\psi) - \psi \sin(\psi - \varphi) + \psi \sin(\psi) + \varphi \sin(\psi - \varphi) \big),
\end{equation*}
\begin{equation*}
\begin{aligned}
u(\varphi) = & \frac{R^2 \cdot L_y \cdot f_z}{16 EA \sin(\psi)} \big[ 4 \cos(\psi - 2\varphi) - 12 \cos(\psi) + 8 \cos(\psi + \varphi) - 4 \cos(3\psi -  2\varphi) \\
    & - 8 \cos(3\psi - \varphi) + 12 \cos(3\psi) + \theta \cdot \big( 8 \sin(\psi - \varphi) - 10 \sin(\psi) + 2 \sin(\psi + \varphi) \\
    & - 6 \sin(3\psi - \varphi) + 6 \sin(3\psi) \big) + \theta^2 \cdot \big( -2 \cos(\psi - \varphi) + \cos(\psi) + \cos(\psi + \varphi) \\
    & + \cos(3\psi - \varphi) - \cos(3\psi) \big) + \theta \cdot \varphi \cdot \big( 4 \cos(\psi - \varphi) - 2 \cos(\psi + \varphi) - 2 \cos(3\psi - \varphi) \big) \\
    & + \varphi \cdot \big( 32 \sin(\psi) + 4 \sin(\psi + \varphi) + 4 \sin(3\psi - \varphi) \big) \big] \\
    & + \frac{R^4 \cdot L_y \cdot f_z}{16 EI \sin(\psi)} \big[ 5 \cos(\psi - 2\varphi) - 9 \cos(\psi) + 4 \cos(\psi + \varphi) - 5 \cos(3\psi - 2\varphi) \\
    & - 4 \cos(3\psi - \varphi) + 9 \cos(3\psi) + \theta \cdot \big( \sin(\psi - 2\varphi) + 4 \sin(\psi - \varphi) - 5 \sin(\psi) \\
    & - \sin(3\psi - 2\varphi) - 4 \sin(3\psi - \varphi) + 5 \sin(3\psi) \big) + \theta^2 \cdot \big( -2 \cos(\psi - \varphi) + \cos(\psi) \\
    & + \cos(\psi + \varphi) + \cos(3\psi - \varphi) - \cos(3\psi) \big) + \theta \cdot \varphi \cdot \big( 4 \cos(\psi - \varphi) - 2 \cos(\psi + \varphi) \\
    & - 2 \cos(3\psi - \varphi) \big) + \varphi \cdot \big( -2 \sin(\psi - 2\varphi) + 16 \sin(\psi) + 4 \sin(\psi + \varphi) \\
    & + 2 \sin(3\psi - 2\varphi)  + 4 \sin(3\psi - \varphi) \big) \big],
\end{aligned}
\end{equation*}
\begin{equation*}
\begin{aligned}
w(\varphi) = & \frac{R^2 \cdot L_y \cdot f_z}{16 EA \sin(\psi)} \big[ 4 \sin(\psi - 2\varphi) + 4 \sin(\psi) - 8 \sin(\psi + \varphi) - 4 \sin(3\psi - 2\varphi) \\
    & - 8 \sin(3\psi - \varphi) + 12 \sin(3\psi) + \theta \cdot \big( -8 \cos(\psi - \varphi) + 6 \cos(\psi) + 2 \cos(\psi + \varphi) \\
    & + 6 \cos(3\psi - \varphi) - 6 \cos(3\psi) \big)  + \theta^2 \cdot \big( -2 \sin(\psi - \varphi) + 3 \sin(\psi) - \sin(\psi + \varphi) \\
    & + \sin(3\psi - \varphi) - \sin(3\psi) \big) + \theta \cdot \varphi \cdot \big( 4 \sin(\psi - \varphi) + 16 \sin(\psi) + 2 \sin(\psi + \varphi) \\
    & - 2 \sin(3\psi - \varphi) \big) + \varphi \cdot \big( -4 \cos(3\psi - \varphi) + 4 \cos(\psi + \varphi) \big) - 16 \varphi^2 \sin(\psi) \big] \\
    & + \frac{R^4 \cdot L_y \cdot f_z}{16 EI \sin(\psi)} \big[ -4 \sin(\psi + \varphi) - \sin(\psi) + 9 \sin(3\psi) + 5 \sin(\psi - 2\varphi) \\
    & - 4 \sin(3\psi - \varphi) - 5 \sin(3\psi - 2\varphi) + \theta \cdot \big( -4 \cos(\psi - \varphi) - \cos(\psi - 2\varphi) \\
    & + 4 \cos(3\psi - \varphi) + \cos(3\psi - 2\varphi) + 5 \cos(\psi) - 5 \cos(3\psi) \big) + \theta^2 \cdot \big( -\sin(\psi + \varphi) \\
    & + 3 \sin(\psi) - \sin(3\psi) - 2 \sin(\psi - \varphi) + \sin(3\psi - \varphi) \big) + \theta \cdot \varphi \cdot \big( 2 \sin(\psi + \varphi) \\
    & + 4 \sin(\psi) + 4 \sin(\psi - \varphi) - 2 \sin(3\psi - \varphi) \big) + \varphi \cdot \big( 2 \cos(\psi - 2\varphi) - 4 \cos(3\psi - \varphi) \\
    & - 2 \cos(3\psi - 2\varphi) + 4 \cos(\psi + \varphi) \big) - 4 \varphi^2 \sin(\psi) \big],
\end{aligned}
\end{equation*}
with $\psi = \frac{\theta}{2}$, $A = t \cdot L_y$, $I = \frac{t^3 \cdot L_y}{12}$, and $\varphi(\vek{x}) = \frac{\theta + \pi}{2} - \arctan({\frac{z}{x}})$.
